\documentclass[twocolumn,useAMS,usenatbib]{mn2e} \usepackage{natbib}

\usepackage{amsmath,latexsym,amssymb} \usepackage{epsfig,graphicx} \topmargin-1cm
\usepackage{epstopdf}
\usepackage{float}
\usepackage{fixltx2e}
\usepackage{color}
\usepackage{verbatim}
\usepackage{url}
\graphicspath{ {./figs_v2png/}}
\DeclareGraphicsExtensions{.eps,.png}
\def\reff@jnl#1{{\rm#1\/}}

\def\aj{\reff@jnl{AJ}}                  
\def\araa{\reff@jnl{ARA\&A}}            
\def\apj{\reff@jnl{ApJ}}                        
\def\apjl{\reff@jnl{ApJ}}               
\def\apjs{\reff@jnl{ApJS}}              
\def\apss{\reff@jnl{Ap\&SS}}            
\def\aap{\reff@jnl{A\&A}}               
\def\aapr{\reff@jnl{A\&A~Rev.}}         
\def\aaps{\reff@jnl{A\&AS}}             
\def\baas{\reff@jnl{BAAS}}              
\def\jcap{\reff@jnl{JCAP}}              
\def\jrasc{\reff@jnl{JRASC}}            
\def\memras{\reff@jnl{MmRAS}}           
\def\mnras{\reff@jnl{MNRAS}}            
\def\physrep{\reff@jnl{Phys.Rep.}}
\def\pra{\reff@jnl{Phys.Rev.A}}         
\def\prb{\reff@jnl{Phys.Rev.B}}         
\def\prc{\reff@jnl{Phys.Rev.C}}         
\def\prd{\reff@jnl{Phys.Rev.D}}         
\def\prl{\reff@jnl{Phys.Rev.Lett}}      
\def\pasp{\reff@jnl{PASP}}              
\def\pasj{\reff@jnl{PASJ}}              
\def\skytel{\reff@jnl{S\&T}}            
\def\solphys{\reff@jnl{Solar~Phys.}}    
\def\sovast{\reff@jnl{Soviet~Ast.}}     
\def\ssr{\reff@jnl{Space~Sci.Rev.}}     
\def\nat{\reff@jnl{Nature}}             

\newcommand{\hmpc}{\ensuremath{h^{-1}\mathrm{Mpc}}}
\newcommand{\hkpc}{\ensuremath{h^{-1}\mathrm{kpc}}}
\newcommand{\hMsun}{\ensuremath{h^{-1}M_{\odot}}}

\newcommand{\Msun}{M_{\odot}}

\newcommand{\ia}{intrinsic alignments}

\newcommand{\wgp}{\ensuremath{w_{g+}}}
\newcommand{\wdp}{\ensuremath{w_{\delta+}}}
\newcommand{\wgx}{\ensuremath{w_{g\times}}}

\newcommand{\mpch}{\ensuremath{h^{-1}\text{Mpc}}}

\newcommand{\beq}{\begin{equation}}
\newcommand{\eeq}{\end{equation}}
\newcommand{\beqa}{\begin{eqnarray}}
\newcommand{\eeqa}{\end{eqnarray}}

\title[MB-II intrinsic alignments]{Intrinsic alignments of galaxies in the MassiveBlack-II simulation: analysis of two-point statistics}
\author[Tenneti et al.]
{Ananth Tenneti$^1$\thanks{\tt vat@andrew.cmu.edu},
Sukhdeep Singh$^1$\thanks{\tt sukhdees@andrew.cmu.edu},
Rachel Mandelbaum$^1$\thanks{\tt rmandelb@andrew.cmu.edu},
Tiziana Di Matteo$^1$\thanks{\tt tiziana@phys.cmu.edu},
\newauthor Yu Feng$^1$, 
Nishikanta Khandai$^{2,3}$
\\$^1$McWilliams Center for Cosmology, Department of Physics, Carnegie Mellon University, Pittsburgh, PA 15213, USA
\\$^2$Department of Physics, Brookhaven National Laboratory, Upton, NY 11973, USA
\\$^3$School of Physical Sciences, National Institute of Science Education and Research, Bhubaneswar, Odisha 751005, India}

\date{\today}
\begin{document}
\maketitle

\begin{abstract}
  The intrinsic alignment of galaxies with the large-scale density
  field is an important astrophysical contaminant in upcoming weak
  lensing surveys. We present detailed measurements of the galaxy intrinsic
  alignments and associated ellipticity-direction (ED) and projected
  shape ($w_{g+}$) correlation functions for galaxies in the
  cosmological hydrodynamic MassiveBlack-II (MB-II) simulation. We
  carefully assess the effects on galaxy shapes, misalignment of the stellar component with the dark matter shape and
  two-point statistics of iterative weighted (by mass and luminosity) definitions of the (reduced and unreduced) inertia
  tensor. We find that iterative procedures must be adopted for a
  reliable measurement of the reduced tensor but that luminosity versus
  mass weighting has only negligible effects. Both ED and $w_{g+}$ correlations
  increase in amplitude with subhalo mass (in the range of $10^{10} -
  6.0\times 10^{14}\hMsun$), with a weak redshift dependence (from
  $z=1$ to $z=0.06$) at fixed mass. At $z \sim 0.3$, we predict
  a $w_{g+}$ that is in reasonable agreement with SDSS LRG measurements
  and that decreases in amplitude by a factor of $\sim 5$--18 for
  galaxies in the LSST survey. We also
  compared the intrinsic alignments of centrals and satellites, with
  clear detection of satellite radial alignments within their host
  halos. Finally, we show that $w_{g+}$ (using subhalos as tracers of density) and $w_{\delta+}$ (using dark
  matter density) predictions
  from the simulations agree with that of non-linear alignment models
  (NLA) at scales where the 2-halo term dominates in the correlations
  (and tabulate associated NLA fitting parameters). The 1-halo term
  induces a scale dependent bias at small scales which is not modeled
  in the NLA model.
\end{abstract}

\begin{keywords}
cosmology: theory -- methods: numerical -- hydrodynamics -- gravitational lensing: weak -- galaxies: star formation
\end{keywords}
\section{Introduction} \label{S:intro}

Upcoming cosmological surveys such as the Large Synoptic Survey Telescope
(LSST)\footnote{\url{http://www.lsst.org/lsst/}},
Euclid\footnote{\url{http://sci.esa.int/euclid/}}, and
WFIRST-AFTA\footnote{\url{http://wfirst.gsfc.nasa.gov}} have the
potential to 
constrain cosmological parameters such as the dark
energy equation of state to percent levels (or better) using weak gravitational
lensing. The sensitivity of weak gravitational lensing to both
luminous and dark matter
\citep{{2004PhRvD..70l3515B},{2004ApJ...600...17B},{2002PhRvD..65b3003H},{2010GReGr..42.2177H},{2004PhRvD..69h3514I},{2004ApJ...601L...1T}}
makes it a powerful way to probe the nature of dark matter, dark energy and
modified theories of gravity
\citep{{2006astro.ph..9591A},2013PhR...530...87W}. However, the
potential to constrain cosmological parameters to sub-percent levels
can only be realized if the systematic errors in lensing
surveys are even smaller than that.

An important astrophysical systematic that contaminates 
weak lensing measurements is the intrinsic alignment of
galaxies
\citep[e.g.,][]{{2000MNRAS.319..649H},{2000ApJ...545..561C},{2002MNRAS.335L..89J},{2004PhRvD..70f3526H}}. Weak
lensing analysis is based on the assumption that the intrinsic shapes
and orientations of galaxies are randomly aligned. In reality, the
galaxy shapes are correlated with each other and with the underlying
density field, mimicking the same coherent shape alignments that are
the signature of weak gravitational lensing. This systematic, called the intrinsic alignment of
galaxies, if ignored, can cause a deviation of $\sim$ $25\%$ when
estimating the dark energy equation of state parameter
\citep{2011A&A...527A..26J}. While several schemes for mitigating
intrinsic alignments have been proposed, such as nulling \citep{2008A&A...488..829J},
self-calibration \citep{2010ApJ...720.1090Z}, and joint modeling of cosmological
parameters and weak lensing \citep[e.g.,]{2005A&A...441...47K}, the methods that remove the least
amount of cosmological information often involve 
 modeling the intrinsic alignments as a function of scale,
redshift, luminosity and environment.

The complex nature of the physics of galaxy formation makes it very
difficult to model the intrinsic alignments analytically. Popular
analytic models include 
the linear alignment model \citep{2004PhRvD..70f3526H},
modifications of it based on the non-linear power spectrum
\citep{2007NJPh....9..444B}, and the halo model
\citep{2010MNRAS.402.2127S}, which makes assumptions about the
alignment of centrals and satellites. Numerical studies based on
$N$-body simulations have studied intrinsic alignments by populating
the halos with galaxies and assigning a misalignment angle
\citep{2006MNRAS.371..750H} or by using semi-analytic models
\citep{2013MNRAS.436..819J}. In general, methods designed to remove
intrinsic alignments from observational data
\citep{2008A&A...488..829J,
  2009A&A...507..105J,2007NJPh....9..444B,2010A&A...523A...1J,2012JCAP...05..041B}
are based on these models or require accurate redshift information
which leads to considerable loss of cosmological information. A
further understanding of intrinsic alignments requires the use of
cosmological numerical simulations that include the physics of galaxy
formation to validate the theoretical
predictions.

 Here, we make use of a large volume, high-resolution
cosmological hydrodynamic simulation, MassiveBlack-II (MB-II) 
\citep{2014arXiv1402.0888K} to directly study the intrinsic alignment
due to the stellar matter component in galaxies. Recent hydrodynamic
simulations of comparable volume that form galaxies include the
Horizon-AGN \citep{2014arXiv1402.1165D} and Illustris
\citep{2014Natur.509..177V}. In a previous paper
\citep{2014MNRAS.441..470T}, we studied the shapes of stellar matter
component in galaxies and their alignment with the shape of the host dark
matter subhalo using MassiveBlack-II. We extend this work further in
this paper, by studying the two-point correlation functions. This 
study allow us to both (a)  compare our results from MB-II with
observational measurements at high luminosity, to validate the use of
these simulations for intrinsic alignment studies; and (b) to predict
intrinsic alignment signals for lower luminosity galaxies that will be
used in upcoming weak lensing surveys.

The intrinsic alignments of galaxies in the simulation are based on
the shapes and orientations of stellar matter component in
galaxies. The shape of a galaxy is determined by the radial weighting
used for measuring the inertia tensor, 
and also the mass or luminosity weighting given to each star particle
while calculating the inertia tensor. We previously studied the
distributions of shapes determined by dark matter and stellar matter
component in galaxies using the unweighted inertia tensor by weighting
each star particle by its mass \citep{2014MNRAS.441..470T}. Using
$N$-body simulations, \cite{2012JCAP...05..030S} found radial
dependence in the axis ratios of the shapes of dark matter
halos. \cite{2012MNRAS.420.3303B} studied the axis ratios of dark
matter halos in $N$-body simulations using different definitions of
the inertia tensor. In this paper, we extend our previous work to investigate the dependence of
axis ratio distributions of the shapes of stellar matter determined
using the unweighted and reduced forms of inertia tensor (defined in
Sec.~\ref{shapedef}). We also consider the effect of weighting star
particles by their luminosity instead of mass, which is more
appropriate for comparison with observations.  In addition to studying
shape distributions, we check the impact of choices made when
calculating the per-galaxy inertia tensor on the predicted intrinsic
alignment two-point functions.

The main focus of this paper is the investigation of two-point
correlation functions using the shapes of stellar matter component in
galaxies. For comparison with previous results based on $N$-body simulations,
we can study the position angle statistics, while the projected shape
correlations are necessary for comparison with many observational
results. The position-angle statistics study the correlation of shapes
by considering only their orientation. Using $N$-body simulations,
\cite{2008MNRAS.389.1266L} and \cite{2005ApJ...618....1H} investigated the mass
dependence and redshift evolution of the alignment of halos with each
other. Due to the mass dependence of the misalignment angle of the
shape of stellar matter component of a galaxy with its host subhalo
shape, we have to investigate this dependence by using the shapes of
stellar matter. \cite{2014arXiv1406.4668C} used the Horizon-AGN
simulation to understand intrinsic alignments of simulated galaxies at
redshift $z=1.2$ using the spin of stellar matter component.

As we know both the ellipticity and orientation of stellar matter
component in galaxies, it is possible to compute the cross
correlations of the projected shapes with each other or the underlying
density field statistic. We investigate the mass
and redshift dependence of the intrinsic shape-density
cross-correlation function in the subhalo mass range of
$10^{11}-10^{14}\hMsun$ and at redshifts  $z=1.0,0.3$, and $0.06$. The
availability of spectral energy distributions (SED) of star particles
in the simulation \citep{2014arXiv1402.0888K} also allows us to
calculate the luminosities of each galaxy in a given band and study
intrinsic alignments for galaxy samples selected with a luminosity
threshold. It is possible to divide the galaxies in the simulation
into centrals and satellites and calculate the intrinsic alignment
separately, for comparison in a given mass bin. The dependence of intrinsic alignments on the color of galaxies (red
and blue) has been investigated observationally, for example by
\cite{2007MNRAS.381.1197H} and \cite{2011MNRAS.410..844M}. These results indicate
larger intrinsic alignments for red galaxies. Here, we will use SEDs to determine colors that we can use to
approximately divide our sample of galaxies into red and blue types,
to confirm the consistency with the observational findings on the
importance of color in determining intrinsic alignments.             


This paper is organized as follows. In Section~\ref{S:methods}, we
describe the simulation, MB-II, used in
this study and the different methods adopted to obtain the shapes and
orientations of the stellar matter component in subhalos. In
Section~\ref{SS:twopoint}, we define the two-point correlation
functions analyzed in this paper. In Section~\ref{shapedefres}, we
show how the axis ratios and two-point
correlation functions depend on the choices made when computing the
inertia tensor, while Section~\ref{lumweighting} discusses the effect of using
luminosity weighted inertia tensor. In
Section~\ref{ia_color}, we analyze the color dependence of shapes and
two-point correlation functions by dividing the galaxy sample into red
and blue types. In Section~\ref{ia_mdrs}, we
investigate the mass and redshift dependence of intrinsic alignment two-point correlation
functions. A comparison of intrinsic alignments in centrals and
satellites is made in Section~\ref{ia_censat}.  In Section~\ref{ia_comvabn}, we compare our results
with observations and make predictions for intrinsic alignments in
upcoming weak lensing surveys. Finally, our conclusions are summarized
in Section~\ref{ia_conc}. In addition, we also provide fitting functions for the intrinsic alignment signals in different mass and luminosity bins at different redshifts in Appendix~\ref{appn:fit_results}.

\section{Methods}\label{S:methods}
\subsection{MassiveBlack-II Simulation}
In this study, we used the MassiveBlack-II (MB-II) hydrodynamic
simulation to predict the intrinsic alignment of the shapes of stellar
matter component in galaxies. MB-II is a state-of-the-art high
resolution, large volume, cosmological hydrodynamic simulation of
structure formation. This simulation has been performed with {\sc
  p-gadget}, which is a hybrid version of the parallel code, {\sc
  gadget2} \citep{2005MNRAS.361..776S} upgraded to run on Petaflop
scale supercomputers. In addition to gravity and smoothed-particle
hydrodynamics (SPH), the {\sc
  p-gadget} code also includes the physics of multiphase ISM model
with star formation \citep{2003MNRAS.339..289S}, black hole accretion
and feedback
\citep{2005MNRAS.361..776S,2012ApJ...745L..29D}. Radiative cooling and
heating processes are included \citep[as in][]{1996ApJS..105...19K},
as is photoheating due to an imposed ionizing UV background. The
details of this simulation can be found in \cite{2014arXiv1402.0888K}.

MB-II contains $N_\mathrm{part} = 2\times 1792^{3}$ dark matter and gas
particles in a cubic periodic box of length $100$\hmpc\ on a side,
with a gravitational smoothing length $\epsilon = 1.85$\hkpc\ in
comoving units. A single dark matter particle has a mass $m_\text{DM} =
1.1\times 10^{7}\hMsun$ and the initial mass of a gas particle is
$m_\text{gas} = 2.2\times 10^{6}\hMsun$, with the mass of each star
particle being $m_\text{star} = 1.1\times 10^{6}\hMsun$. The cosmological
parameters used in the simulation are as follows: amplitude of matter
fluctuations $\sigma_{8} = 0.816$, spectral index $\eta_{s} = 0.96$,
mass density parameter $\Omega_{m} = 0.275$, cosmological constant
density parameter $\Omega_{\Lambda} = 0.725$, baryon density parameter
$\Omega_{b} = 0.046$, and Hubble parameter $h = 0.702$ as per WMAP7
\citep{2011ApJS..192...18K}.

Halo catalogs of particles in the simulation are generated using
the friends of friends (FoF) halo finder algorithm
\citep{1985ApJ...292..371D}. The FoF algorithm identifies halos on the
fly using a linking length of $0.2$ times the mean interparticle
separation. The subhalo catalogs are generated using the {\sc
  subfind} code \citep{2001MNRAS.328..726S} on the halo catalogs. The
subhalos are defined as locally overdense, self-bound particle
groups. In this paper, we will be concerned with the analysis of 
shapes and their two-point correlation functions. Groups of particles
are identified as subhalos if they have at least $20$ gravitationally
bound particles; however, based on convergence tests in
\cite{2014MNRAS.441..470T}, we only use their measured shapes if there
are $\ge 1000$ particles. In this paper, we identify the galaxies to be the subhalos and only consider the shape defined by the stellar component while computing 1-point and 2-point statistics as it is directly relevant to observational measurements. A comparison of the properties of galaxies identified by different subfinder codes (such as Subfind, Structure finder, etc.) in cosmological simulations that include baryonic physics can be found in \cite{2013MNRAS.428.2039K}. They find that various galaxy properties agree among the different subfinder codes. However, the impact on shapes in high resolution cosmological simulations is not investigated yet.
                                                    
\subsection{Shapes of galaxies and dark matter halos}\label{shapedef}
In this section, we give the details of the different methods adopted
to find the shape defined by the dark matter and stellar matter
component in subhalos. We model the shapes of the dark matter and stellar
matter components of subhalos as ellipsoids in three dimensions by using
the eigenvalues and eigenvectors of the inertia tensor, which describes
the mass or luminosity distribution. In the interest of comparison
with observations, we also project the halos and subhalos onto the
$XY$ plane and model the shapes as ellipses. These are needed to
compute projected shape correlation functions, which we will define 
 in Sec.~\ref{SS:pjshap}. In 3D, consider the eigenvectors of the
inertia tensor to be ${\hat{e}_{a},\hat{e}_{b},\hat{e}_{c}}$ with the
corresponding eigenvalues being
${\lambda_{a},\lambda_{b},\lambda_{c}}$, where $\lambda_{a} >
\lambda_{b} > \lambda_{c}$. The eigenvectors represent the principal
axes of the ellipsoid with the lengths of the principal axes
$(a,b,c)$ given by the square roots of the eigenvalues
$(\sqrt{\lambda_{a}},\sqrt{\lambda_{b}},\sqrt{\lambda_{c}})$. The 3D
axis ratios are defined as
\begin{equation} \label{eq:axisratios}
q = \frac{b}{a}, \,\, s = \frac{c}{a}
\end{equation}

In 2D, the eigenvectors are ${\hat{e}_{a}',\hat{e}_{b}'}$ with the
corresponding eigenvalues ${\lambda_{a}',\lambda_{b}'}$, where
${\lambda_{a}' > \lambda_{b}'}$. The lengths of major and minor axes
are $a' = \sqrt{\lambda_{a}'}$, $b' = \sqrt{\lambda_{b}'}$ with axis
ratio $q' = \frac{b'}{a'}$.

We explore several different ways of computing the inertia tensor based on the
mass or luminosity, and the radial weighting given to each particle. The
unweighted inertia tensor (used for all results in \citealt{2014MNRAS.441..470T}) is given by
\begin{equation} \label{eq:uwinertensor}
 I_{ij} = \frac{\sum_{n} m_{n}x_{ni}x_{nj}}{\sum_{n} m_{n}},
\end{equation}
where $m_{n}$ represents the mass of the $n^{th}$ particle and
$x_{ni},x_{nj}$ represent the position coordinates of the $n^{th}$
particle with $ 0 \leq i,j \leq 2$ in 3D and $0 \leq i,j \leq 1$ in
2D. Here all particles are given equal weight irrespective of their
distance from the center of a subhalo. We can also use the reduced
inertia tensor, which gives more weight to particles which are closer
to the center:
\begin{equation} \label{eq:redinertensor}
\widetilde{I}_{ij} = \frac{\sum_{n} m_{n}\frac{x_{ni}x_{nj}}{r_{n}^{2}}}{\sum_{n} m_{n}}
\end{equation}
where
\begin{equation} \label{eq:rn2}
 r_{n}^{2} = \sum_{i}x_{ni}^{2}
\end{equation}

Unlike for $N$-body simulations where it is natural to let each
equally-weighted dark matter particle contribute equally to the
inertia tensor, for simulated galaxies it is natural to consider
weighting each particle by its luminosity, considering that flux is
what we actually see when we observe the galaxy.  This results in
another definition for the inertia tensor:
\begin{equation} \label{eq:luminertensor}
 I_{ij}^{(\text{lum})} = \frac{\sum_{n} l_{n}x_{ni}x_{nj}}{\sum_{n} l_{n}},
\end{equation}
where each stellar particle is weighted by its luminosity, $l_{n}$
instead of its mass. The definition presented here refers to the
luminosity-weighted form of unweighted inertia tensor given in
Eq.~\ref{eq:uwinertensor}. In our analysis, we also use the shapes
obtained using the luminosity-weighted form of reduced inertia tensor
(Eq.~\ref{eq:redinertensor}) defined analogously.

Instead of determining axis ratios with a single calculation, we can
also adopt iterative methods for finding the shapes using unweighted
and reduced inertia tensors. In the unweighted iterative and reduced
iterative methods, we first determine the axis ratios by the standard
definitions of the corresponding inertia tensors using all the
particles of a given type in the subhalo. Keeping the enclosed volume
constant \citep[as in][]{2012JCAP...05..030S}, the lengths of the
principal axes of ellipsoids are rescaled accordingly\footnote{Note
  that some authors instead keep the length of the major axis fixed
  \citep[e.g.,][]{2006MNRAS.367.1781A,2012MNRAS.420.3303B}}.  After
this rescaling, we determine the shapes again, discarding particles
outside the ellipsoidal volume. This process is repeated until
convergence is reached. Our convergence criterion is that the
fractional change in axis ratios must be below 1\%. It is to be noted
here that although we only use subhalos that initially have at least
1000 dark matter and star particles to calculate shapes, the use of
iterative methods results in some low mass subhalos having fewer than
1000 particles in the enclosed volume. But, since this is a very low
fraction (less than 0.5\%) and the number of particles remaining is
very close to 1000, we include them for further analysis.
 
We will investigate the dependence of using these different
definitions on the probability distributions of axis ratios and the
two-point correlation functions. Having outlined the differences, we
will present the rest of our predictions from the simulation based on the reduced
iterative inertia tensor alone.

\subsection{Misalignment angle}

To study the relative orientation between the shapes defined by
dark matter and stellar matter component in subhalos, we compute 
the probability distributions of misalignment angles as in \cite{2014MNRAS.441..470T}. Let
$\hat{e}_{da}$ and $\hat{e}_{ga}$ be the major axes of the shapes
defined by dark matter and stellar matter components respectively. We
can then define the misalignment angle by
\begin{equation} \label{eq:misalignangle}
 \theta_{m} = \arccos(\left|\hat{e}_{da} \cdot \hat{e}_{ga}\right|).
\end{equation}

\section{Two-point correlation functions}\label{SS:twopoint}

Here we define the intrinsic alignment two-point correlation functions
that we use in this work. Intrinsic alignments can arise due to the correlation of intrinsic shapes of galaxies with each other (II term) or the correlation of the gravitational shear and intrinsic ellipticity (GI term). The two-point statistics discussed in this paper concern the GI term.

\subsection{Position angle statistics}\label{SS:pes}

The position angle statistics, Ellipticity-Ellipticity (EE) and
Ellipticity-Direction (ED) correlation functions, are useful to quantify
the correlations between the position angles of galaxies or halos with each other
and with the large-scale density field as a function of mass and
redshift.  These can then 
be compared against results for halos in $N$-body simulations. We follow the notation
of \cite{2008MNRAS.389.1266L} to define the EE and ED correlations.

If $\hat{e}_{a}(\textbf{x})$ is the direction of the major axis of the
shape of the dark matter or stellar matter component of a subhalo centered
at position \textbf{x}, then the EE correlation function in 3D,
$\eta(r)$, is given by
\begin{equation} \label{eq:EE3d}
 \eta(r) = \langle \mid \hat{e}_{a}(\textbf{x}) \cdot \hat{e}_{a}(\textbf{x+r})\mid^{2} \rangle - \frac{1}{3}.
\end{equation}
Here, $\langle . \rangle$ means an average over pairs of galaxies separated by a distance, $r$. For galaxies or halos randomly oriented according to a uniform distribution, the expectation value of this
quantity is zero.

The ED correlation function cross-correlates the orientation of the
major axis of the shape of a subhalo with the large-scale density
field. For a subhalo centered at position \textbf{x} with major axis
direction $\hat{e}_{a}$, let the unit vector in the direction of the tracer of the matter
density field 
at a distance $r$ be $\hat{\textbf{r}} =
\textbf{r}/r$. Then the ED cross-correlation function is
given by
\begin{equation} \label{eq:ED3d}
 \omega(r) = \langle \mid \hat{e}_{a}(\textbf{x})\cdot \hat{\textbf{r}}(\textbf{x}) \mid^{2} \rangle - \frac{1}{3}
\end{equation}
which is again zero in the case of no intrinsic alignments.

We can represent the tracers of the matter density field using either the positions of 
dark matter particles (in which case the correlation function is
denoted by the symbol $\omega_{\delta}$) or the positions of subhalos
(in which case it includes a factor of the subhalo bias, and is simply
denoted $\omega$).

\subsection{Projected shape correlation functions}\label{SS:pjshap}

The projected shape correlation functions are computed to directly
compare our results from simulations with observations. Here, we
follow the notation of \cite{2006MNRAS.367..611M} to give formulae for
the calculation of galaxy-intrinsic shear correlation function
($\hat{\xi}_{g+}(r_{p},\Pi)$) and the projected statistic, $w_{g+}$. Here, $r_{p}$ is the comoving transverse separation of a pair of galaxies in the $XY$ plane and $\Pi$ is their separation along the $Z$ direction.

If $q' = \frac{b'}{a'}$ is the axis ratio of the projected shape of
the 
dark matter or stellar matter component of a subhalo and $\phi$ is the
position angle of the major axis of the ellipse, the components of the
ellipticity are given by
\begin{equation} \label{eq:ellipticity}
 (e_{+},e_{\times}) = \frac{1 - q'^{2}}{1 +
    q'^{2}}\left[\cos{(2\phi)},\sin{(2\phi)}\right],
\end{equation}
where $e_{+}$ refers to the radial component of ellipticity and $e_{\times}$ is the component at $45^{\circ}$ rotation. 
The galaxy-intrinsic shear correlation function cross-correlates the
ellipticity of 
galaxies with the density field. The ``shape sample'' denoted by $S_{+}$
is selected on the basis of a threshold or binning in subhalo mass,
stellar mass, band luminosity and other properties of the galaxies in
the simulation, while all the subhalos are used to trace the density
field, forming a ``density sample'' denoted by $D$. The
cross-correlation function is then computed using 
\begin{equation} \label{eq:gicorr}
 \hat{\xi}_{g+}(r_{p},\Pi) = \frac{S_{+}D - S_{+}R}{RR}
\end{equation}
where $r_{p}$ is the transverse separation of the galaxy points and
$\Pi$ is the radial red-shift space separation (here, it is the
separation along the $Z$ direction), and $S_{+}D$ is the sum over all
pairs with separations $r_{p}$ and $\Pi$:
\begin{equation} \label{eq:SpD}
 S_{+}D = \sum_{i\neq j\mid r_{p},\Pi}\frac{e_{+}(j\mid
   i)}{2\mathcal{R}},
\end{equation}
where $e_{+}(j | i)$ is the $+$ component of the ellipticity of a
galaxy ($j$) from the shear sample relative to the direction of a tracer of
density field ($i$) selected from the density sample. Here, $\mathcal{R}
= (1 - e_{rms}^{2})$ is the shear responsivity that converts from distortion
to shear with $e_{rms}$, the rms ellipticity per component of the shape sample. Alternatively, we can also define the ellipticity by $e = \frac{1-q'}{1+q'}$, in which case we do not have to take the responsivity correction into account. However, using this definition decreases the intrinsic alignment signal by only about $\sim 6\%$. So, in the rest of this paper, we employ the former definition as it makes it easier for comparison with observations. $S_{+}R$ is defined by a
similar equation for the correlation of the data sample with a random
density field distribution to remove observational systematics in the
shear estimates, and hence we can neglect this term here. 
The projected
correlation function, 
$w_{g+}(r_{p})$ is now
given by

\begin{equation} \label{eq:wgp}
 w_{g+}(r_{p}) =
 \int_{-\Pi_\text{max}}^{+\Pi_\text{max}}\hat{\xi}_{g+}(r_{p},\Pi)\,\mathrm{d}\Pi
\end{equation}
We calculated the correlation functions over the whole length
of the box ($100\hmpc$) with $\Pi_\text{max} = 50\hmpc$, in 25 bins of size $4\hmpc$
each. The projected correlation functions are obtained by summing over
the galaxy-intrinsic and intrinsic-intrinsic shear correlation
functions with the integrand replaced by a summation. Note
  that the $w_{g+}$ signal can also be calculated using projected
  shapes along some other plane instead of XY. However, we did not
  observe significant differences in the signal for $w_{g+}(r_p)$
  calculated by projecting along YZ and XZ planes. Thus, all reported
  results use shapes projected on the XY plane.

An alternative way to trace the density field for the calculation of
$w_{g+}$ is to use the positions of all dark matter particles in the
simulation instead of subhalos. The correlation function obtained in
this way 
is denoted by $w_{\delta +}$. The former is what we can compare with
observations, but we can use the latter to test the standard
conversion that is used between the two (dividing the observational
signals by the linear galaxy bias).

The observable, $w_{g+}$ is related to the GI term which is
  discussed further in the section below. We do not discuss the
  intrinsic shear-shear correlation functions,
  ($\hat{\xi}_{++}(r_{p},\Pi)$, $\hat{\xi}_{\times
    \times}(r_{p},\Pi)$) and their corresponding projected statistics,
  ($w_{++}$, $w_{\times \times}$) in this paper due to their being
  extremely noisy. Moreover, it has been shown in multiple theoretical
  studies \citep[e.g.,][]{{2004PhRvD..70f3526H}} that if intrinsic
  alignments are caused by something like the tidal alignment model,
  the II contamination to cosmic shear signals will be quite
  subdominant to the GI contamination. Given that all measurements to
  date of strong intrinsic alignments on large scales have been made
  with red galaxies, and are consistent with the tidal alignment model
  \citep[e.g.,][]{{2011JCAP...05..010B}}, we  mainly focus on the
  GI-type intrinsic alignment contamination here. As a practical
  matter, there is additional motivation to focus on measuring
  $w_{g+}$ rather than $w_{++}$, because for alignments consistent
  with the tidal alignment model, the signal-to-noise ratio for the
  former will be higher than for the latter (see section~4.1 of
  \citealt{2014arXiv1411.1755S}). Finally, for this type of alignment,
  measurements of GI provide a unique prediction for II, so our
  measurements are equally informative about both given that they
  appear completely consistent with the tidal alignment.

	\subsection{Formalism: Linear Alignment Model}\label{ssec:nla}
        The linear alignment model is the standard formalism used to
        study \ia{} of galaxy shapes at large scales
        \citep{Catelan2001,2004PhRvD..70f3526H,2011JCAP...05..010B,Chisari2013}. The
        observational measurements of \ia{} on large scales can be
        reproduced using this model. In this section, we briefly
        describe the main features of the model.
			
			The linear alignment model is based on the assumption that the intrinsic shear of galaxies is determined by the tidal field at the time of formation of the galaxy \citep[assumed to be during matter domination,][]{Catelan2001}. Thus we can write the intrinsic shear in terms of the primordial potential as 
			\begin{equation}
				\gamma^I=(\gamma^I_+,\gamma^I_\times)=-\frac{C_1}{4\pi G}(\partial^2_x-\partial^2_y,\partial_x\partial_y)\phi_p
				\label{eqn:gamma_phi}
			\end{equation} 
			\cite{2004PhRvD..70f3526H} derived the 2-point matter-\ia{} power spectra, relating them to the linear matter power spectrum, $P_\delta^{\text{lin}}$ 
			\begin{align}
				P_{g+}(\vec{k},z)&=A_I b \frac{C_1\rho_{\text{crit}}\Omega_m}{D(z)} \frac{k_x^2-k_y^2}{k^2} P_\delta^{\rm lin} (\vec{k},z)\label{eqn:LA+}\\
				P_{++}(\vec{k},z)&=\left(A_I \frac{C_1\rho_{\text{crit}}\Omega_m}{D(z)} \frac{k_x^2-k_y^2}{k^2} \right)^2P_\delta^{\rm lin} (\vec{k},z)\label{eqn:LA++}\\
				P_{g\times}(\vec{k},z)&=A_I b\frac{C_1\rho_{\text{crit}}\Omega_m}{D(z)}\frac{k_x k_y}{k^2}P_\delta^{\rm lin}(\vec{k},z)\label{eqn:LAx}
			\end{align} 
			Following \cite{Joachimi2011}, we fix $C_1\rho_{\text{crit}}= 0.0134$ and use the arbitrary constant $A_I$ to describe the amplitude of \ia{} for different samples. $D(z)$ is the linear growth factor, normalized to unity at 
			$z=0$.
			
			\cite{Bridle2007} suggested using the full non-linear matter power spectrum $P_\delta^{\rm nl}$ in Eq.~\eqref{eqn:LA+} to extend the linear alignment model to quasi-linear scales. This model is called the non-linear linear alignment 
			model (NLA). In this work, we will use the non-linear matter power spectrum generated with the CAMB software \citep{2002PhRvD..66j3511L}, with a fixed WMAP7 cosmology \citep{2013ApJS..208...19H}.
						
			Fourier transforming Eq.~\eqref{eqn:LA+} and integrating over line of sight separation $\Pi$, we get the two point correlation function
			\begin{align}
				&w_{g+}(r_p)=\frac{A_I b_D C_1 \rho_{\text{crit}} \Omega_m}{\pi^2}\int dz \frac{W(z)}{D(z)}\int_0^{\infty} dk_z\int_0^{\infty}\nonumber \\
				&dk_{\perp}\frac{k_\perp^3}{(k_\perp^2+k_z^2)k_z} P_\delta^{\rm nl}(\vec{k},z)\sin(k_z\Pi_{max})J_2(k_\perp r_p)\left(1+\beta\mu^2\right)\label{eqn:wgp}
			\end{align}
			$b_D$ is the bias for density sample, $\mu=k_z/k$ and $\beta$ is the linear redshift distortion parameter with $(1+\beta\mu^2)$ accounting for the effects of redshift-space distortions \citep[RSD,][]{1987MNRAS.227....1K,2014arXiv1411.1755S}. $\beta (z)=f(z)/b$, where $f(z)$ is the logarithmic growth rate at redshift $z$; in $\Lambda$CDM, $f(z)\sim \Omega_m(z)^{0.55}$.  
			\wgx{} is expected to be zero by symmetry. 

\section{The impact of using different inertia tensor definitions}\label{shapedefres}
In this section, we compare 
the axis ratio distributions and misalignment angle distributions (as
presented in \citealt{2014MNRAS.441..470T}) when using the different definitions of
inertia tensor defined in Sec.~\ref{shapedef}. We also consider how
the two-point correlation functions vary when using different shape
definitions. For convenience, we define three mass bins based on total subhalos mass, M1
($10^{10.0} - 10^{11.5}$\hMsun), M2 ($10^{11.5} - 10^{13.0}$\hMsun),
and M3 ($> 10^{13.0}$\hMsun).

\subsection{Axis ratio distributions}\label{arres}

\begin{figure*}
\begin{center}
\includegraphics[width=2.25in]{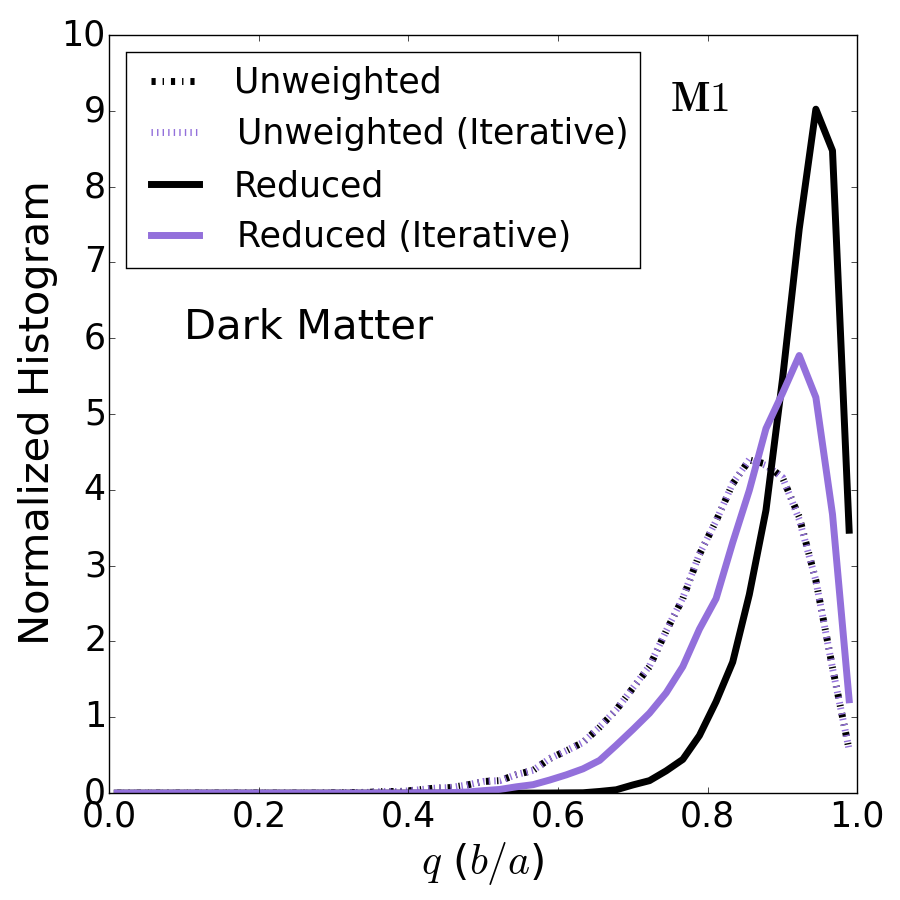}
\includegraphics[width=2.25in]{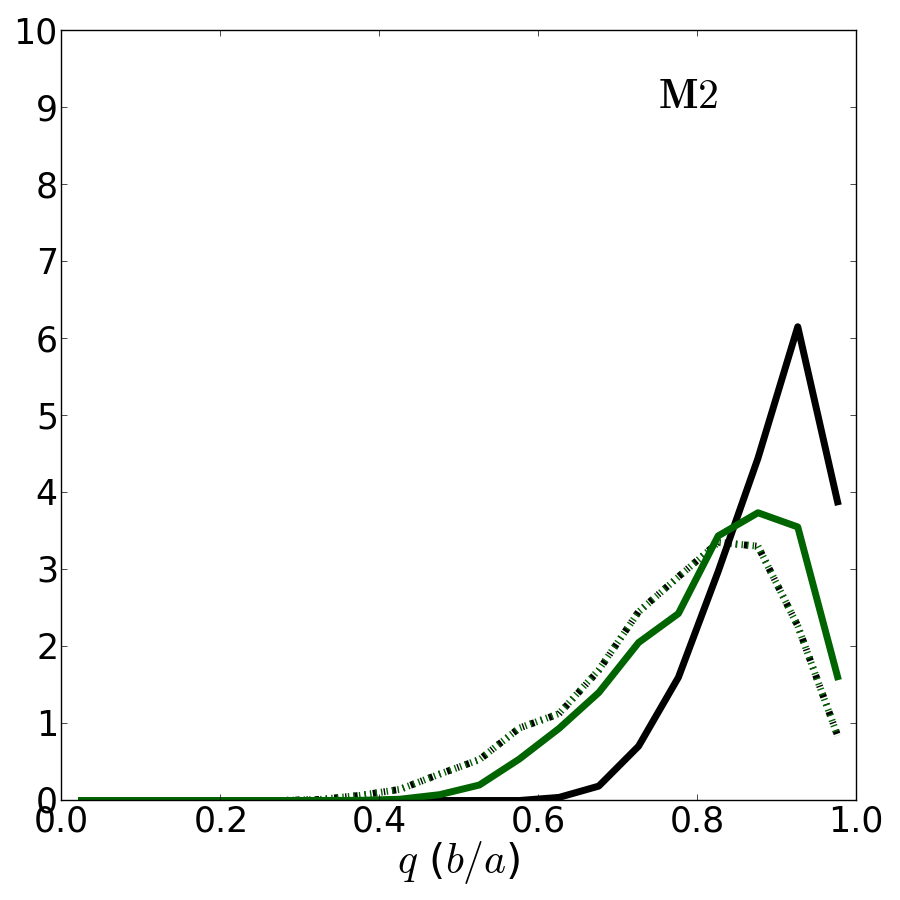}
\includegraphics[width=2.25in]{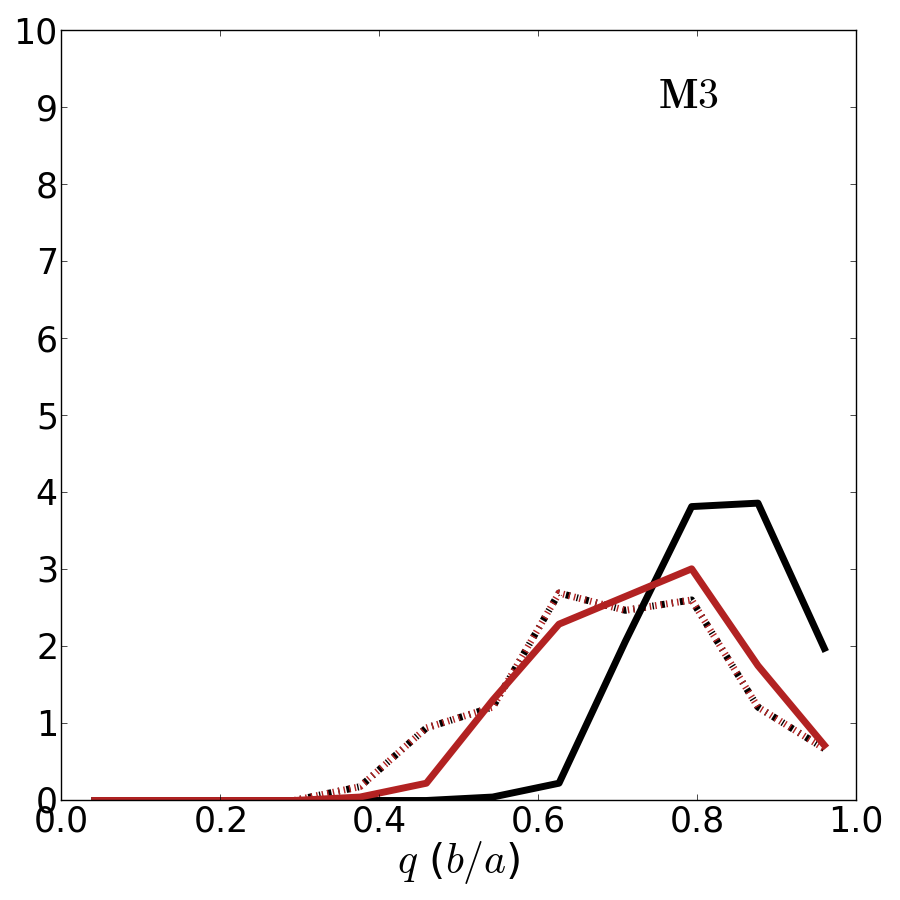}\\ 
\includegraphics[width=2.25in]{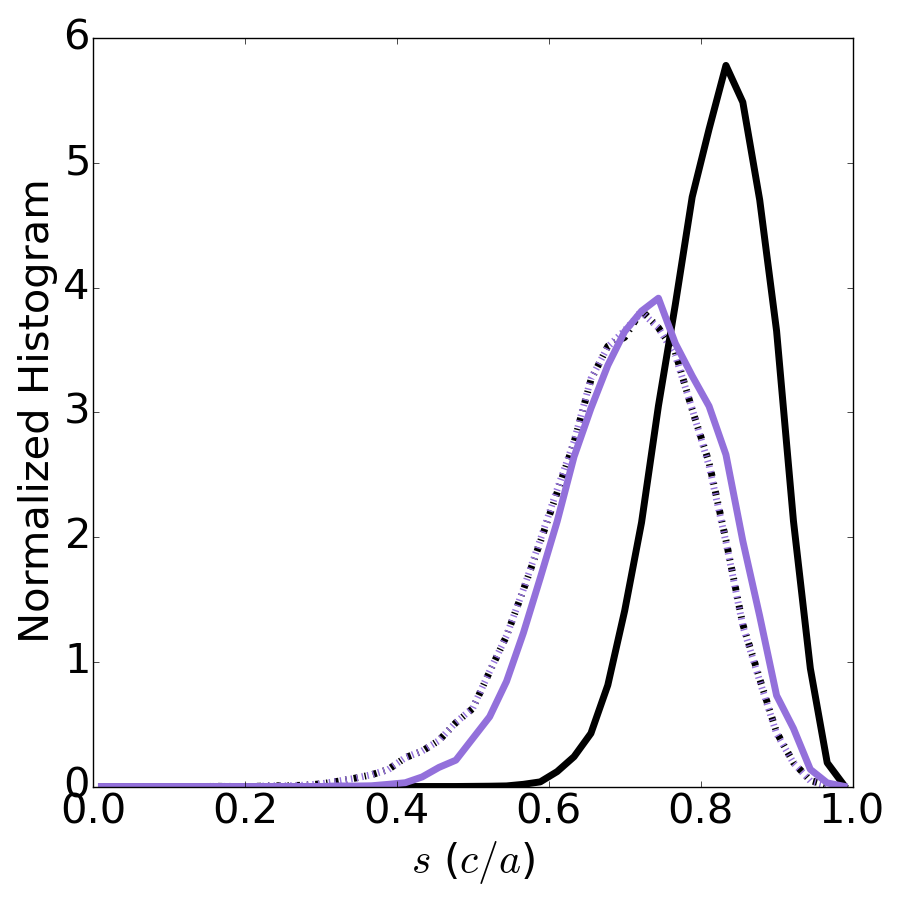}
\includegraphics[width=2.25in]{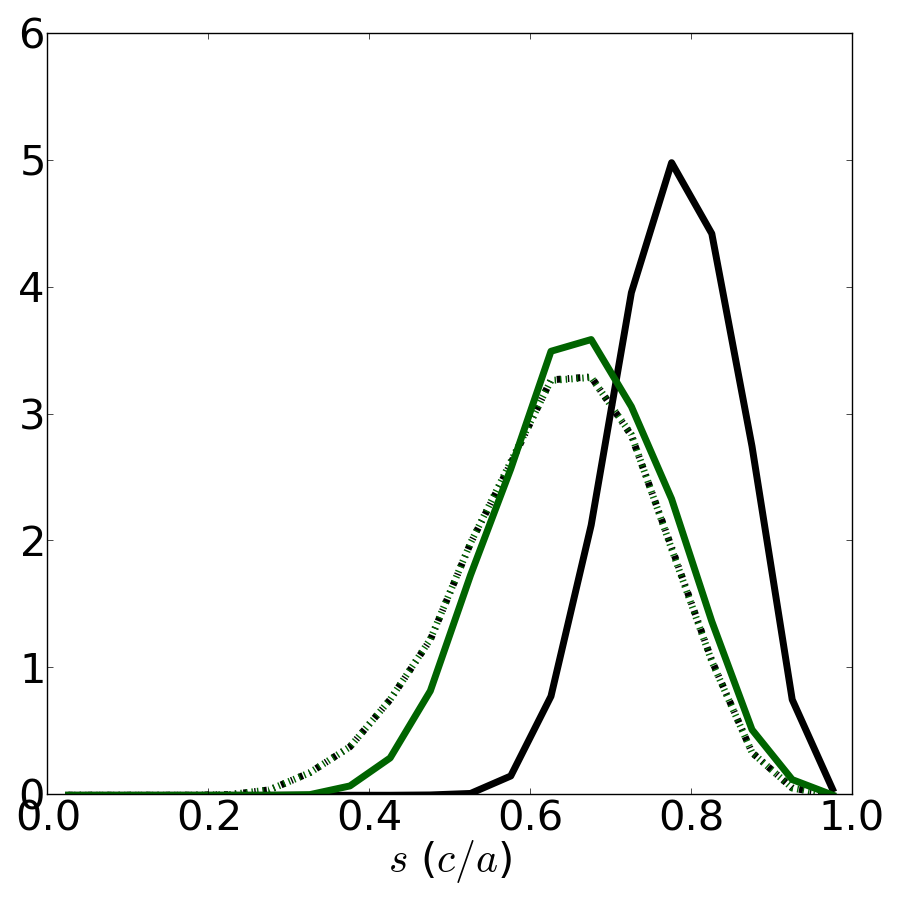}
\includegraphics[width=2.25in]{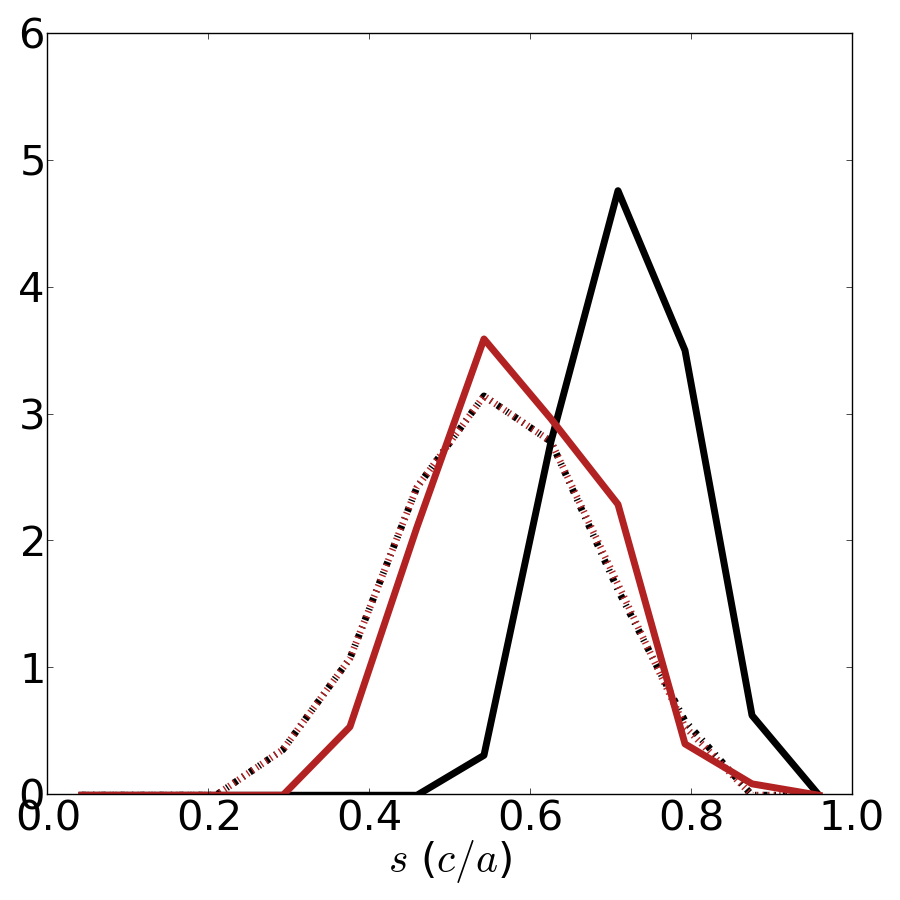}
\caption{\label{F:fig_dm_qs} Normalized histograms of 3D axis ratios of dark matter component in subhalos using different
  definitions of inertia tensor in mass bins M1, M2 and M3 at
  $z=0.3$. The number of galaxies are $38768$, $8438$, and $267$ respectively in mass bins M1, M2 and M3. {\em Top:} $q~(b/a)$; {\em Bottom:} $s~(c/a)$.}
\end{center}
\end{figure*}
\begin{figure*}
\begin{center}
\includegraphics[width=2.25in]{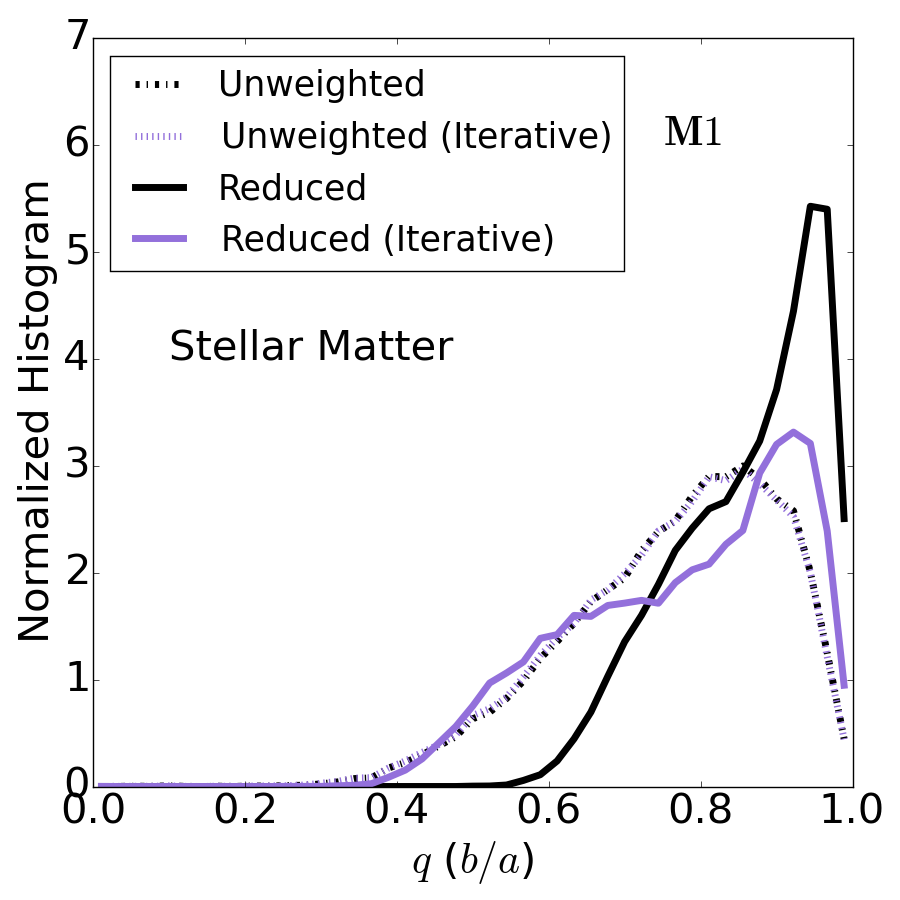}
\includegraphics[width=2.25in]{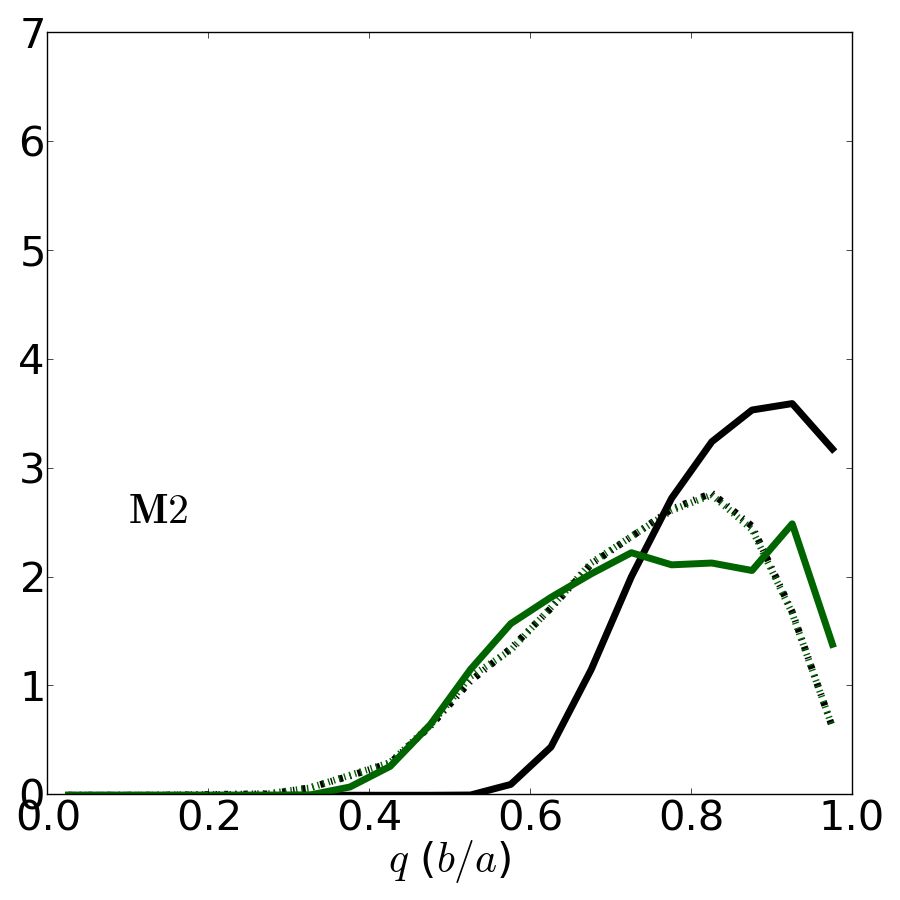}
\includegraphics[width=2.25in]{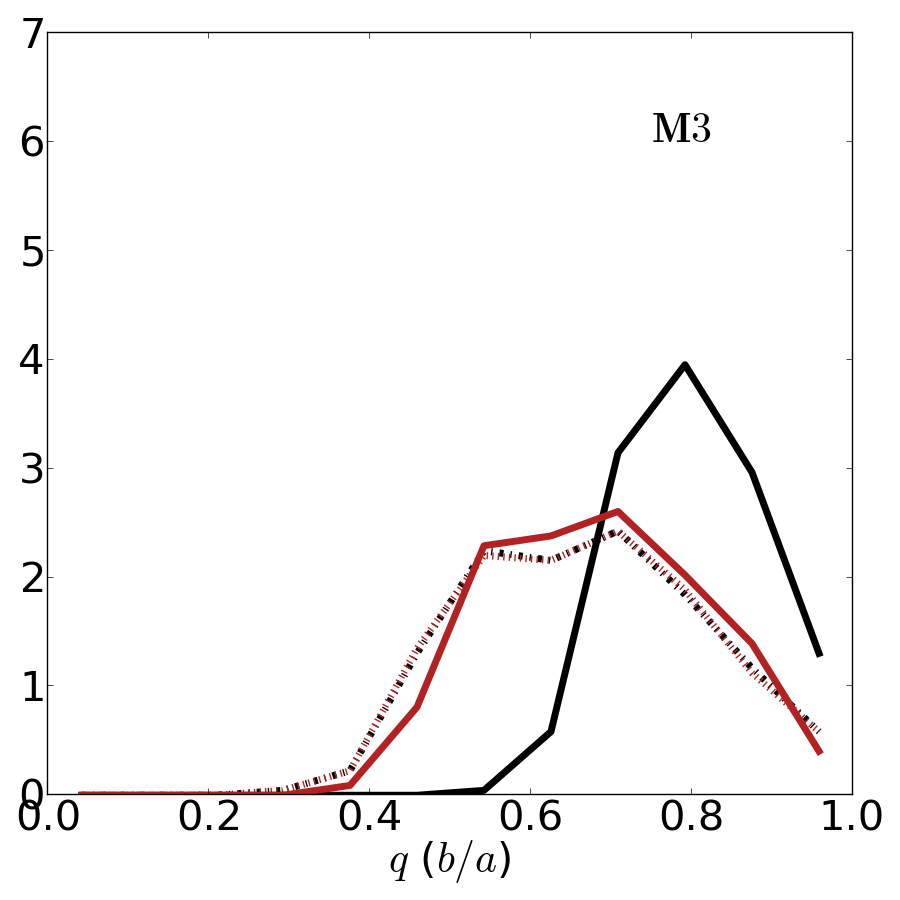}\\ 
\includegraphics[width=2.25in]{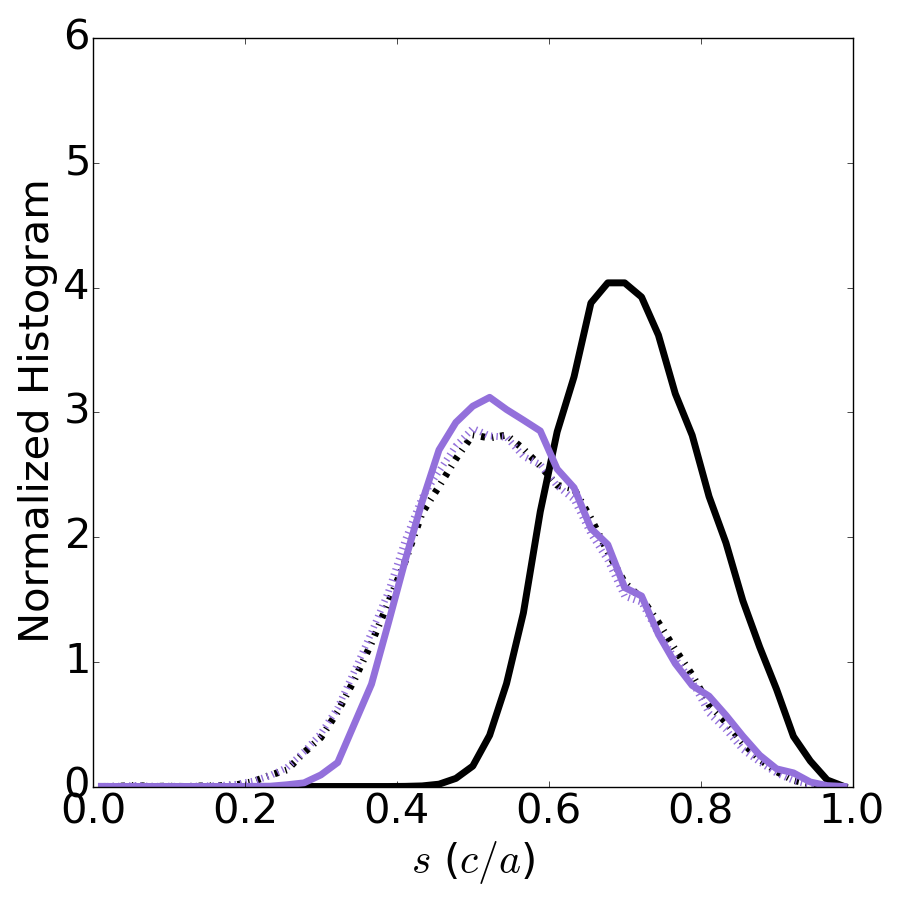}
\includegraphics[width=2.25in]{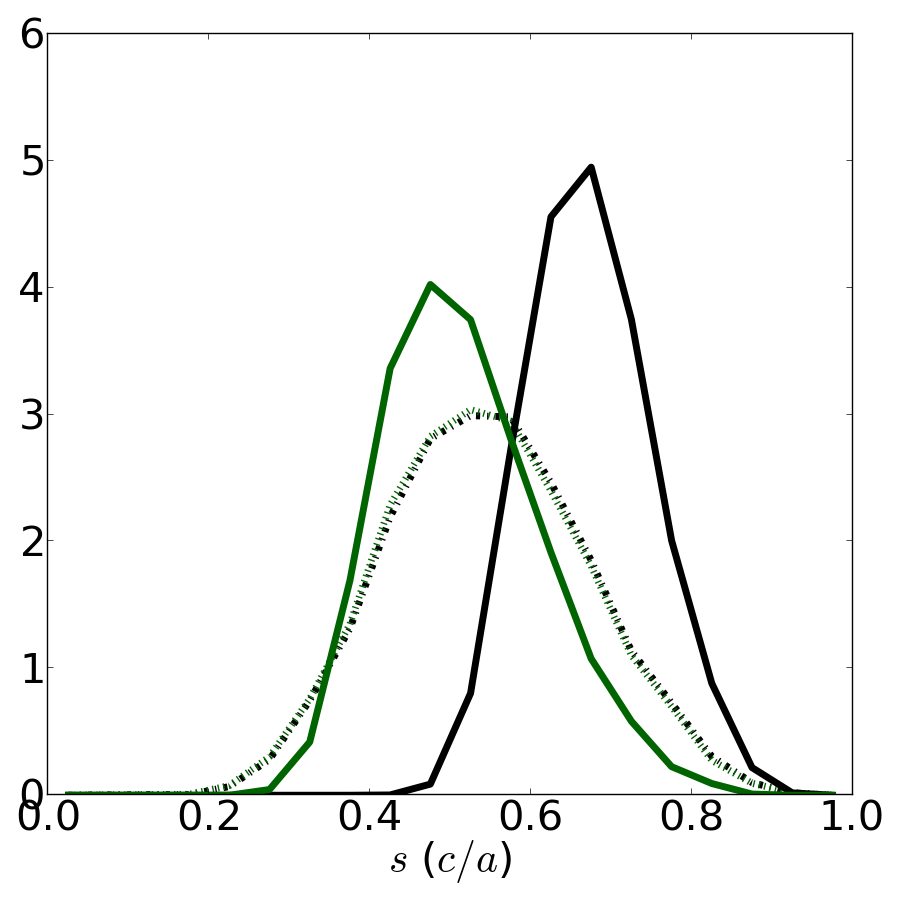}
\includegraphics[width=2.25in]{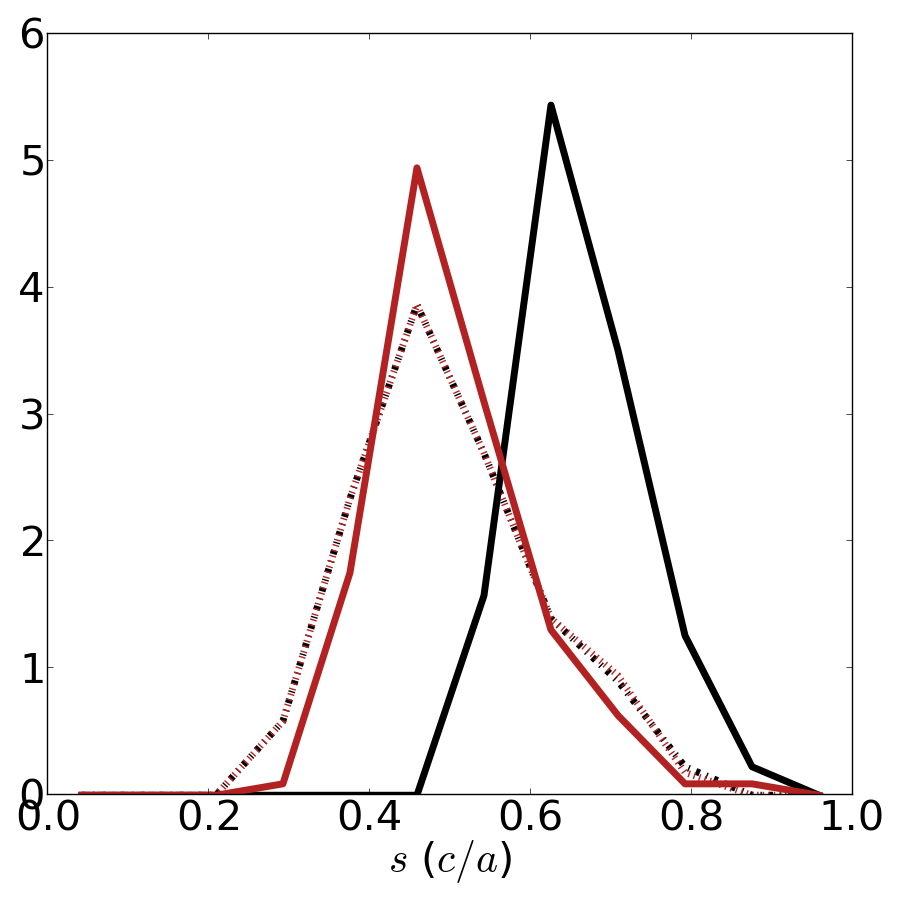}
\caption{\label{F:fig_stellar_qs} Normalized histograms of 3D axis ratios
  of stellar matter component in subhalos using
  different definitions of inertia tensor in mass bins M1, M2 and
  M3 at $z=0.3$. {\em Top:} $q~(b/a)$; {\em Bottom:} $s~(c/a)$.}
\end{center}
\end{figure*}

Here, we compare the axis ratios of shapes obtained using different
definitions of inertia tensor. In Figs.~\ref{F:fig_dm_qs}
and~\ref{F:fig_stellar_qs}, we show the histograms of the axis ratios
of the 3D shapes of dark matter (Fig.~\ref{F:fig_dm_qs} and stellar
(Fig.~\ref{F:fig_stellar_qs}) matter components in subhalos using four
inertia tensor calculations: unweighted and reduced, non-iterative and
iterative. We considered mass bins M1, M2 and M3 with $38768$,
  $8438$, and $267$ galaxies, respectively. From the plots, we can see that the axis ratio
distributions obtained with non-iterative and iterative unweighted
inertia tensors are essentially identical.  For the reduced inertia
tensor, the results for the iterative calculation are uniformly more
flattened than for the non-iterative calculation.  The reason for this
is that the non-iterative reduced calculation implicitly imposes
spherical symmetry (via the $1/r^2$ weighting), which will result in
an overly-rounded shape estimate.  For this reason, we do not consider
the reduced non-iterative calculation to be useful.

Comparing the iterative reduced vs. unweighted results, the axis
ratios of dark matter subhalos are slightly larger (rounder) when
using the reduced inertia tensor than when using the unweighted
one. This finding agrees qualitatively with the findings of
\cite{2012MNRAS.420.3303B} using $N$-body simulations. Additionally,
the inclusion of baryonic physics in hydrodynamic simulations can lead
to more round dark matter shapes in the inner regions of subhalos
\citep[e.g.,][]{2006EAS....20...65K,2013MNRAS.429.3316B}. In future
work, we will directly study the impact of baryonic physics on the
shapes determined by reduced inertia tensor by comparing our results
on shape distributions with those obtained with a dark matter only
simulation.

When considering the stellar shapes, we see that the histograms of
intermediate-to-major axis 
ratio, $q$ $(\frac{b}{a})$, indicate a slight increase for the reduced
inertia tensor compared to
shapes obtained from the unweighted tensor, while the histograms of 
minor-to-major axis ratio, $s$ $(\frac{c}{a})$, show a decrease in axis
ratio. Thus the shape distributions with the reduced inertia tensor
are more oblate than the ones with the unweighted inertia
tensor. In a previous study \citep{2014MNRAS.441..470T}, we found that the projected shapes of stellar matter determined using the unweighted inertia tensor are slightly smaller, but compare favorably with observational measurements using the RMS ellipticity statistic. We note here that the projected shapes with reduced inertia tensor will have a smaller value of the RMS ellipticity statistic.
  
\subsection{Misalignment angle distributions}\label{mares}

\begin{figure*}
\begin{center}
\includegraphics[width=2.25in]{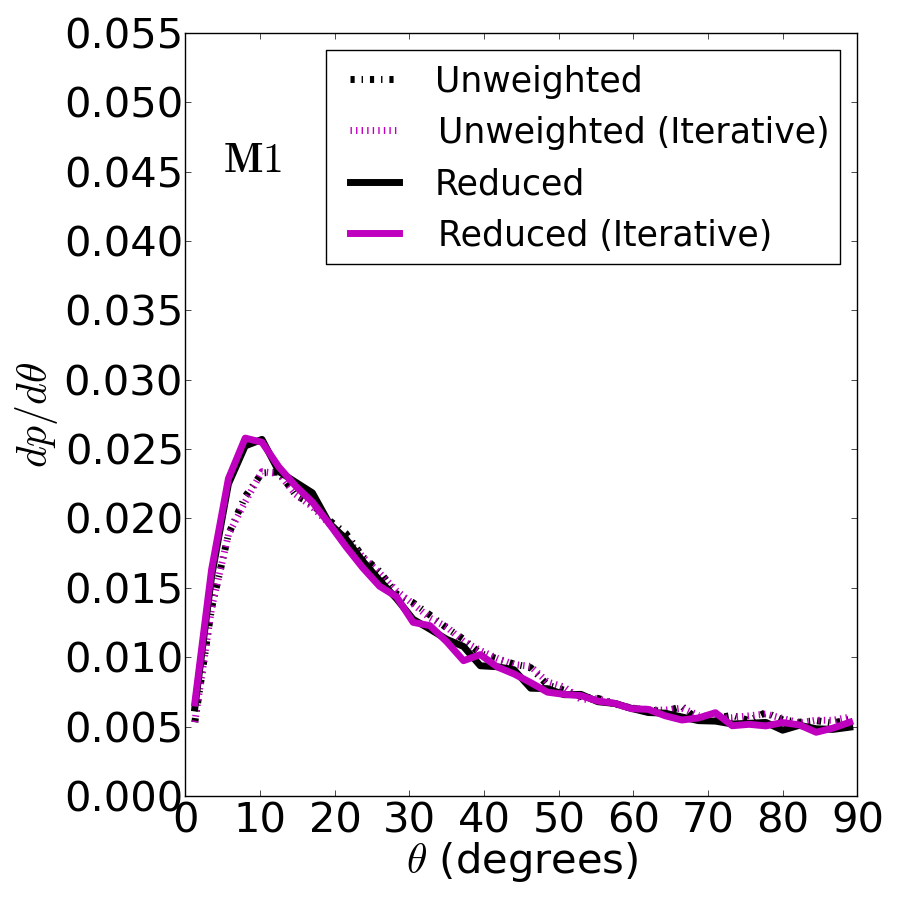}
\includegraphics[width=2.25in]{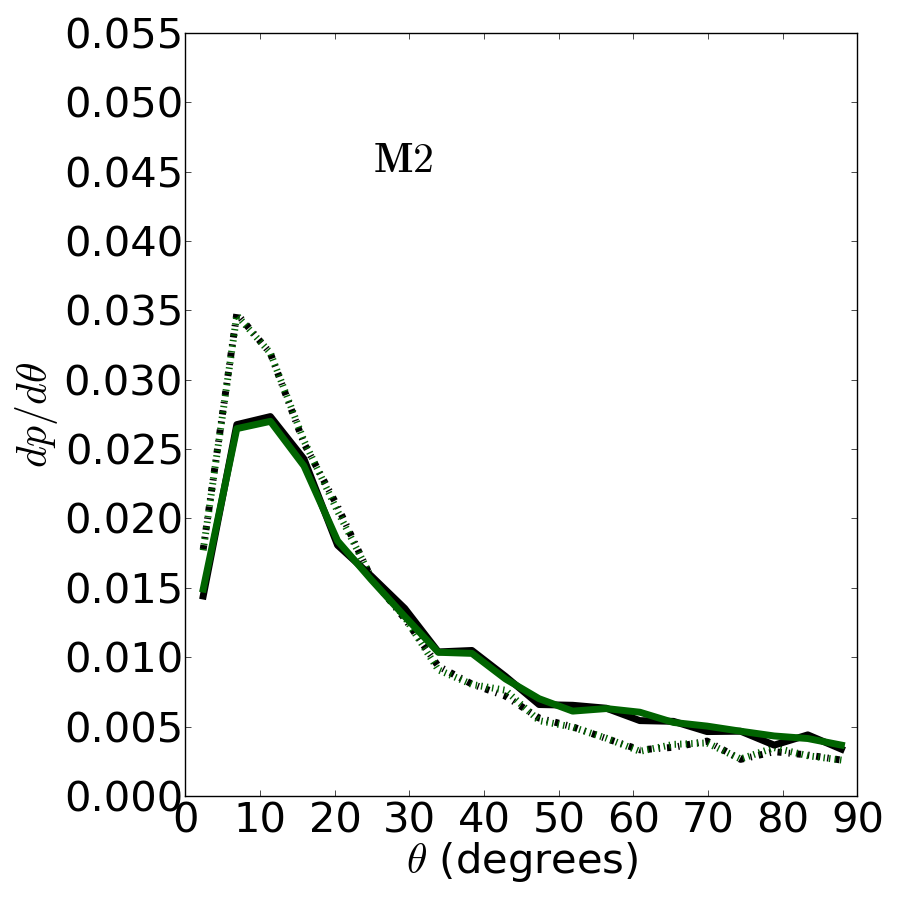}
\includegraphics[width=2.25in]{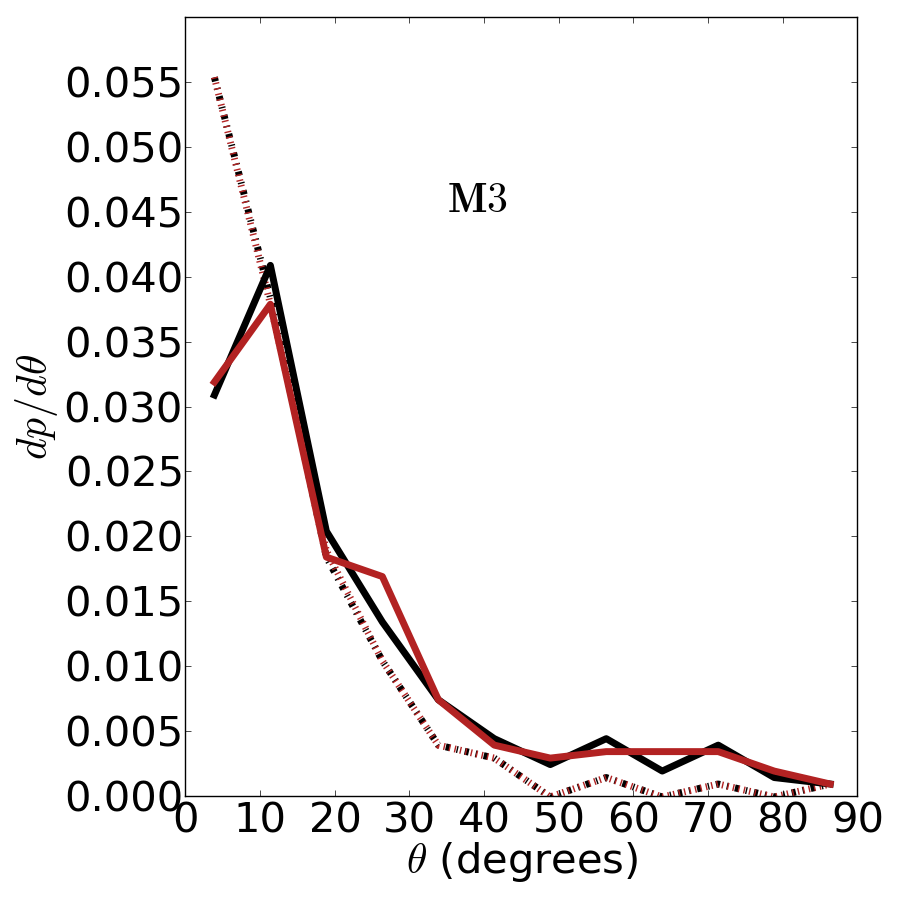}
\caption{\label{F:fig_ma} Normalized histograms of misalignment angles
  between the major axes of 3D shapes defined by the dark matter and stellar matter component in subhalos using different definitions of inertia tensor, in mass bins (M1, M2, and M3) at $z=0.3$. Note that for uniformly distributed misalignment angles in 3D, the probability distribution is proportional to $\sin{\theta}$.}
\end{center}
\end{figure*}
In Fig.~\ref{F:fig_ma}, we plot the normalized histograms of misalignment angles between the shapes defined by
dark matter and stellar matter component in subhalos. The plots show
that there is no significant difference in misalignments if we
adopt an iterative or non-iterative definition of shape tensor, for
both unweighted and reduced cases. For the unweighted inertia tensor,
this result is consistent with the distribution of axis ratios in
Sec.~\ref{arres}, where the histograms are similar for unweighted
non-iterative and iterative definitions.
 For the shapes obtained using the 
reduced inertia tensor, the histograms of misalignment angles seem to
indicate that while the axis ratios change
significantly, the relative shape orientation is not altered much. Comparing
misalignment histograms obtained using unweighted and reduced inertia
tensor, we observe that in the lowest mass bin, M1, the
misalignments are slightly smaller if we use the reduced inertia
tensor to define shapes, while they are slightly higher in mass bins
M2 and M3. 
  
\subsection{Two-point correlation functions}\label{twopointres}

\begin{figure*}
\begin{center}
\includegraphics[width=2.25in]{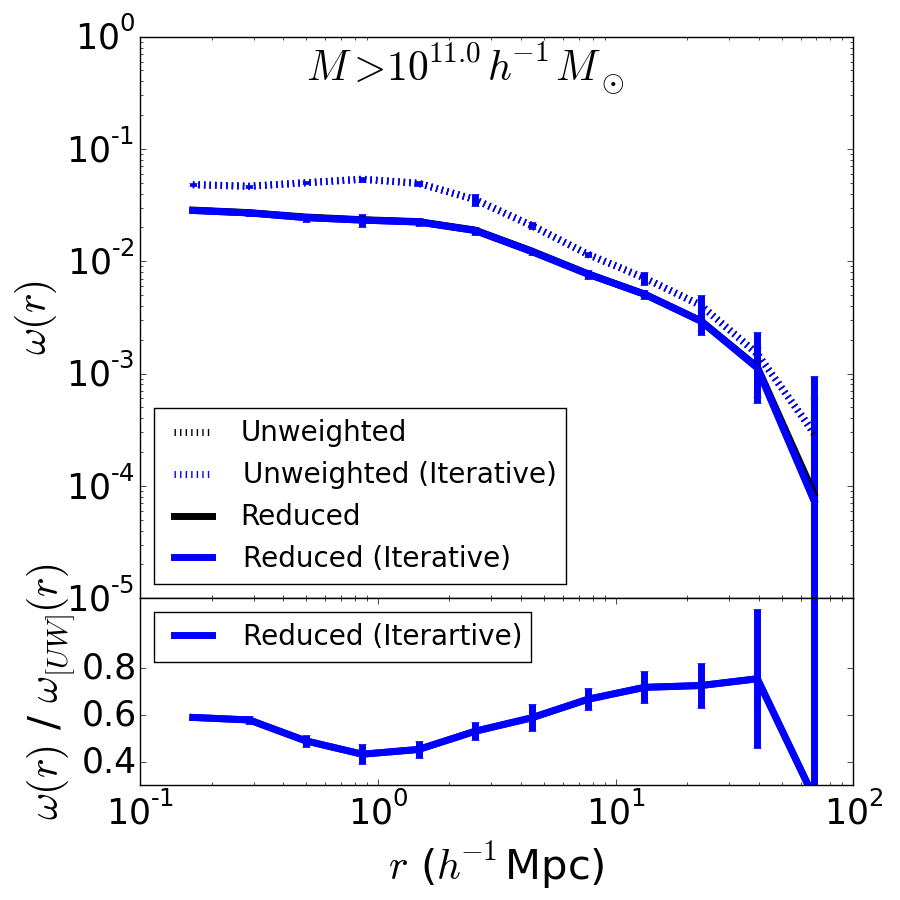}
\includegraphics[width=2.25in]{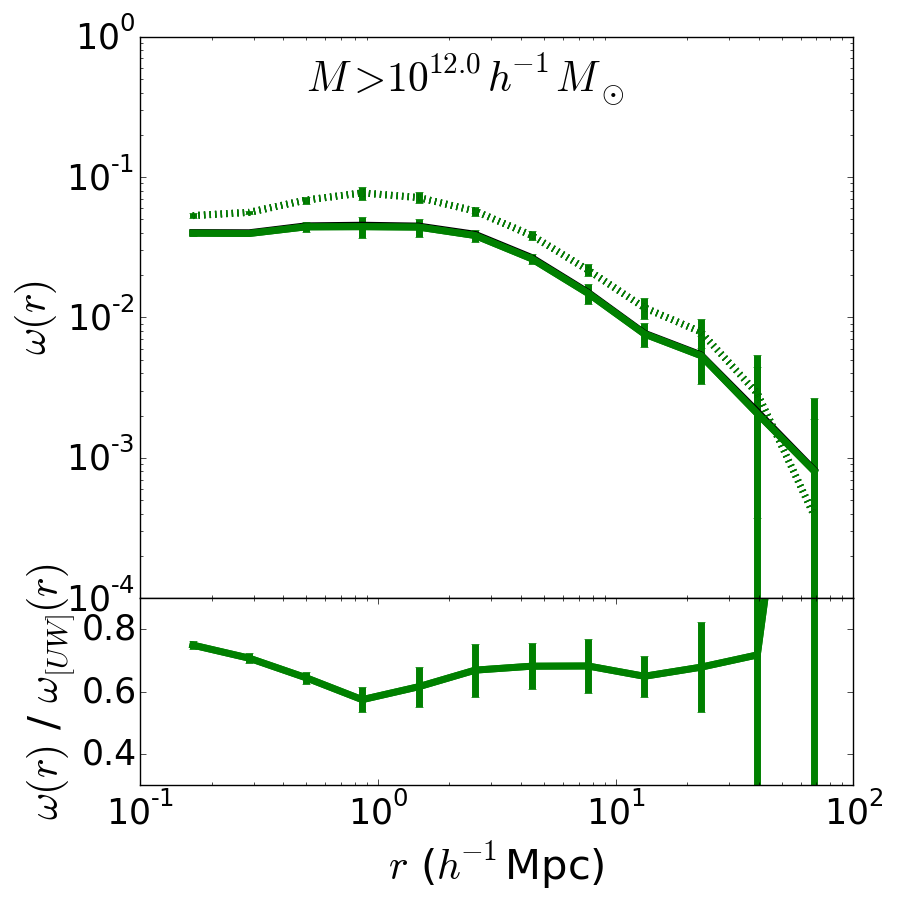}
\includegraphics[width=2.25in]{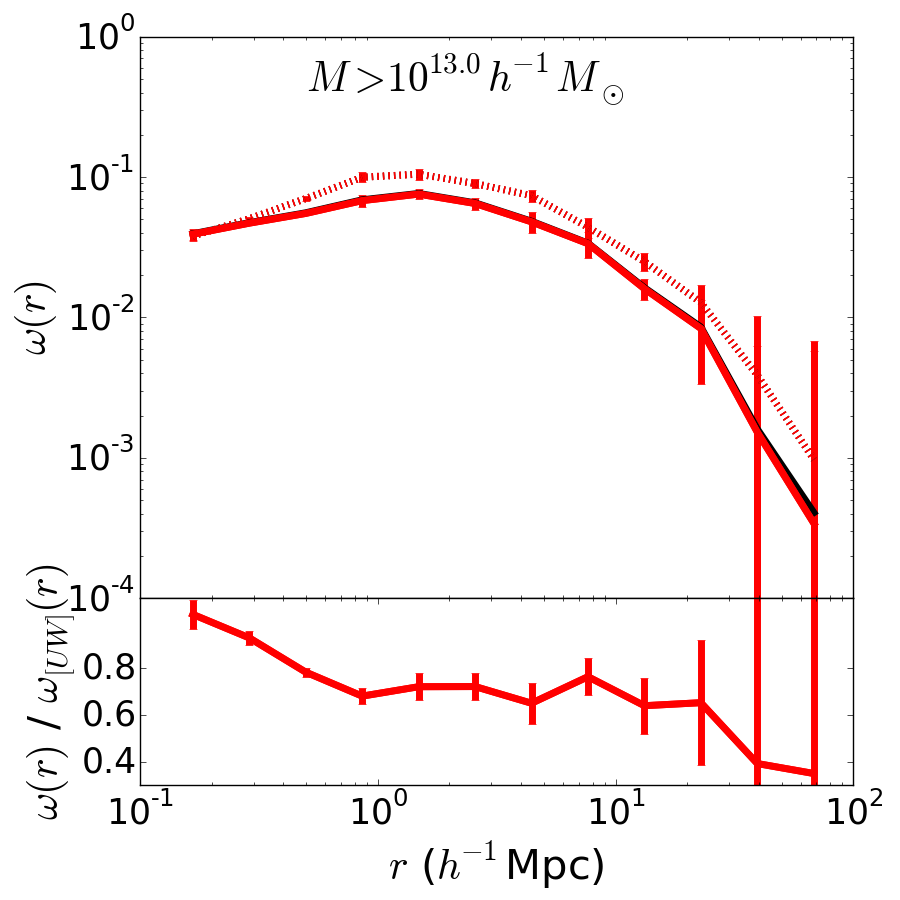}
\caption{\label{F:fig_ed_itensor} ED correlation function, $\omega(r)$,
  for the 3D shapes of stellar matter obtained using different
  definitions of inertia tensor in subhalos selected by a mass
  threshold. The {\em top} panel, shows the ED correlation function and the {\em bottom} panel shows the ratio of the signals obtained using iterative reduced inertia tensor with the unweighted inertia tensor. Note that in the {\em top} panel, the lines labeled Unweighted and Unweighted (Iterative); Reduced and Reduced (Iterative) are close enough that they cannot be easily distinguished. {\em Left:} $M > 10^{11}\hMsun$ ($24648$ galaxies); {\em Middle:} $M >
  10^{12.0}\hMsun$ ($2947$ galaxies); {\em Right:} $M > 10^{13.0}\hMsun$ ($267$ galaxies) at $z=0.3$.}
\end{center}
\end{figure*}
\begin{figure*}
\begin{center}
\includegraphics[width=2.25in]{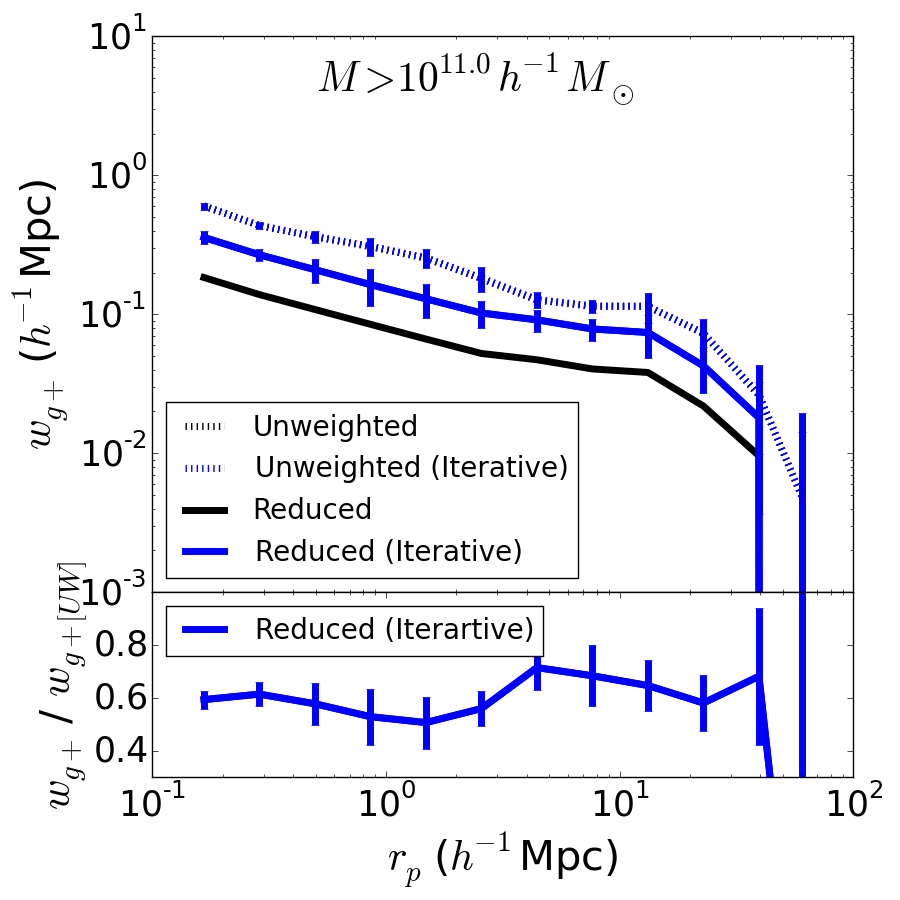}
\includegraphics[width=2.25in]{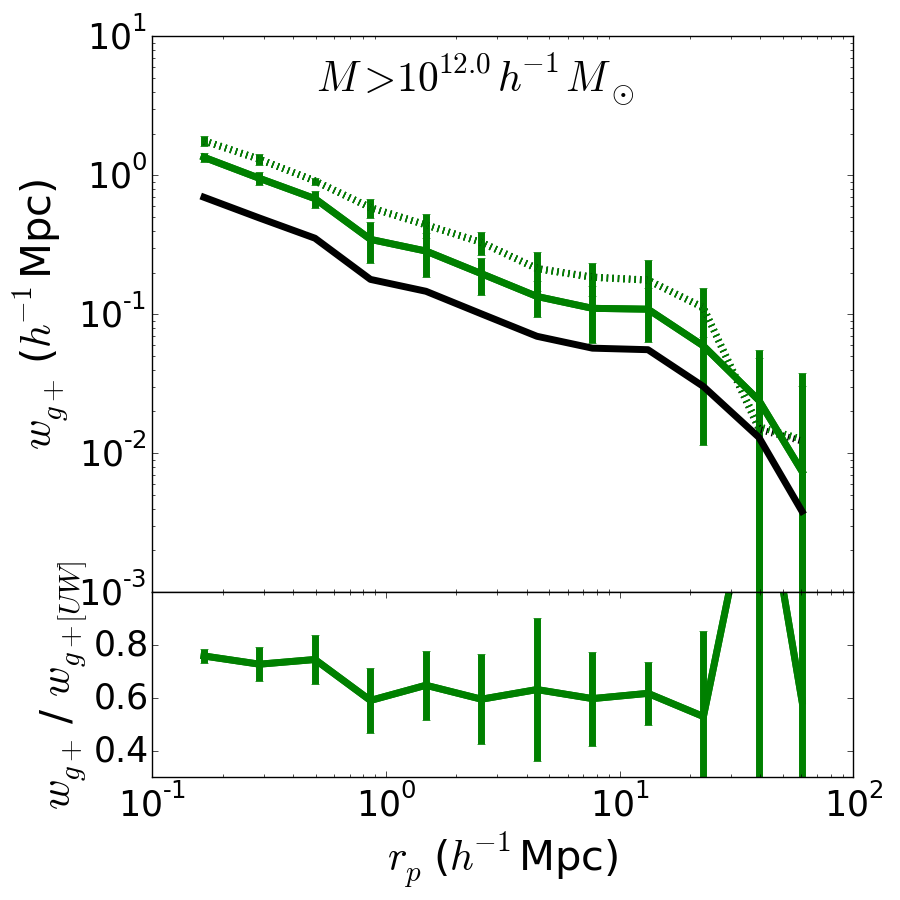}
\includegraphics[width=2.25in]{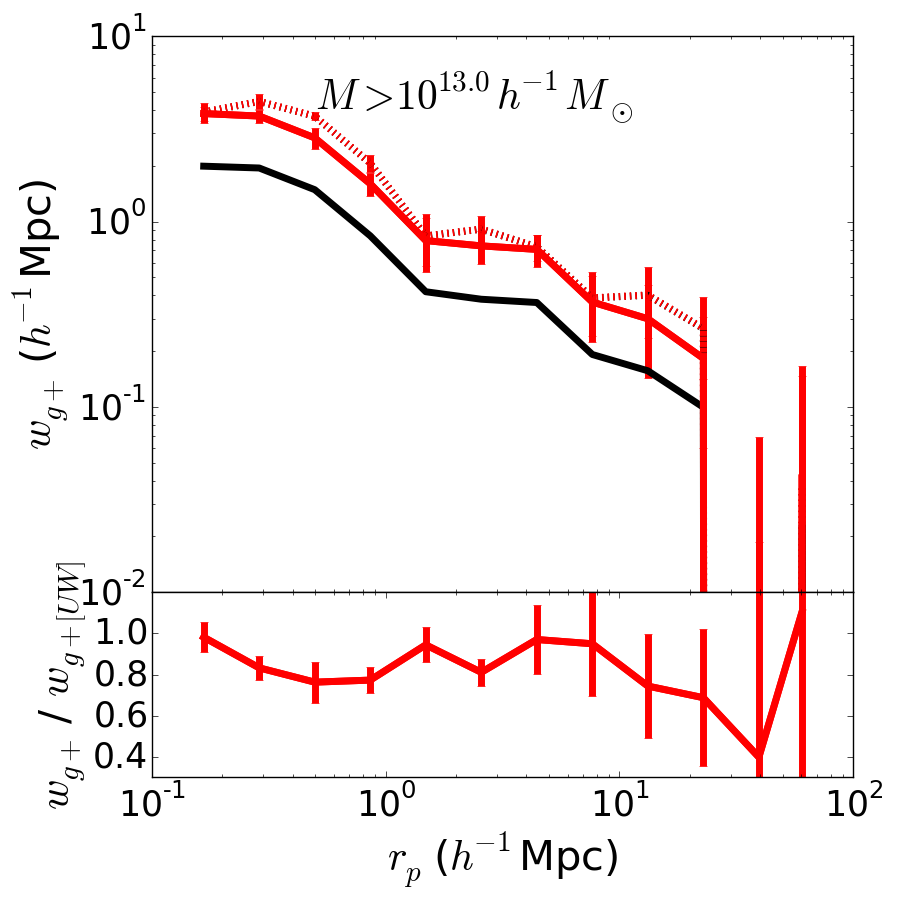}
\caption{\label{F:fig_wgp_itensor} $w_{g+}$ correlation function for
  the projected (2D) shapes of stellar matter obtained using different definitions
  of inertia tensor in subhalos selected by a mass threshold at $z=0.3$. The {\em top} panel, shows the $w_{g+}$ correlation function and the {\em bottom} panel shows the ratio of the signals obtained using iterative reduced inertia tensor with the unweighted inertia tensor. Note that in the {\em top} panel, the lines labeled Unweighted and Unweighted (Iterative) are close enough that they cannot be easily distinguished. {\em
    Left:} $M > 10^{11}\hMsun$; {\em Middle:} $M > 10^{12.0}\hMsun$;
  {\em Right:} $M > 10^{13.0}\hMsun$.
}
\end{center}
\end{figure*}
Here, we consider the dependence of the intrinsic alignment two-point correlation
functions for the shapes of  the stellar matter component in subhalos using
different definitions of inertia tensor. For both ED and $w_{g+}$ correlation functions, the errors bars shown in the plots are obtained using the jackknife variance. 

In Fig.~\ref{F:fig_ed_itensor}, we show the ED correlation function,
$\omega(r)$, for the shapes of the stellar matter component for subhalo
mass thresholds of $10^{11}\hMsun$, $10^{12}\hMsun$, and
$10^{13}\hMsun$. Similar to the histograms of misalignment angles, the
position-angle correlation functions are the same when we use iterative or non-iterative
definitions of inertia tensor. The correlation functions are noticeably
smaller
if we use the reduced inertia tensor to define the shape for all the mass
thresholds.

In Fig.\ref{F:fig_wgp_itensor}, we show the projected shape
correlation function, $w_{g+}(r)$, in different mass bins. We do not
observe any significant difference in the correlation function if we
use the non-iterative vs. iterative unweighted inertia tensor to define
shape. This is consistent with histograms of axis ratios, misalignment
angles and the ED correlation function shown before. Going to the
reduced definition of inertia tensor, it can be seen that $w_{g+}$
is smaller for the shapes obtained from iterative reduced inertia
tensor. This is expected due to the lower ellipticities (or higher axis
ratios) obtained using the reduced inertia tensor. The
values of $w_{g+}$ for the reduced non-iterative shape tensor are even
smaller due to the very high axis ratios, however as mentioned
previously we do not consider this a viable way of measuring shapes.

Our analysis presented in this section shows that the results from
the unweighted non-iterative inertia tensor are quite similar to those
obtained using the iterative tensor, so we do not have to consider this
option separately. It is fair to not consider the results obtained
using non-iterative reduced inertia tensor due to the expectation that
it will produce overly round shapes. Based on these conclusions,
our further analysis is based on the shapes obtained using only the
iterative versions of unweighted and reduced inertia tensors

\subsection{Shapes determined using luminosity weighting}\label{lumweighting}

\begin{figure*}
\begin{center}
\includegraphics[width=3.2in]{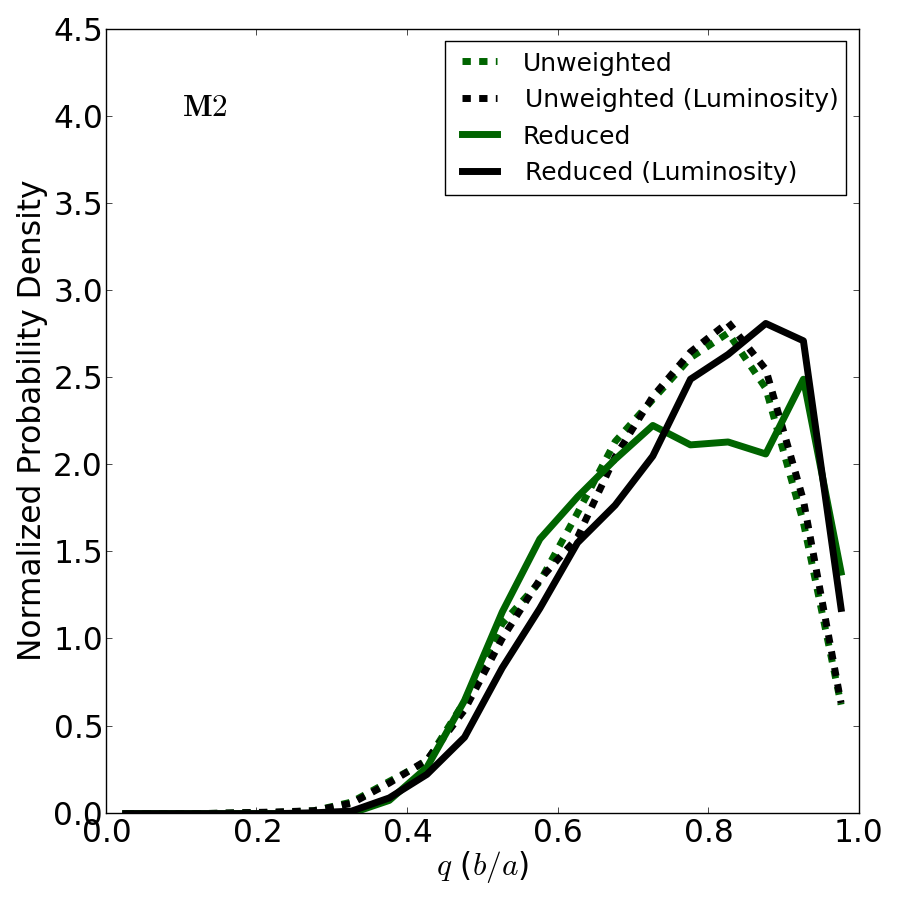}
\includegraphics[width=3.2in]{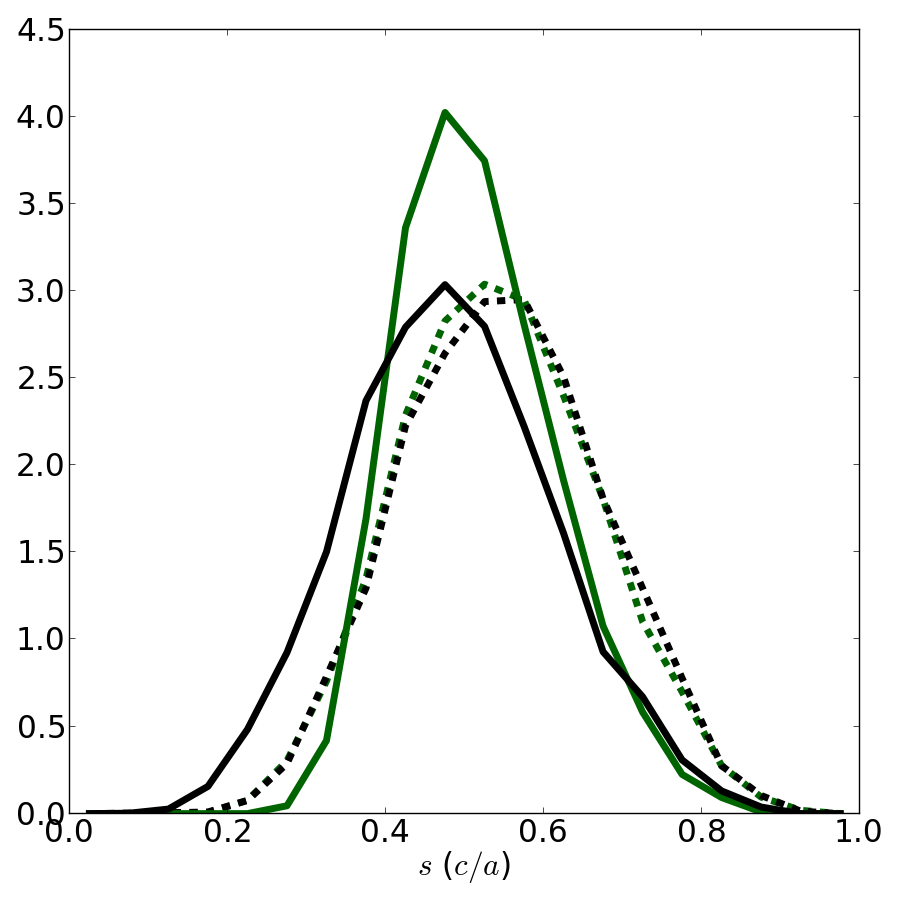}
\caption{\label{F:fig_strlum_qs} Normalized histograms of 3D axis ratios of stellar matter in subhalos using iterative unweighted and
  iterative reduced inertia tensors with each particle weighted by its
  luminosity or mass.  Results are shown only for the mass bin
  M2. {\em Left:} $q~(b/a)$; {\em Right:} $s~(c/a)$.}
\end{center}
\end{figure*}

\begin{figure*}
\begin{center}
\includegraphics[width=2.25in]{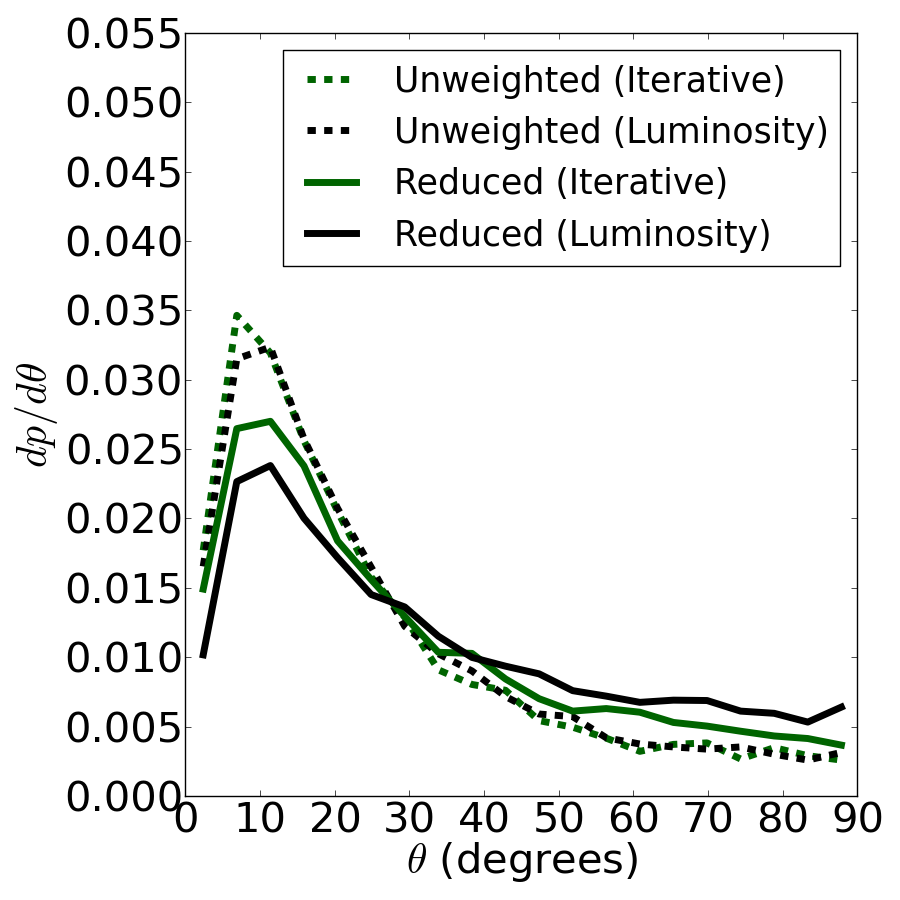}
\includegraphics[width=2.25in]{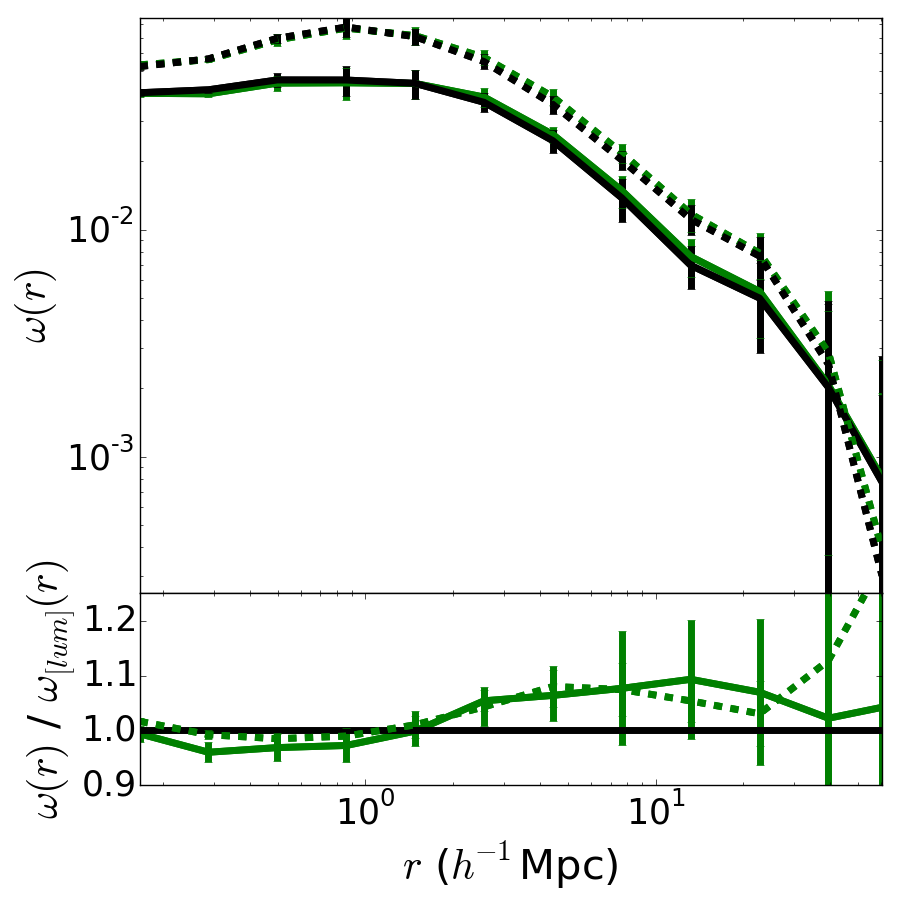}
\includegraphics[width=2.25in]{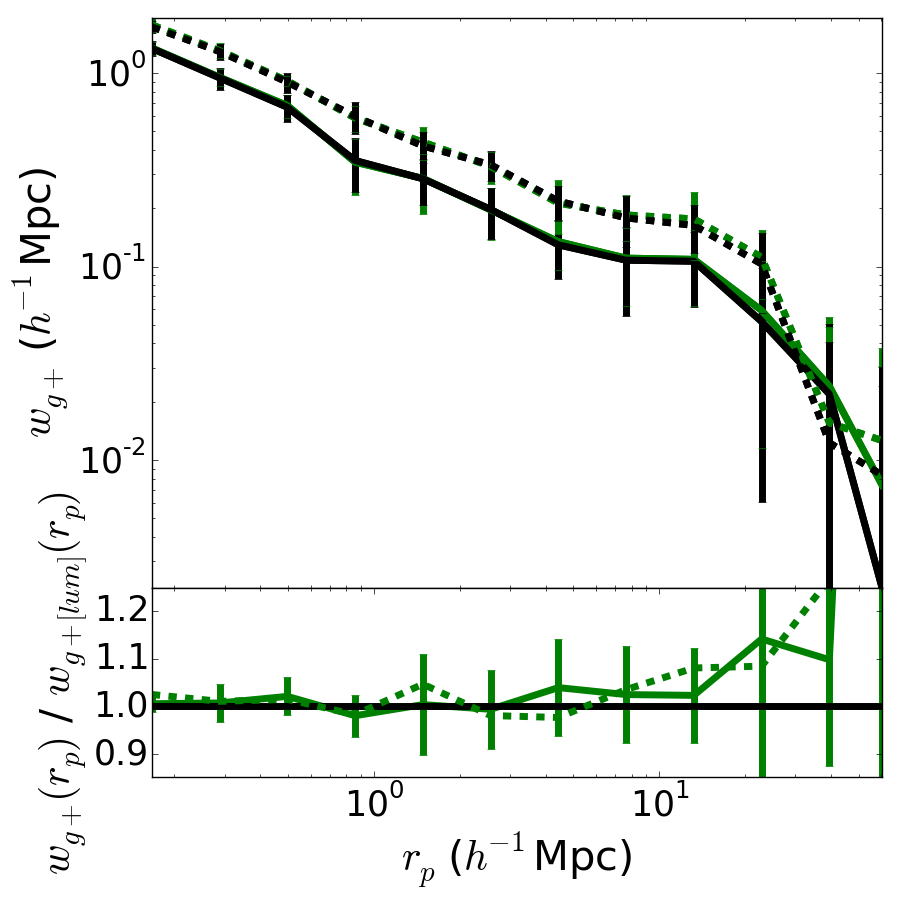}
\caption{\label{F:fig_strlum_ang} {\em Left:} Normalized histogram of misalignment angles using luminosity weighted shapes
  of stellar matter in subhalos in the mass bin, M2: $10^{11.5} -
  10^{13.0}$\hMsun  at $z=0.3$. {\em Middle:} ED correlation of luminosity
  weighted shapes of stellar matter in subhalos for $M >
  10^{12}\hMsun$. {\em Right:} $w_{g+}$ correlation of luminosity
  weighted shapes of stellar matter in subhalos for $M >
  10^{12}\hMsun$. For a direct comparison, the ratio of signals obtained using the mass and luminosity weighted inertia tensors are shown in the {\em bottom} panels.
}
\end{center}
\end{figure*}

In this section, we investigate the effect of weighting 
each stellar particle by its luminosity instead of mass while
computing the inertia tensor. For the unweighted inertia tensor, we
follow Eq.~\ref{eq:luminertensor}; the reduced form
of the luminosity weighted inertia tensor can be inferred from it in a
straight-forward manner. For each star particle, we use the SDSS
$r$-band luminosity from the simulation, and determine shapes
iteratively.

In Fig.~\ref{F:fig_strlum_qs}, we show the histograms of axis ratios
(in the M2 mass bin) of stellar matter in subhalos, computed using
both the mass- and luminosity-weighted form for the unweighted and
reduced inertia tensor. From the plot, we can see that there is no
major change in the distribution of axis ratios due to luminosity
weighting for the unweighted inertia tensor. The histograms of axis
ratios obtained from the reduced inertia tensor show that the
luminosity weighting leads to larger values of $q$ ($\frac{b}{a}$) and
smaller values of $s$ ($\frac{c}{a}$). Thus, the overall shapes are
more oblate when using luminosity weighting. This is expected as the
mass to light ratio is not constant in the inner regions of the
subhalos.

Likewise, we can infer from
the left panel of Fig.~\ref{F:fig_strlum_ang} that luminosity
weighting has no effect on the distribution of misalignment angles in
the unweighted case, while the stellar shapes obtained from reduced
luminosity weighting are more misaligned with the shapes of their host
dark matter subhalos. The middle panel of the same figure shows the ED
correlation function, $\omega(r)$, and the right panel shows the plot
of $w_{g+}(r)$. In the bottom panels, we plot the ratio of the ED and $w_{g+}$ signals obtained using the mass weighted inertia tensor with the ones using luminosity weighted tensor. Both the plots indicate that the amplitude and shape of correlation functions obtained using luminosity-weighted shapes are consistent with the ones obtained using mass-weighted shapes. Similarly, at other mass thresholds, the effect of luminosity weighting on correlation functions is not very significant.
Although the histograms of shapes and misalignment angles obtained by
using the reduced form of luminosity weighted inertia tensor are
different, we do not observe a significant change in the two point
correlation functions, in comparison with the much stronger mass
dependence of the two-point
correlation shown in Sec.~\ref{mass2point}. So, we do not consider
luminosity weighted inertia tensor in the rest of the sections in this
paper. 

\section{Color dependence of intrinsic alignments}\label{ia_color}
In this section, we investigate the color dependence of 
galaxy shape distributions, misalignment angle distributions, and
two-point correlation functions. To 
do this, we roughly divide our entire sample of galaxies into red and
blue types.
\begin{figure}
\begin{center}
\includegraphics[width=3.2in]{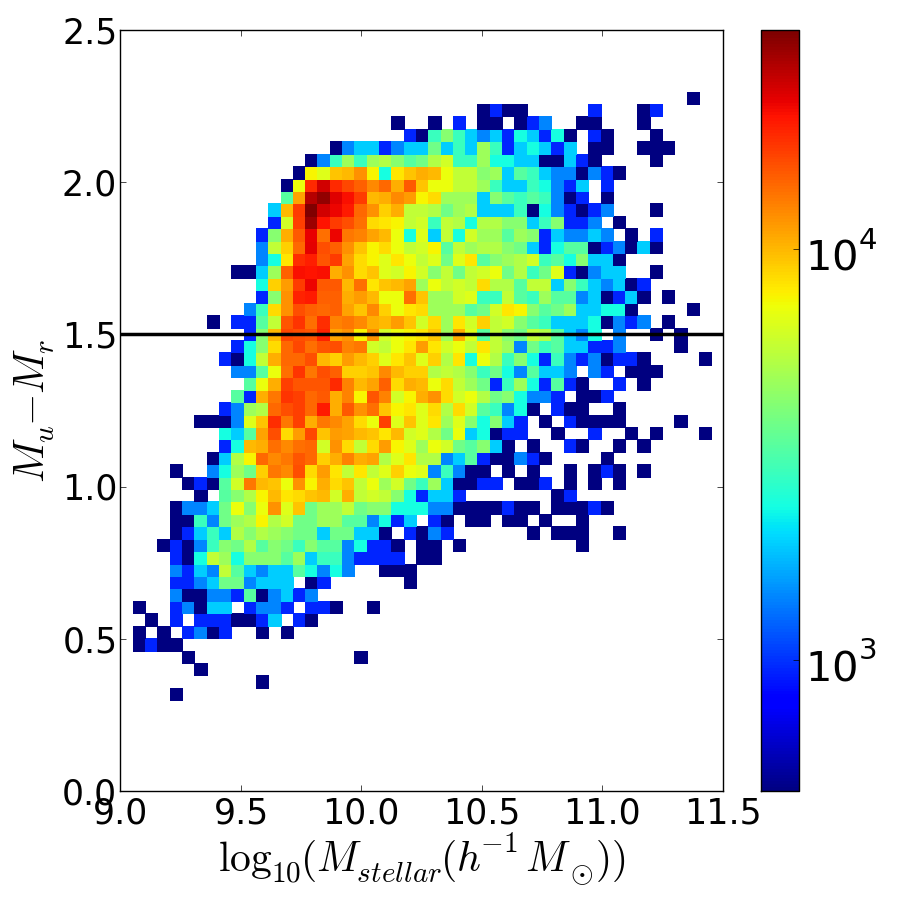}
\caption{\label{F:fig_color} Rest-frame color ($M_{u}-M_{r}$) versus
  stellar mass for galaxies in the simulation at $z=0.3$.}
\end{center}
\end{figure}

\subsection{Division into blue and red galaxies}

The color of a galaxy is obtained by calculating the difference in the
absolute magnitudes in the SDSS $u$-band ($M_{u}$) and $r$-band
($M_{r}$) obtained from the simulation. In Fig.~\ref{F:fig_color}, we
show a 2D histogram of color ($M_{u}-M_{r}$) versus the stellar
mass of subhalos at $z=0.3$. Prior to plotting this histogram, we
imposed a magnitude limit by eliminating galaxies with $M_{r} < -18$ and eliminated galaxies with very
bright AGNs.  

Our colors
do not exactly match those from observations, which have a clear
bimodal distribution in the color-mass contour plot. So, we choose the
median of $M_{u}-M_{r}$ to roughly divide our sample of galaxies in
the simulation into blue and red types. It is important to bear in
mind that because of the procedure we have used, this might not be
exactly analogous to the blue vs. red divisions used in studies of observed  galaxies \citep[e.g.,][]{2007MNRAS.381.1197H}. Together with the fact that
color and morphology are not perfectly correlated, this implies that our color based division is not
same as a division into bulge-dominated and disk-dominated galaxies.

\cite{2014arXiv1406.4668C}
used a similar definition. They used the $u-r$ rest-frame colors to
divide their sample of galaxies in the simulation into three equal
bins consisting of blue, red/blue and red types. \cite{2014MNRAS.444.1518V} divided the sample of galaxies in the Illustris simulation using the $u-i$ color into blue, green and red types based on a star formation rate threshold, but they only produce a slightly bimodal distribution in colors that is not comparable with observations. 

Here, we only consider the
shapes obtained from the iterative reduced inertia tensor for our
analysis in this section. We obtain similar results using the unweighted inertia tensor. 

\subsection{Axis ratio distributions}

\begin{figure*}
\begin{center}
\includegraphics[width=3.2in]{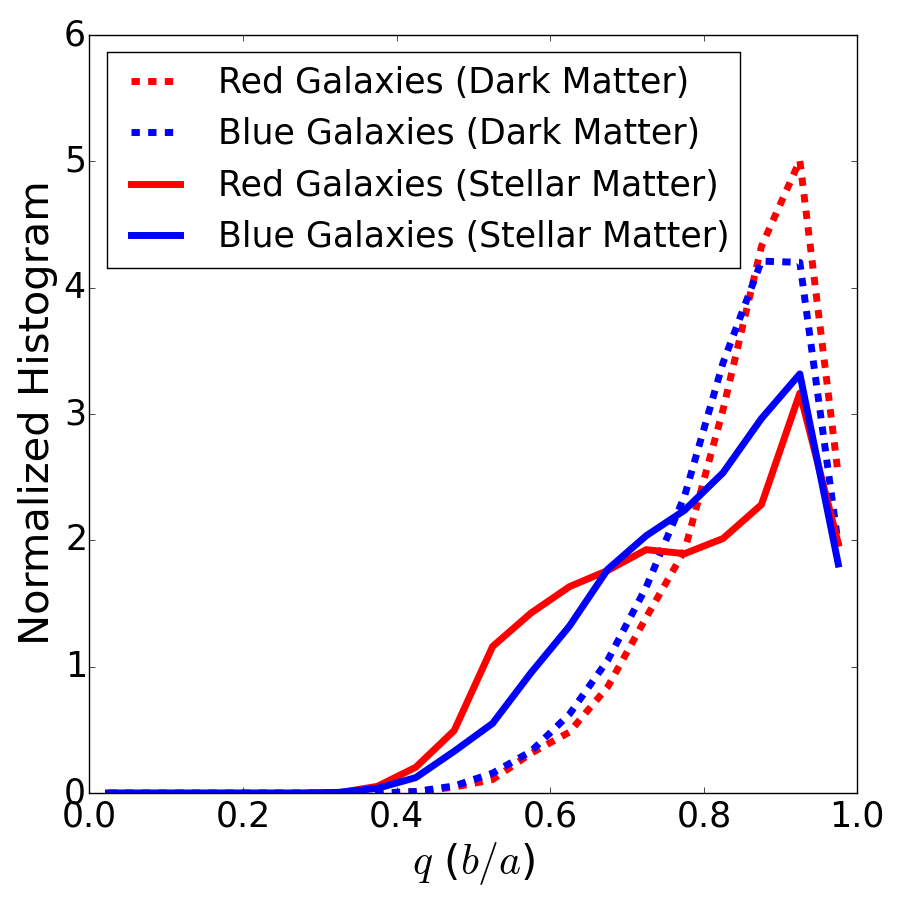}
\includegraphics[width=3.2in]{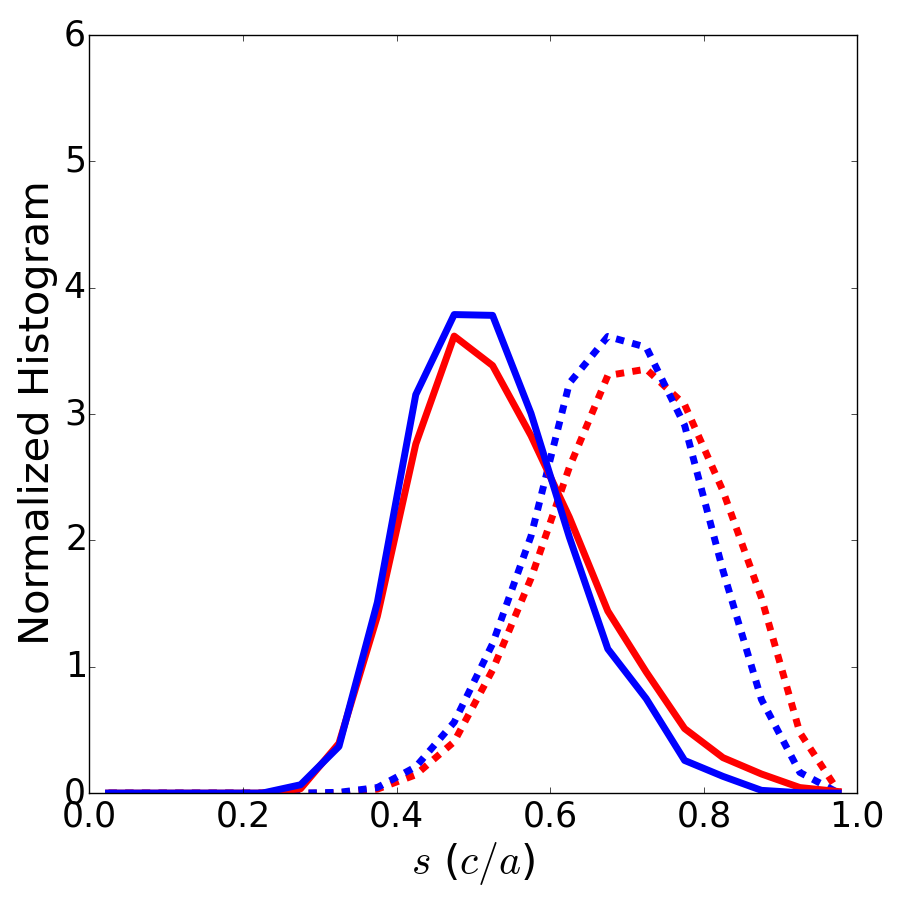}
\caption{\label{F:fig_clrdm_qs} Normalized histograms of axis ratios ($q$, $s$)
  of dark matter and stellar matter component in subhalos for blue ($6343$ galaxies) and red ($6343$ galaxies) galaxies at
  $z=0.3$. {\em Left:} $q~(b/a)$; {\em Right:} $s~(c/a)$.}
\end{center}
\end{figure*}

The histograms of axis ratios of dark matter and stellar
matter component in subhalos for the red and blue galaxies are shown in
Fig.~\ref{F:fig_clrdm_qs}. The plots show that the red galaxies have slightly
higher (rounder) axis ratios for the shapes defined by dark matter. For the
shapes defined by stellar matter, the blue galaxies have slightly
higher values of $q~(\frac{b}{a})$ and lower values of
$s~(\frac{c}{a})$, indicating more oblate or disk-like shapes, as we
would expect.

\subsection{Misalignment angles and two-point correlation functions}

\begin{figure*}
\begin{center}
\includegraphics[width=2.25in]{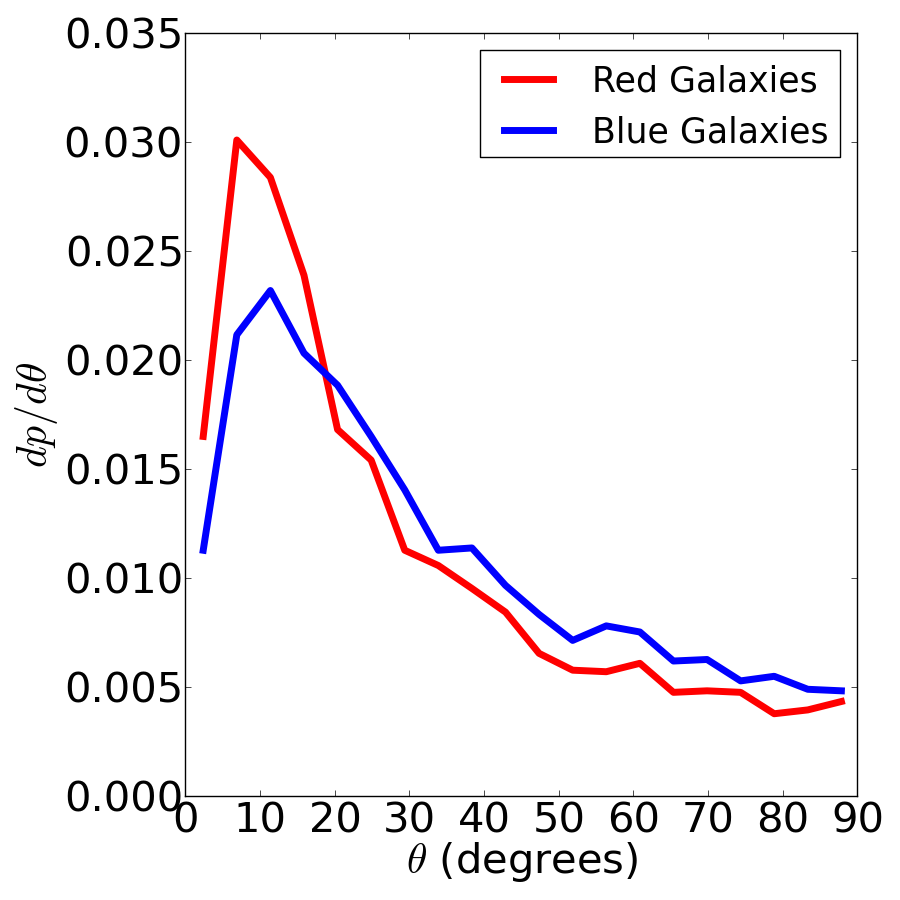}
\includegraphics[width=2.25in]{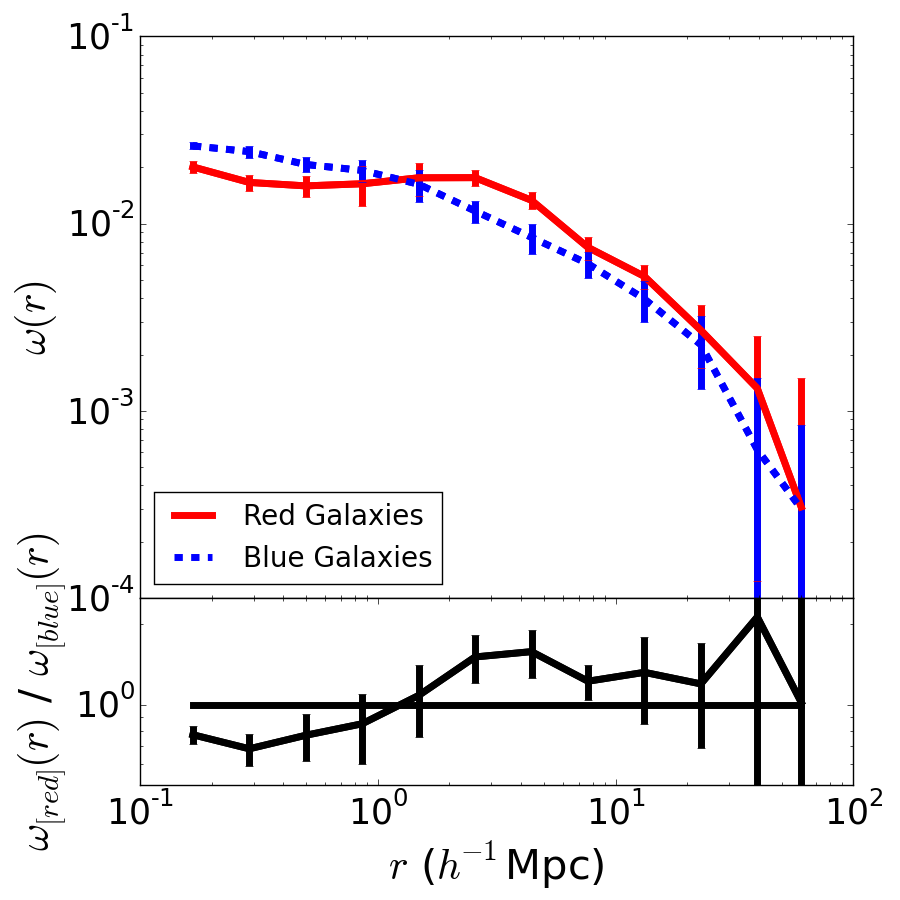}
\includegraphics[width=2.25in]{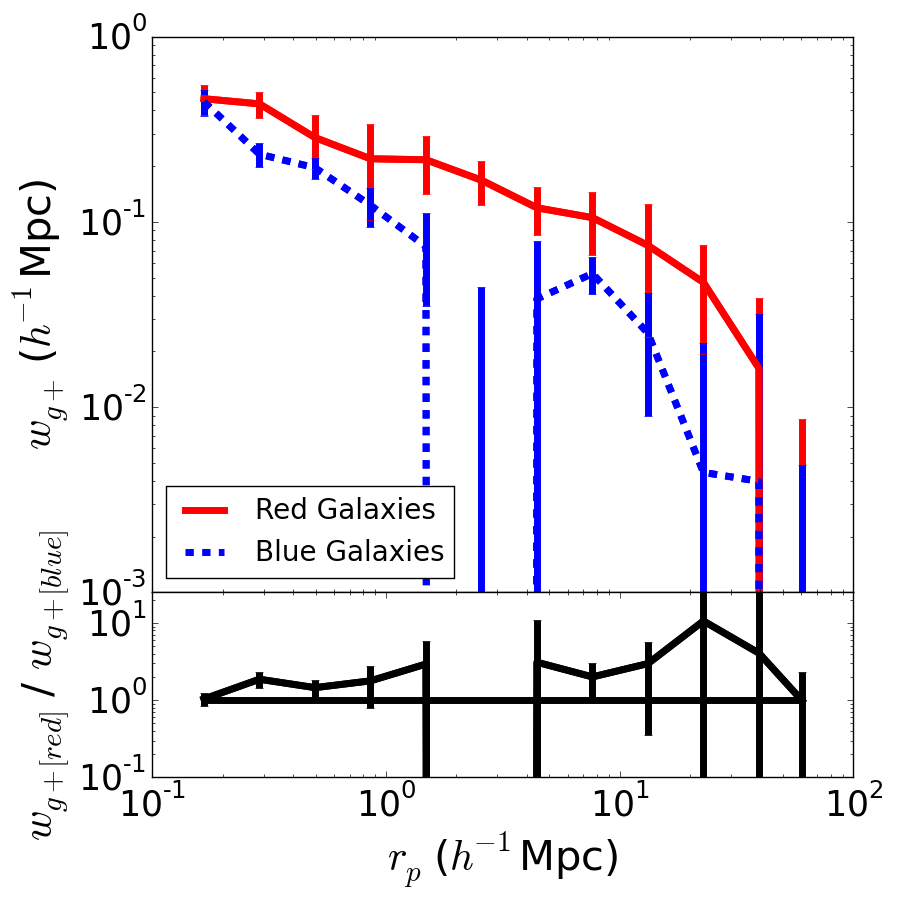}
\caption{\label{F:figclr_red} Comparison of misalignment angles and
  two-point correlation function in red and blue galaxies at $z=0.3$. {\em Left:}
  Histogram of misalignment angles; {\em Middle} ED position angle
  statistic; {\em Right} $w_{g+}$ projected shape correlation
  function. At around $\sim 1\hmpc$, the correlation function becomes negative for the blue galaxies.}
\end{center}
\end{figure*}

The histogram of misalignment angles shown in the left panel of
Fig.~\ref{F:figclr_red} indicates a larger misalignment between dark
matter halo and galaxy shapes in blue
galaxies. The mean misalignment angles are $29^{\circ} \pm 0.3^{\circ}$ and $33^{\circ} \pm 0.3^{\circ}$ respectively for red and blue galaxies. If we wish to interpret these differences, we
have to consider other factors that might change the distribution of misalignment angles, the most important of which is the mass.  
The mean masses of the sample of red and blue
galaxies are similar. The red (blue) sample has a mean subhalo mass of
$8.0$ ($7.9$) $\times 10^{11}\hMsun$.  Given the nearly consistent
masses, the larger alignment for the red sample is not entirely due to mass.

In the middle panel of Fig.~\ref{F:figclr_red} we show the ED correlation for red and blue galaxies. The $w_{g+}$ signals are plotted in the right panel. From the ratio plots of the intrinsic alignment  signals for red and blue galaxies shown in the bottom panel, we conclude that there is no significant difference for our sample of red and blue galaxies.  

\section{Mass and redshift dependence of two-point correlation functions}\label{ia_mdrs}
In this section, we show the results for intrinsic alignment two-point correlation
functions for shapes defined by the stellar matter component in
subhalos. We focus in particular on the mass dependence and redshift
evolution of the ED and $w_{g+}$ correlation function.
\subsection{Mass dependence}\label{mass2point}

\begin{figure*}
\begin{center}
\includegraphics[width=3.2in]{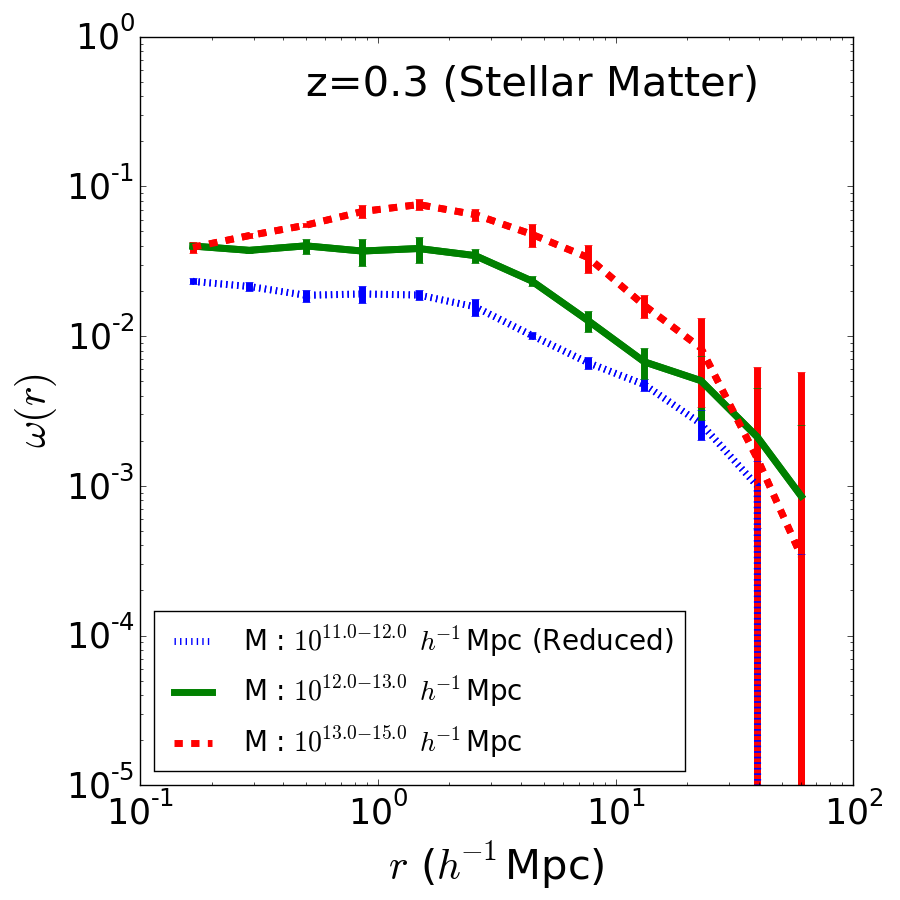}
\includegraphics[width=3.2in]{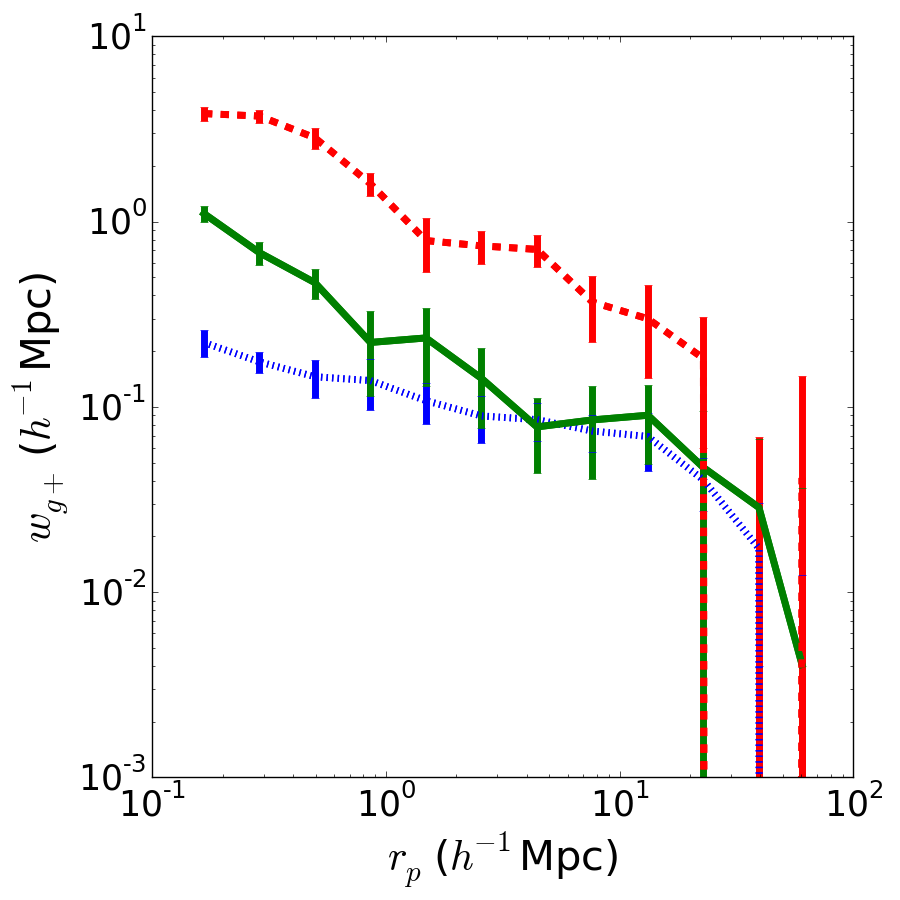}
\caption{\label{F:fig_md_edwgp} Mass dependence of two-point
  correlation functions for shapes defined by stellar matter in
  subhalos at $z=0.3$ using iterative reduced inertia
  shape tensors. {\em Left:} Position angle statistic, ED
  correlation function; {\em Right:} Projected shape-correlation
  function, $w_{g+}$.}
\end{center}
\end{figure*}

\begin{figure*}
\begin{center}
\includegraphics[width=3.2in]{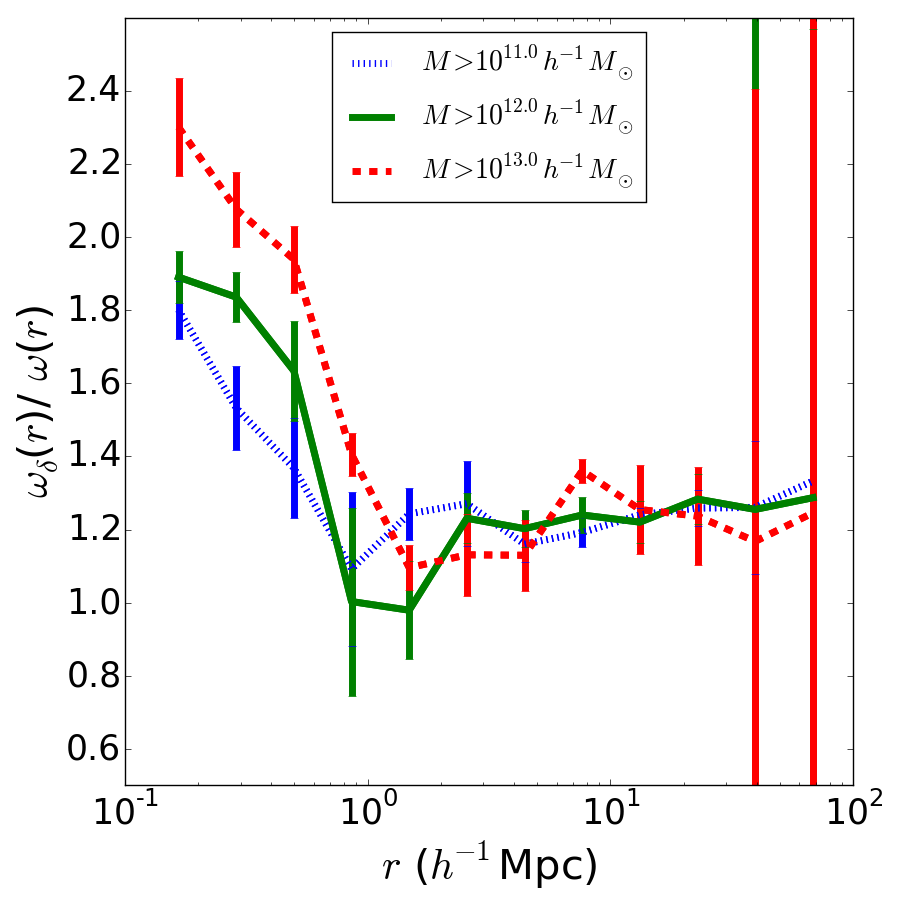}
\includegraphics[width=3.2in]{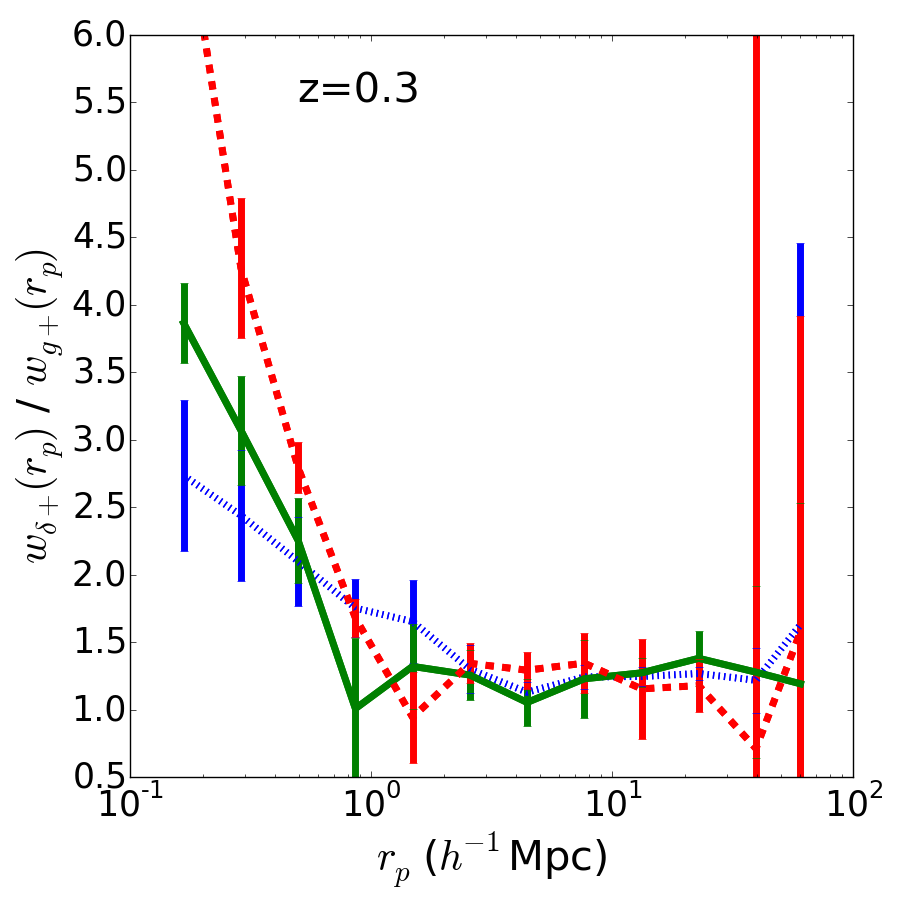}
\caption{\label{F:fig_sgdm} {\em Left:} Ratio of
  $\omega_{\delta}(r)$(density field traced by dark matter particles)
  to $\omega(r)$ {\em Right:} Ratio of $w_{\delta +}(r_{p})$
  correlation function, where the density field is traced by dark
  matter particles, to $w_{g+}(r_{p})$ (density field traced by
  subhalos).}
\end{center}
\end{figure*}

In Fig.~\ref{F:fig_md_edwgp}, we consider the mass dependence of
two-point correlation functions for shapes defined by stellar matter in
subhalos. The left panel shows the ED correlation function
for shapes obtained using iterative reduced inertia
tensors. The galaxy samples here are selected based on total subhalo mass in the mass bins, $M : 10^{11-12}\hMsun$, $10^{12-13}\hMsun$ and $10^{13-15}\hMsun$. We
observe a substantial increase in the amplitude of these correlation functions with
increasing mass for both ED and $w_{g+}$. 
For $M :
10^{13-15}\hMsun$, the correlation function dips at small scales,
possibly indicating a slightly random alignment of satellite subhalos with the
orientation of the central galaxy. In a previous study, \cite{2008MNRAS.389.1266L} investigated the ED 
correlation functions from $N$-body simulations for
shapes defined by dark matter. However, 
we know that the shape defined by stellar matter in galaxies is
misaligned with the shape of host dark matter subhalo
\citep{2014MNRAS.441..470T}. This can significantly change 
the ED correlation function of shapes obtained with stellar matter when compared with results from an $N$-body simulations. For instance, previous studies have noted that there is a suppression in the intrinsic alignment signal due to misalignment of galaxy shape with the host dark matter shape \citep{2006MNRAS.371..750H,2009ApJ...694..214O,2011JCAP...05..010B}. Qualitatively similar to our results of ED
correlation using shapes of stellar matter component,
\cite{2008MNRAS.389.1266L} also found that the correlation of dark matter
shapes with the density field increases with halo mass at all
scales. However, \cite{2008MNRAS.389.1266L} only measure the signal starting at $r > 1\hmpc$. So, we cannot directly compare our results at small scales for the mass bin, $M : 10^{13-15}\hMsun$, where we observe a dip in the correlation. 

In the right panel of Fig.~\ref{F:fig_md_edwgp}, we considered 
the mass dependence in $w_{g+}$ using the same
mass bins. For this correlation function, the different
ellipticities and orientation of shapes defined by stellar matter in
galaxies can lead to a different correlation function, when compared
with that obtained using dark matter shapes. For $w_{g+}$, we
observe an increase in the amplitude of correlations with increasing
subhalo mass threshold. The increase in intrinsic ellipticity-density
correlation signal with halo mass is also predicted from $N$-body
simulations \citep{2006MNRAS.371..750H} and semi-analytic models
\citep{2013MNRAS.436..819J}. Unlike the ED correlation in 3D, and
consistent with observations of $w_{g+}$ for real galaxies \citep[e.g.,][]{2007MNRAS.381.1197H}, we do not observe a dip in $w_{g+}$ for $M>10^{13}\hMsun$ at small scales.

Although we do not show the mass dependence of intrinsic alignments for the shapes obtained using the iterative unweighted inertia tensor, it can be inferred from the plots shown in \ref{F:fig_ed_itensor} and \ref{F:fig_wgp_itensor} that $ED$ and $w_{g+}$ correlation functions have similar mass dependence using the unweighted tensor. However, for comparison with observations, we expect that the iterative 
reduced inertia tensor might be a better choice as it gives more
weight to the particles in the inner regions of subhalos. Hence, in the rest of this paper, we only present the two-point statistics using shapes obtained from the iterative reduced inertia tensor.  

\subsubsection{Comparison of $w_{g+}$ and $w_{\delta +}$} 

The projected correlation function, $w_{g+}$, includes a factor of the
galaxy bias due to the correlation with galaxy positions. In observational data, it is necessary to estimate a large-scale galaxy bias and use
the linear bias approximation to remove this galaxy bias dependency,
an approach which should fail on small to intermediate scales. 
In order to take the effect of subhalo bias at large scales into
consideration, here we considered the ratio of two-point correlation
functions using the dark matter particles to trace the density field
with those obtained by using the subhalos to trace the density
field. In the left panel of Fig.~\ref{F:fig_sgdm}, we plotted the
$\omega_{\delta}(r)/\omega(r)$ correlation function at $z=0.3$ for
shapes defined by stellar matter in galaxies, for
$M>10^{11}\hMsun$, $M>10^{12}\hMsun$, and $M>10^{13}\hMsun$ 
using the reduced iterative inertia tensor to calculate shapes. At
small scales, we observe that the $\omega_{\delta}(r)$ is a factor of
1.6--2 larger than $\omega (r)$, and the ratio is larger at higher mass
thresholds. This result indicates that the shapes of massive galaxies
are better aligned with the shape of the dark matter field than with
the positions of other galaxies.  The right panel shows a similar
plot for the projected shape correlation function ($w_{\delta
  +}$). Again, we observe a larger $w_{\delta +}$ at small scales, and
the ratio increases with mass threshold. In both the plots, the ratio
is nearly constant at large scales for all mass thresholds, and is
inversely proportional to the large-scale bias of the density tracer
sample (all subhalos in the simulation).  Since the simulation
includes relatively low mass subhalos, their average bias is $<1$ and
hence the ratio that is plotted is slightly above 1.

\subsection{Redshift evolution}\label{rs2point}

\begin{figure*}
\begin{center}
\includegraphics[width=3.2in]{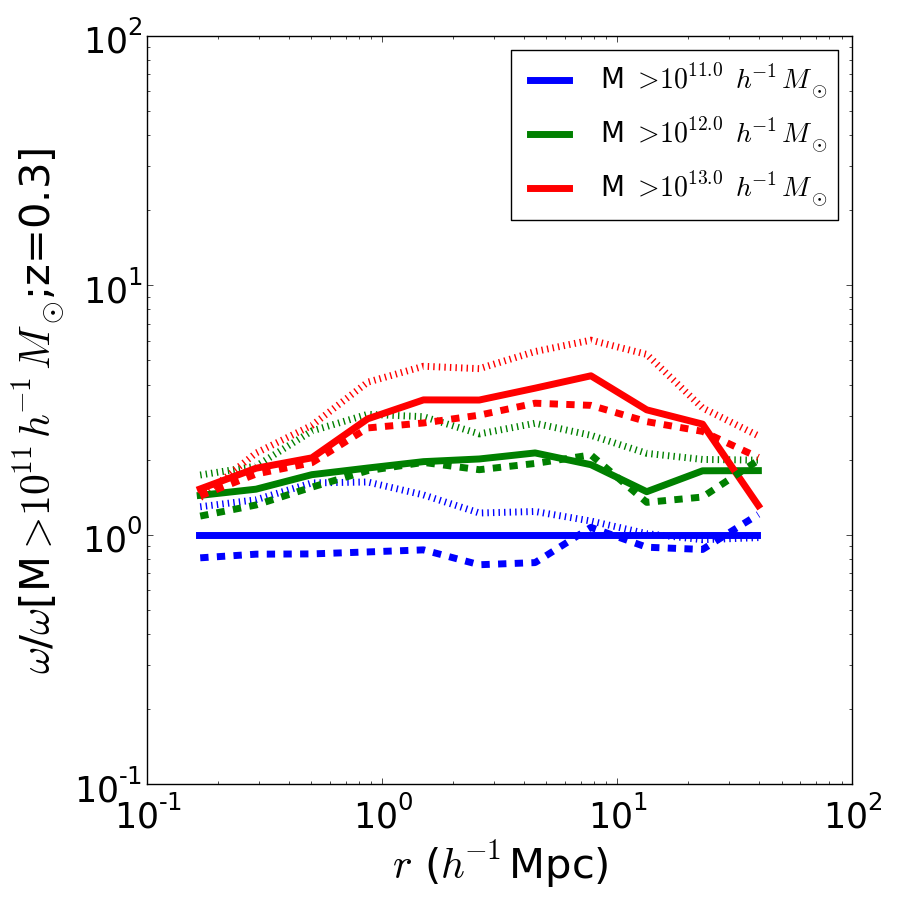}
\includegraphics[width=3.2in]{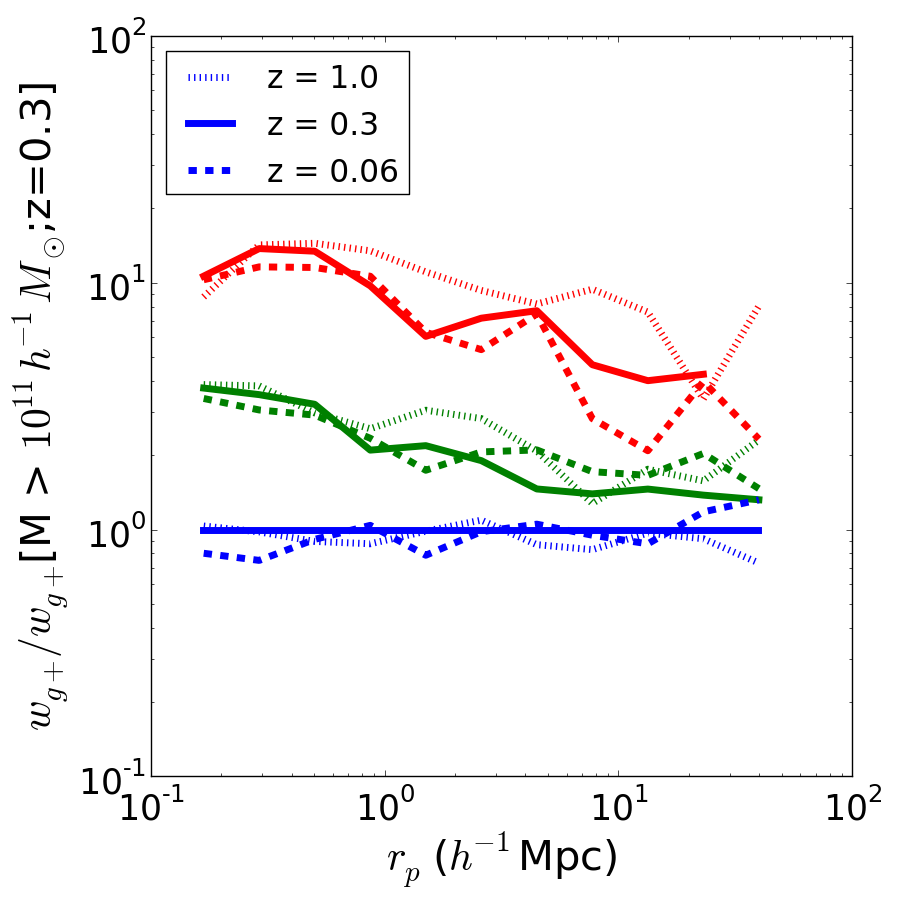}
\caption{\label{F:fig_rsew} Ratio plot of two-point correlation
  functions at redshifts $z=1.0$, $0.3$ and $0.06$ and various
  mass thresholds to the corresponding value at $z=0.3$ for
  $M>10^{11.0}\hMsun$. {\em Left:} ED; {\em Right:} $w_{g+}$.}
\end{center}
\end{figure*}

We show the redshift evolution of intrinsic alignment two-point
correlation functions by plotting the ratios of ED and $w_{g+}$ at
$z = 1.0$, $0.3$, and $0.06$ to the corresponding quantities at $z=0.3$ for
$M>10^{11}\hMsun$. In the left (right) panel of Fig.~\ref{F:fig_rsew}, we
show this ratio for the ED ($w_{g+}$) correlation functions with three subhalo mass
threshold values. We
observe that (for fixed mass threshold) the amplitude of the ED correlation function decreases
significantly at all scales and for all mass thresholds as we go to lower
redshifts. Using $N$-body simulations, \cite{2008MNRAS.389.1266L} also
found that the amplitude of the ED correlation function decreases at 
lower redshifts. However, the $w_{g+}$
correlation function does not exhibit a strong dependence on
redshift.  This is due to a difference in
the shape distributions that compensates for the redshift evolution of
position angle alignments shown by the results for $\omega$.

The linear alignment model predicts that for the range of redshifts
considered here, $w_{\delta +}$
varies roughly as $(1+z)^{-0.7}$ \citep{2007MNRAS.381.1197H}. We do not detect any significant
redshift evolution of $w_{\delta +}$ for most of our samples. 
However, this particular test for redshift evolution based on mass (or luminosity) threshold samples may not be fair for \ia{} evolution, since we are not comparing results for a high-redshift sample of progenitors of the low-redshift sample at a given mass threshold (due to additional mergers and mass accretion). We defer the exploration of this effect to future work.

\section{Two point correlation functions: centrals and satellites}\label{ia_censat}
A central subhalo is located at or near the potential minimum of 
its host halo. The remaining subhalos of that  host halo are
satellites. Here, we investigate the intrinsic alignment two-point
correlation functions for central
and satellite subhalos separately, by looking at the projected shape
correlation in various mass bins. 
\subsection{Alignments of central and satellite galaxies}\label{censatalign}
\begin{figure*}
\begin{center}
\includegraphics[width=3.2in]{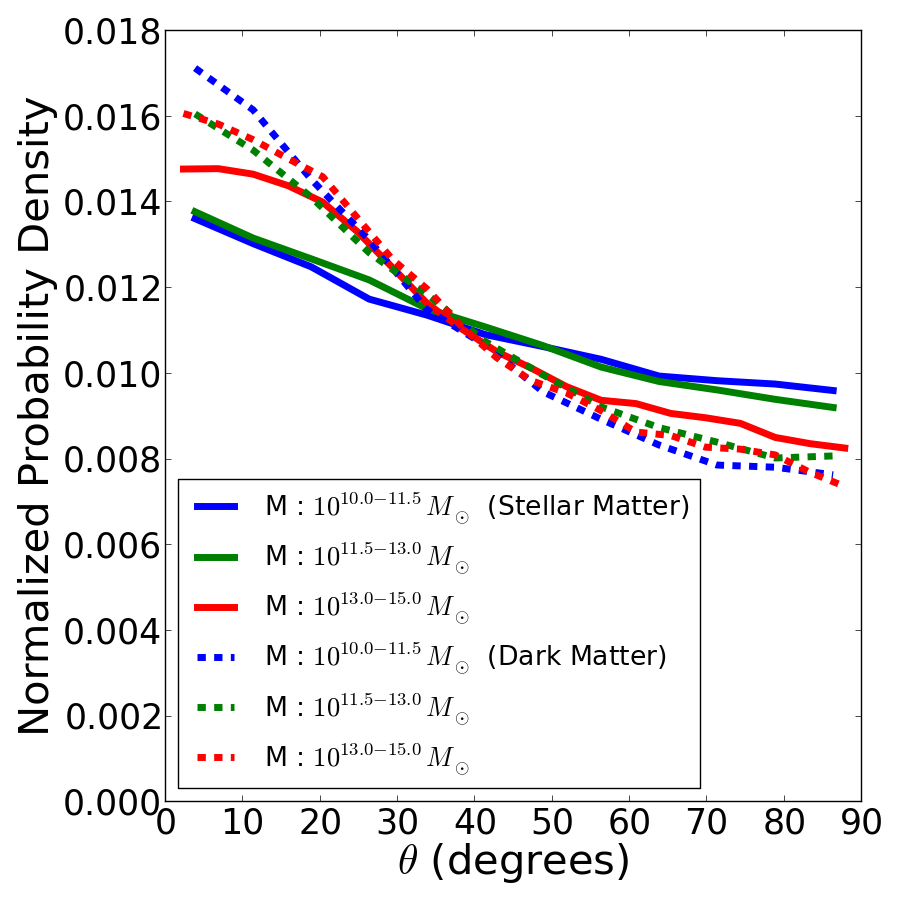}
\includegraphics[width=3.2in]{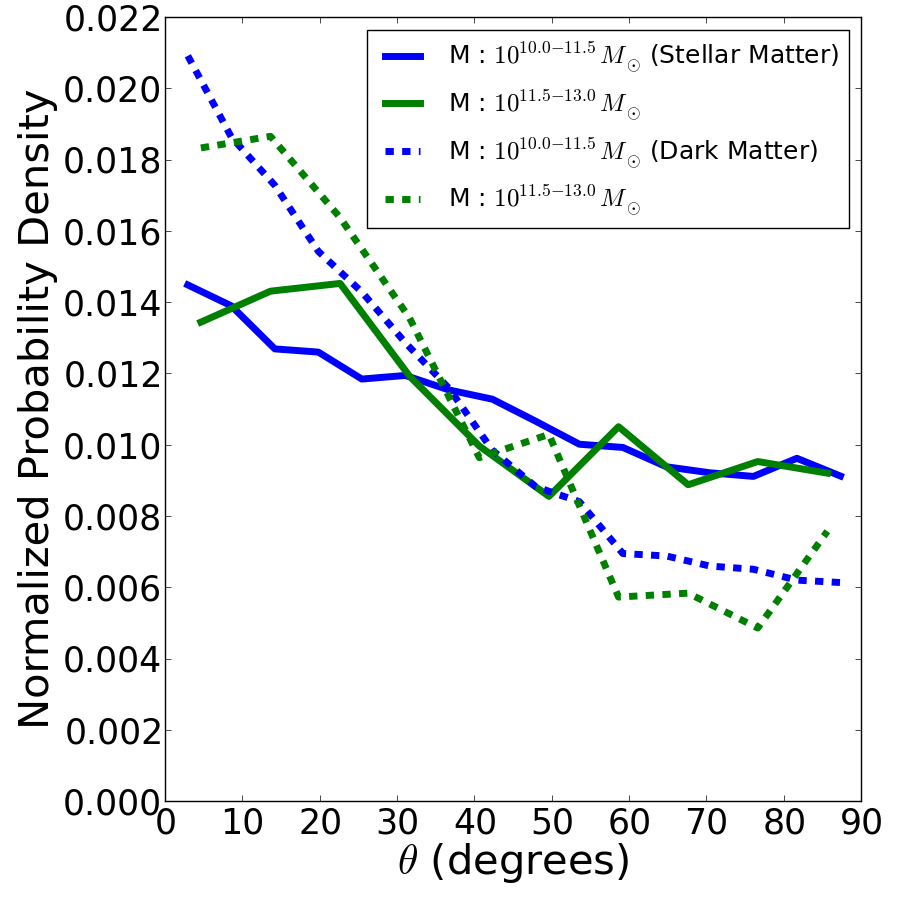}
\caption{\label{F:fig_censat_align} {\em Left:} Normalized histogram
  of alignment angle of the major axis of the 2D stellar shape of a
  central galaxy with satellite subhalos in mass bins, $M1$, $M2$ and $M3$ of central subhalo mass at $z=0.3$. {\em Right:} Normalized
  histogram of alignment angles of the major axis of the 2D stellar
  shape of satellite galaxies with host halo in mass bins, $M1$ and $M2$ of satellite subhalo mass at $z=0.3$.}
\end{center}
\end{figure*}

Observationally, the distribution of satellites around central galaxies has been found to be anisotropic, with more satellites along the major axis of the central galaxy \citep[e.g.,][]{2004MNRAS.348.1236S,2005ApJ...628L.101B,2006MNRAS.369.1293Y,2012ApJ...752...99N,2013ApJ...770L..12L,2014MNRAS.442.1363W,2014ApJ...791...15L}. This has also been studied through $N$-body simulations \citep{2008ApJ...675..146F,2010ApJ...709.1321A,2014ApJ...786....8W} and hydrodynamic simulations of smaller volume \citep{2007MNRAS.374...16L,2011MNRAS.415.2607D}. However, $N$-body simulations overestimate the strength of the alignment signal if it is assumed that the shape of central galaxy follows the shape of dark matter halo \citep{2007MNRAS.378.1531K,2010ApJ...709.1321A}. In a recent paper,
\cite{2014arXiv1407.6708D} used a large volume hydrodynamical simulation without AGN feedback to study this problem.

Here, we explore the distribution of the location of satellite
subhalos with respect to the major axis of the central subhalo in the
host halo. The left panel of Fig.~\ref{F:fig_censat_align} shows the
histogram of angle between the major axis of shapes of dark matter and
stellar matter of a central subhalo with the line joining the
satellite subhalos. From the plot, we can conclude that the satellite
subhalos are more concentrated along the major axis of its central
galaxy. Our results are qualitatively consistent with results from
$N$-body simulations by
\cite{2008ApJ...675..146F,2014ApJ...786....8W}. Using hydrodynamic
simulations of smaller volume, \cite{2011MNRAS.415.2607D} found that
the satellites are more distributed along the axis of the shape
determined by dark matter component of a central subhalo, when
compared with that of stellar matter. We also confirm this finding
with a large statistical sample. From the plot, we can observe that
there is no significant mass dependence in the distribution of
satellites along the major axis of the subhalo with shape determined
using dark matter particles. On the other hand, for shapes defined by
stellar matter of galaxies, we observe that the alignment increases
with increasing subhalo mass with the mean alignment angles being
$42.0^{\circ}$, $41.5^{\circ}$ and $39.6^{\circ}$ in the mass bins,
$M1$, $M2$ and $M3$ respectively. This is due to a greater
misalignment angle between the shapes defined by the dark matter and
stellar matter in less massive central galaxies. \cite{2014arXiv1407.6708D} also studied the spatial distribution of
satellite galaxies with respect to the orientation of their host
central galaxy using a large volume hydrodynamical simulation. They found more alignment in massive halos with mean alignment angles varying from $45^{\circ} - 40^{\circ}$ in the mass range, $10^{11}-10^{14}\hMsun$ which agrees qualitatively with our findings.

We also investigate the orientation of satellite galaxies with respect
to the location of its central subhalo. The right panel of
Fig.~\ref{F:fig_censat_align} shows the alignment of the major axis of
the shapes of satellite galaxies with the direction to their central
subhalo. This is the radial alignment signal, which has been studied
for dark matter component of satellites using $N$-body simulations
\citep{2007ApJ...671.1135K,2008ApJ...672..825P,2008ApJ...675..146F}
and hydrodynamic simulations \citep{2010MNRAS.405.1119K}. These
studies found that the orientation of satellite subhalos is not
random, but point more towards the center of their host halo. Here, we
observe that the shapes of stellar matter in satellites are also more
aligned with the direction to their host halo. Recent observational
measurements of \cite{2013MNRAS.433.2727S,2014arXiv1406.5196S} have
not detected radial alignment of satellite galaxies with their host
halo.

\subsection{$w_{g+}$ and $w_{\delta +}$ for centrals and satellites}
\begin{figure*}
\begin{center}
\includegraphics[width=2.25in]{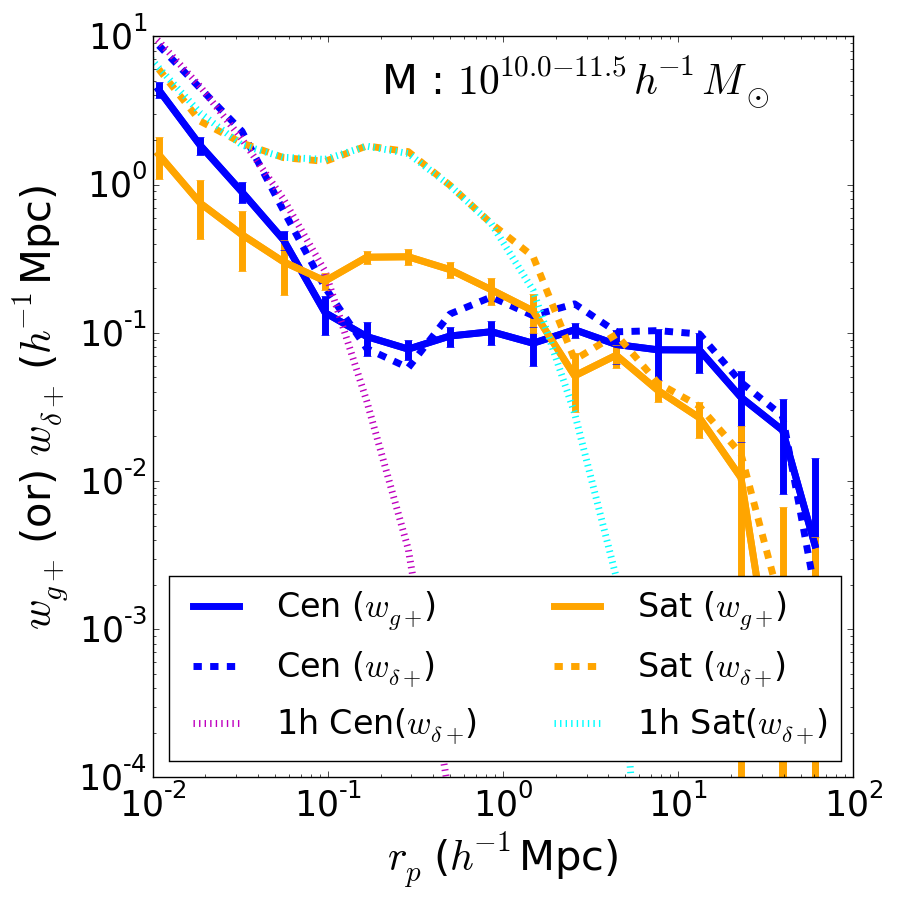}
\includegraphics[width=2.25in]{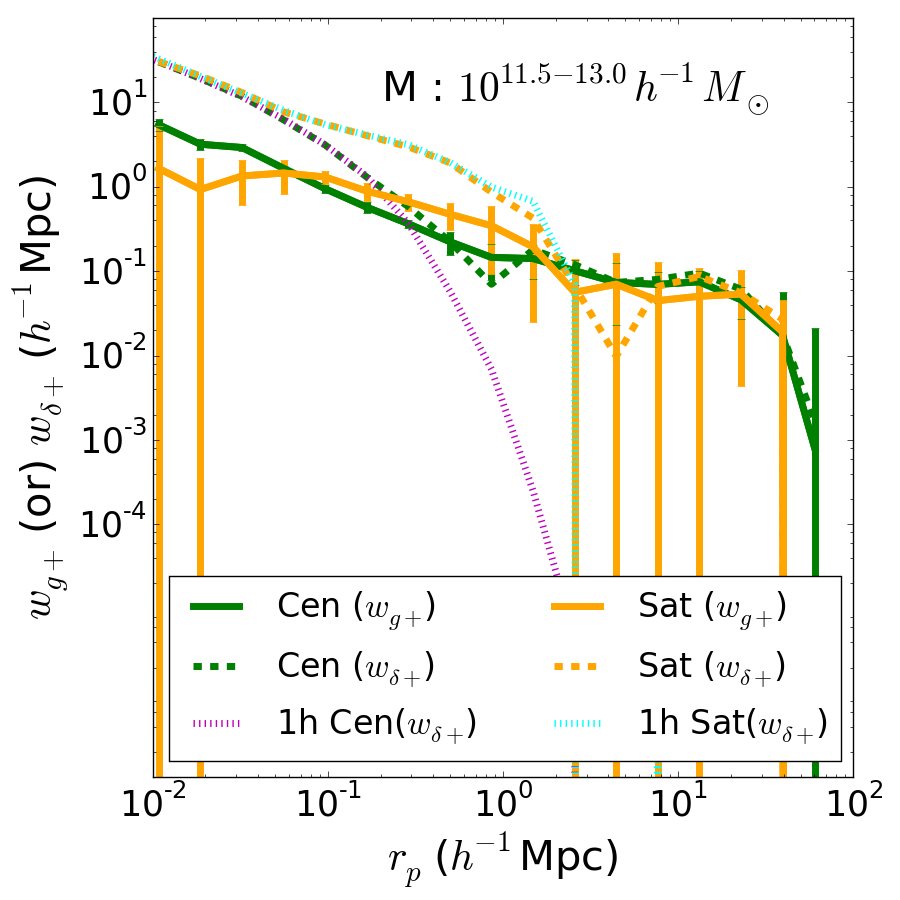}
\includegraphics[width=2.25in]{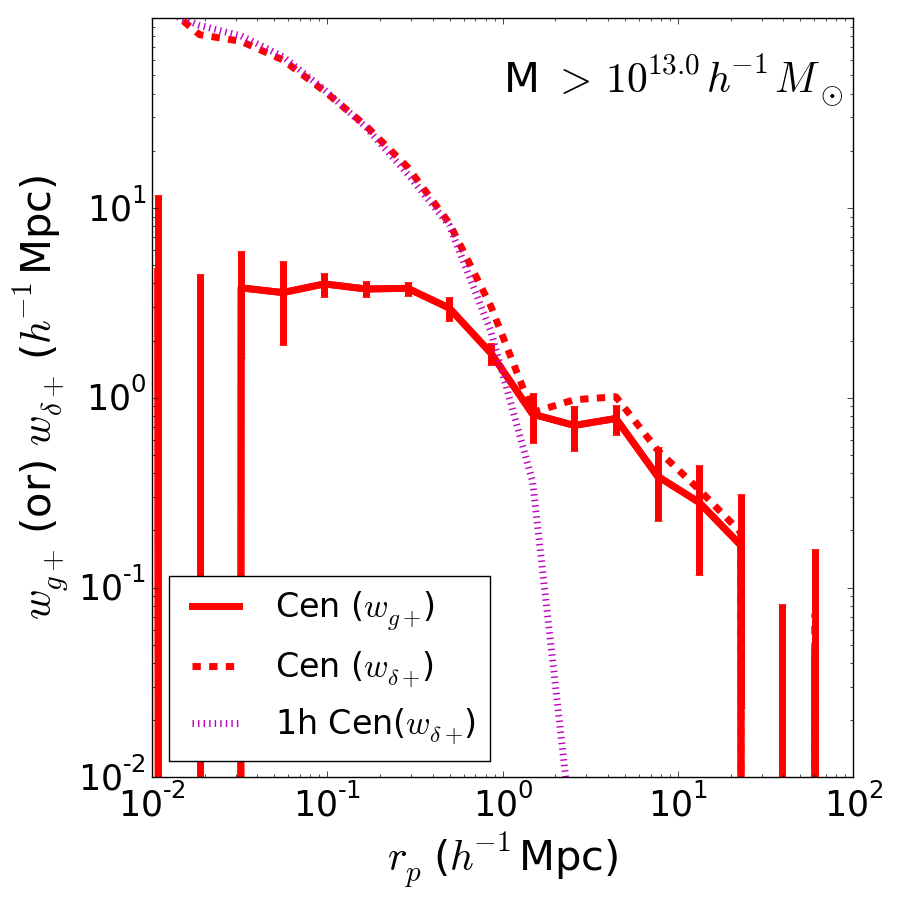}
\caption{\label{F:fig_wgpcensat} $w_{g+}$ and $w_{\delta +}$ correlation function
  for centrals and satellites  at $z=0.3$. {\em Left:} M1: $10^{10-11.5}\hMsun$;
  {\em Middle:} M2: $10^{11.5-13.0}\hMsun$; {\em Right:} M3:
  $>10^{13.0}\hMsun$. The labels ``Cen'' and ``Sat'' refer to the correlation functions ($w_{g+}$, $w_{\delta +}$) of centrals and satellites respectively. Similarly, ``1h Cen'' and ``1h Sat'' refer to the 1-halo term of $w_{\delta +}$ for central and satellite subhalos respectively. The number of central galaxies is $23014$, $7415$ and $255$ in mass bins M1, M2 and M3 respectively.}
\end{center}
\end{figure*}
 
In Fig.~\ref{F:fig_wgpcensat}, we show the $w_{g+}$ and $w_{\delta +}$
correlation function for centrals and satellites in mass bins, M1 and
M2.  In the highest mass bin, M3, the signal for satellites is not
shown due to lack of sufficient number of satellite subhalos.  The
figure shows that at small scales, the \wgp{} and \wdp{} signal for
satellites is larger than that for centrals for subhalos in the mass bin M1. This is interesting and could be
due to following possibilities: 1. Satellite subhalos have stronger
alignments with the local tidal fields than the central subhalos. Note
that within a halo, tidal fields are predominantly radial, consistent
with the radial alignments of satellites. More generally, since
central subhalos are in reality the innermost subhalo, this could
imply some radial dependence of \ia{}. 2. Another possibility is that
satellite and central \ia{} are not very different, but the overall
\ia{} signal depends on the host halo mass. In this case, more massive
halos with more satellite subhalos will get higher weight in satellite
correlations but not in central correlations. This could also push up
the \wgp{} and \wdp{} signal for satellites. We speculate that final
result is likely to be combination of these two effects, with radial
dependence being the more dominant factor. In mass bin M2, the plot shows that there is no statistically significant differences in the intrinsic alignments of centrals and satellites at any scale.
     
At large
scale, it is expected that the intrinsic alignment signal due to
satellites goes to zero in the halo model
\citep{2010MNRAS.402.2127S}, based on the assumption that
the satellite subhalos are uniformly distributed throughout the host halo
pointing towards the center. However, the latter assumption is not
quite true in reality. As shown
in Sec.~\ref{censatalign}, the satellite subhalos have a tendency to
be distributed more along the major axis of the central galaxy and are also radially aligned. Hence, they ``inherit'' the large-scale intrinsic alignments of the host
halo at some level. This could be the explanation for the fact that
the satellite $w_{g+}$, while dropping on large scales, is still
non-zero.

From Fig.~\ref{F:fig_wgpcensat}, we can also see that as we go to higher masses, the
amplitude of intrinsic alignments in central subhalos increases. 

In addition, the transverse separation, $r_{p}$, where we observe a slight dip or change in the shape of the correlation function shifts to smaller values as we go to lower masses of central subhalos. This change of shape indicates a region of transition from the 1-halo term at small scales to the 2-halo term at large scales. To further illustrate our point, we also show the 1-halo term of $w_{\delta +}$ for central and satellite subhalos in these mass bins. This is directly calculated by correlating the shape of a galaxy with the location of dark matter particles that belong to its host halo. As seen from the plot, the 1-halo term follows the shape of $w_{\delta +}$ at small scales and drops to zero at large distances (on scales comparable to the virial radius), where the 2-halo term is becoming more significant.

\section{Modeling, Comparisons and Predictions}\label{ia_comvabn}

In this section, we present the results of fitting the non-linear alignment (NLA) model to the MB-II intrinsic alignment two-point correlation functions. The NLA model has been shown to describe realistic galaxy intrinsic alignments. Comparing the results from the simulation with the NLA model will help us understand on what scales the NLA model describes the alignments in MB-II. Additionally, these fits are a much more compact way to represent our predictions, encapsulating all the information about the scale-dependence of the signal as a single amplitude parameter and a well-known physical model. On small-scales, the NLA model does not describe the signals well, so we provide simple power-law fits for these scales. We also 
compare the intrinsic
alignment two-point correlation functions in MB-II with those in real
data.  There are two
purposes of this comparison. The first is simply to confirm that
MB-II gives physically-reasonable results for samples for which
intrinsic alignments have been robustly detected.  The second is to
then make predictions for samples that will be used for lensing by
upcoming surveys.

	\subsection{Fitting models to MB-II correlation functions}\label{sec:fit_results}
\begin{figure}
\begin{center}
\includegraphics[width=3.2in]{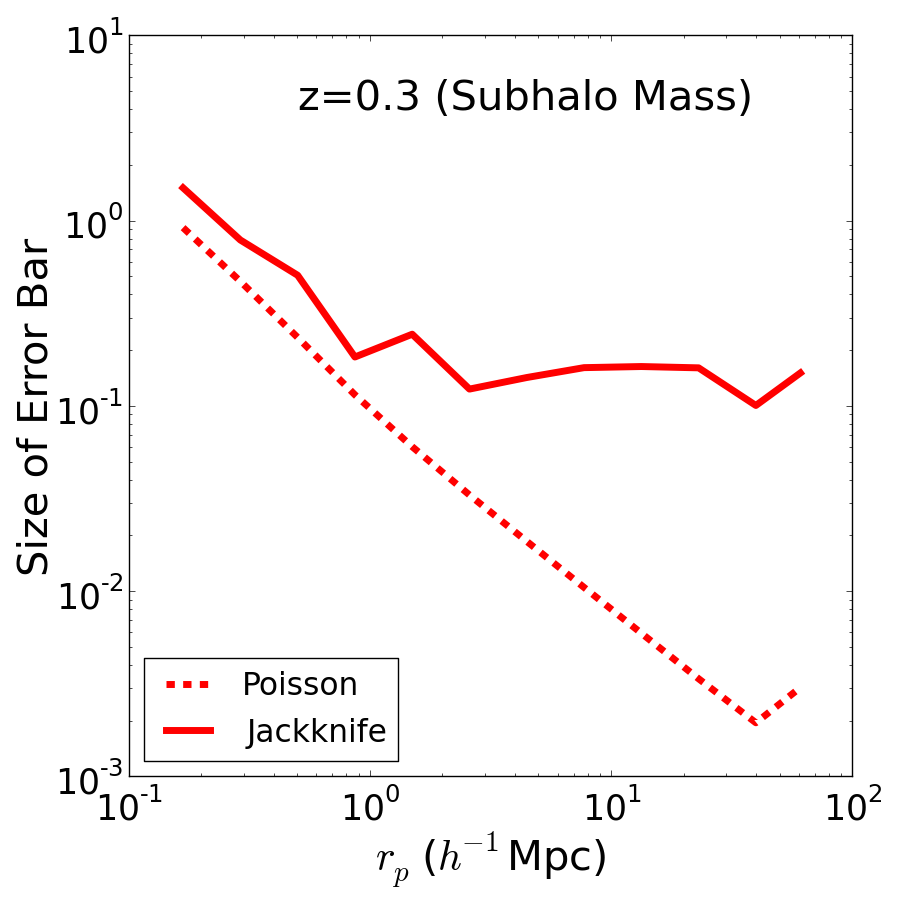}
\caption{\label{F:fig_wgp_err} Comparison of the size of Poisson and
  jackknife error bars in the calculation of $w_{g+}(r)$ for subhalo
  mass-selected samples.} 
\end{center}	
\end{figure}

\begin{figure}
			\begin{center}
				\includegraphics[width=1.0 \columnwidth]{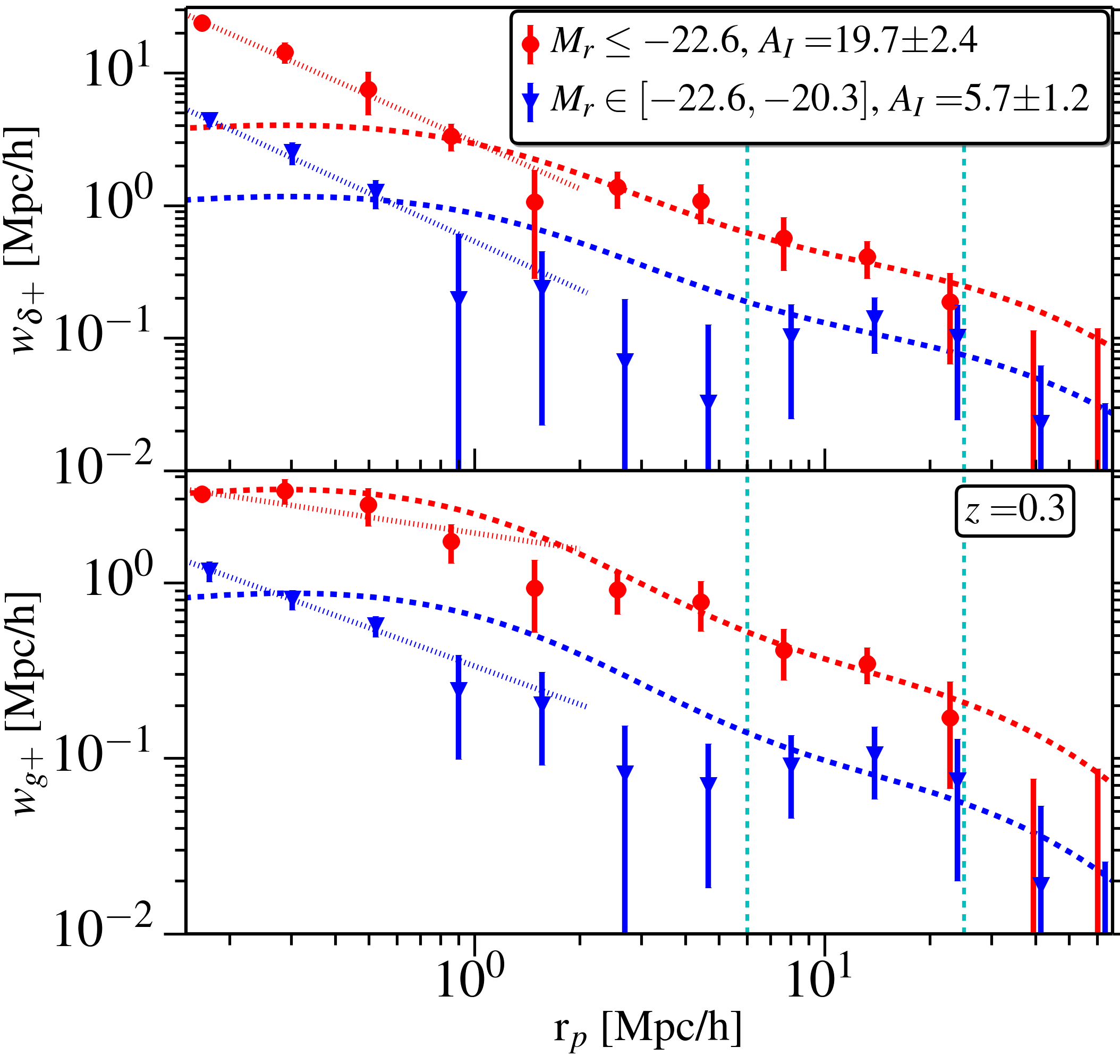}
				\caption{NLA and power law fitting to \wdp{} (top) and
                  \wgp{} (bottom) for two different samples defined by luminosity bins. Vertical lines show the range over which we fit the NLA model ($6\mpch<r_p<25\mpch$). Note that the power law is fitted only for
                  $r_p<1\mpch$, though the function is shown out to
                  $r_p\sim2\mpch$.}
				\label{fig:lum_bin2_z03}
			\end{center}
		\end{figure}

	Here we present results of fitting NLA and power
        law functions to our predictions for \wgp{} and \wdp{} from MB-II. Fig.~\ref{fig:lum_bin2_z03} shows an example of models fitted to the measurements for two 
		different samples defined by luminosity bins, $M_r\le-22.6$ and $M_r\in[-22.6,-20.3]$.
 More examples and tables with fit
        parameters can be found in Appendix~\ref{appn:fit_results}. We
        fit the NLA model in the range $6<r_p<25\mpch$. Beyond
        $25$\mpch, the MB-II predictions are 
		dominated by cosmic variance. We fit \wgp{} and \wdp{}
        simultaneously assuming the same $A_I$ for both, and an
        additional large-scale (constant) subhalo bias $b_D$ for
        \wgp{}. As can be seen in Fig.~\ref{fig:lum_bin2_z03}, the NLA
        model fits the data well 
		in the fitting range and can be extended down to
        $r_p\sim4\mpch$, below which the signal differs in both amplitude and scale 
		dependence. We add a note of caution that we use the simple weighted least squares method to fit the model using only the diagonal terms in the covariance matrix, which underestimates the errors on the parameters when compared with the errors on data points. The errors 
		shown on data points are calculated from jackknife variance, but due to the limited size of our simulation box, the jackknife errors on the maximum scales used are not very reliable. Fig.~\ref{F:fig_wgp_err} shows a comparison of jackknife 
		and Poisson error bars.  The
Poisson errors tend to be very small (and are certainly underestimated
above a few $h^{-1}$Mpc scales, where cosmic variance will be
important). However they are within a factor of 1.5--2 of the
jackknife errors on small scales, which is reasonable. While Poisson errors are underestimated, the scale dependence of jackknife errors suggest that they are cosmic variance dominated and due to the limited size of the simulation box, the covariance 
		matrix is very noisy. Keeping in mind the limitations of our jackknife covariance matrix, we do not attempt a more sophisticated fitting method to get better error estimates on the model parameters.

		On small scales, we fit a power law function
        separately to \wgp{} and \wdp{},  in the range
        $0.1<r_p<1\mpch$. The power law function is of the form:
		\begin{equation}
			\wgp=P_A r_p^{P_I}
			\label{eqn:powerlaw_fit}
		\end{equation} 
		
\begin{figure}
			\begin{center}
			{\includegraphics[width=1.0\columnwidth]{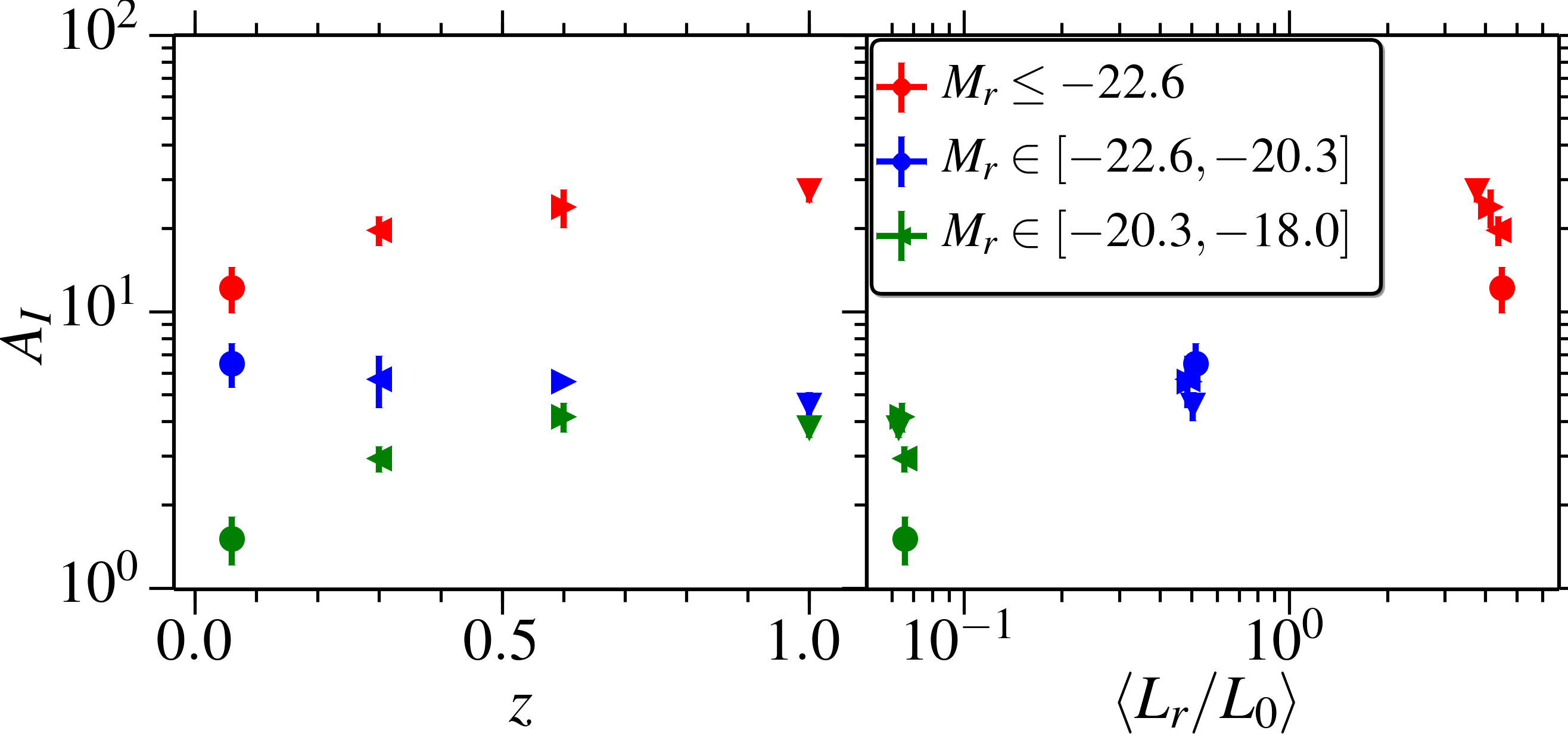} }
			\caption{NLA amplitude, $A_I$, as a function of redshift for different luminosity samples. The horizontal axis indicates the average mass, luminosity or redshift of different samples. Points are colored by sample 
			definition, while markers are set according to the redshift.
		}
		\label{fig:ia_amps_lum}
			\end{center}
		\end{figure}

		Fig.~\ref{fig:ia_amps_lum} shows \ia{} amplitudes for \wgp{} as function of average luminosity and redshift of the sample, for different samples defined with different luminosity bins.  
		We see clear evolution with luminosity and mild evolution with redshift. More luminous objects show stronger \ia{}, qualitatively consistent with LRG observations. Within the NLA model, where it is assumed that \ia{} are set 
		at time of galaxy formation, we do not expect any redshift evolution of $A_I$. This is consistent with LRG observations, where no significant redshift dependence for $A_I$ is detected \citep{Joachimi2011}, admittedly with a 
		narrower redshift range than considered here. LRG are, however, a special 
		population of old, very massive, passively evolving galaxies. Our sample in MB-II is much more diverse in properties and is heavily dominated by much less massive galaxies that will include a variety of formation and 
		evolutionary histories, including recent mergers and accretion. This is expected to change the \ia{} signal at all scales. We see clear redshift evolution in two of the three samples defined by luminosity bins, with the middle 
		bin showing negligible evolution. For the brightest and faintest samples, we observe that the NLA model amplitude decreases at lower redshifts, which suggests that dynamical processes such as galactic mergers play some 
		role in \ia{} evolution at those luminosities.

		To quantify the evolution of \ia{} with redshift, mass and luminosity, we fit the non-linear alignment model amplitude $A_I$ with the following functions:
		
		\begin{equation}
			A_I=A\left(\frac{\langle L_r \rangle}{L_0}\right)^{\alpha_L} (1+z)^{\alpha_z}\label{eqn:L_evolution}
		\end{equation}
		$\langle L\rangle$ is the average $r$-band luminosity, normalized by pivot luminosity $L_0$ corresponding to r-band magnitude $M_r=-22$. Results from the fitting are shown in Table~\ref{tab:amp_fit_lum}. We also show results 
		from similar fitting to power law amplitude and index. 

		Doing a similar fit to $A_I$ in luminosity and redshift to LRG samples, \cite{Joachimi2011} got 
		$\alpha_L=1.13^{+0.25}_{-0.20}$ and $\alpha_z=-0.27^{+0.80}_{-0.79}$ (MegaZ-LRG + SDSS LRG + L4 + L3). Our power law indices are different, with our samples showing weaker luminosity evolution than LRGs. This 
		 is likely due to differences in the samples, since our samples do not include color cuts, and also extend to fainter luminosities. Our results are qualitatively consistent with results of \cite{2013MNRAS.436..819J}, who used semi-analytical approach to populate dark matter halos in Millennium simulation and measured the \ia{} signal. 
		When they measured intrinsic alignments amplitudes as a function of luminosity, they found a shallower luminosity dependence at the faint end than for LRGs.  
		  Our $\alpha_z$ is consistent with zero within $1\sigma$, consistent with \cite{Joachimi2011} and \cite{2013MNRAS.436..819J}. 
\begin{table}
	\caption{Results of fitting different parameters (luminosity bin samples only) to find their mass and luminosity evolution (Eq.~\ref{eqn:M_evolution} and~\ref{eqn:L_evolution}). Different columns are the parameters that go into Eq.~\eqref{eqn:M_evolution} and~\eqref{eqn:L_evolution} while different rows are for different \ia{} model parameters, with $A_I$ being the NLA amplitude, $P_A$ and $P_I$ are power law fits (Eq.~\ref{eqn:powerlaw_fit}) to \wgp{} and $P_A^\delta$ and $P_I^\delta$ are power law fits to \wdp{}.
	}
	\label{tab:amp_fit_lum}
	\begin{tabular}{|@{}c@{}|@{}c|c|c|}\hline
		Parameter & $A$ &$\alpha_L$ & $\alpha_z$ \\ \hline
$A_I$ & 6.7$\pm$1.7 & 0.47$\pm$0.08 & 0.5$\pm$0.5\\ 
$P_A$ & 0.59$\pm$0.08 & 0.48$\pm$0.05 & -0.7$\pm$0.2\\ 
$P_I$ & -0.49$\pm$0.08 & 0.09$\pm$0.06 & 0.5$\pm$0.3\\ 
$P_A^\delta$ & 1.5$\pm$0.3 & 0.6$\pm$0.1 & -1.7$\pm$0.5\\ 
$P_I^\delta$ & -1.1$\pm$0.1 & 0.1$\pm$0.03 & 0.1$\pm$0.2\\
	\end{tabular}
\end{table}

		\begin{figure}
			\begin{center}
				\includegraphics[width=1.0 \columnwidth]{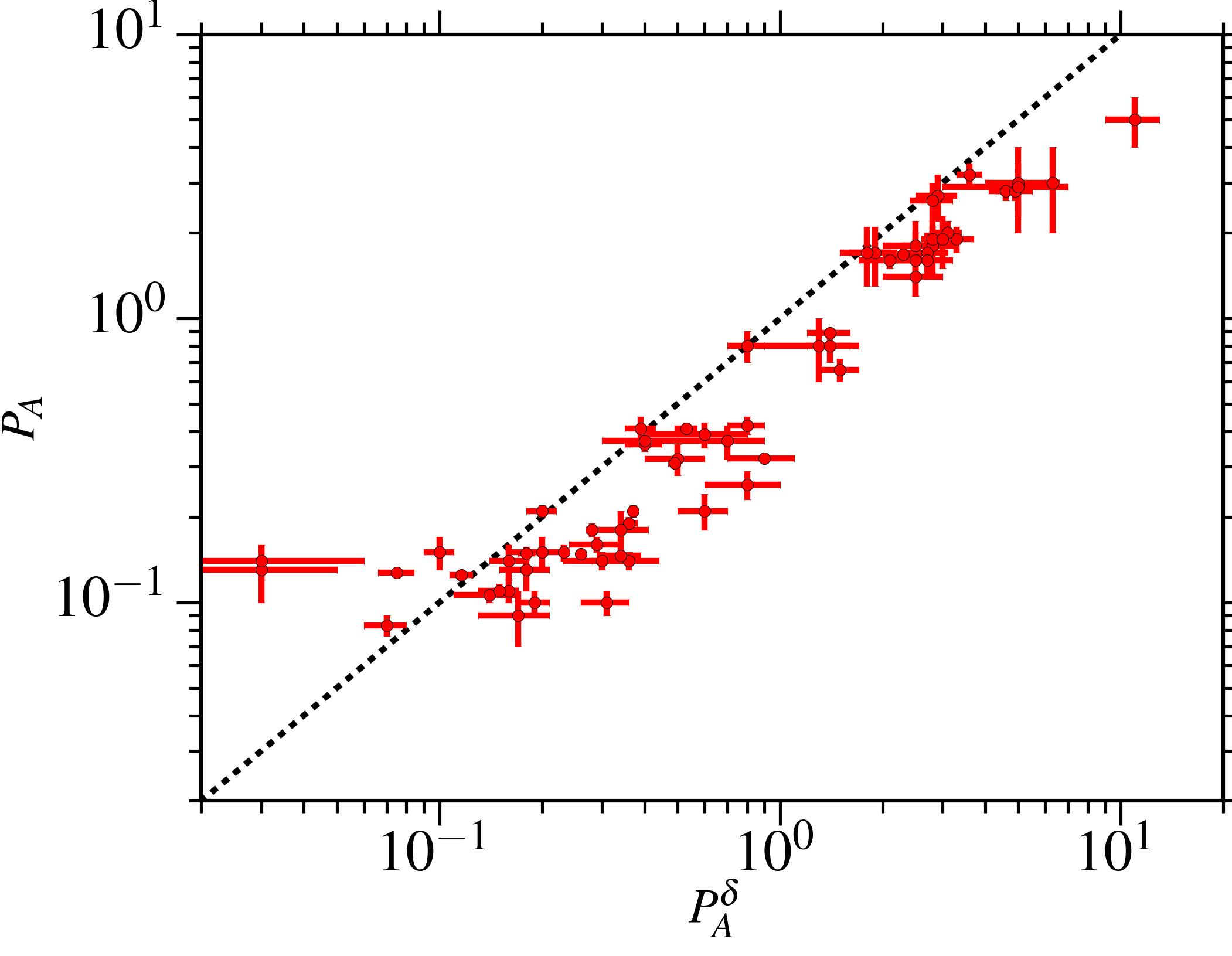}
				\caption{Comparison of power law amplitude for \wgp{} ($P_A$) and \wdp{} ($P_A^{\delta}$) for various samples used in this work. The dotted line shows the $x=y$ relation. \wdp{} is observed to have systematically higher amplitude than \wgp{} for separations below $1$\hmpc.
				 }
				\label{fig:wgp_wdp_amp}
			\end{center}
		\end{figure}
		\begin{figure}
			\begin{center}
				\includegraphics[width=1.0 \columnwidth]{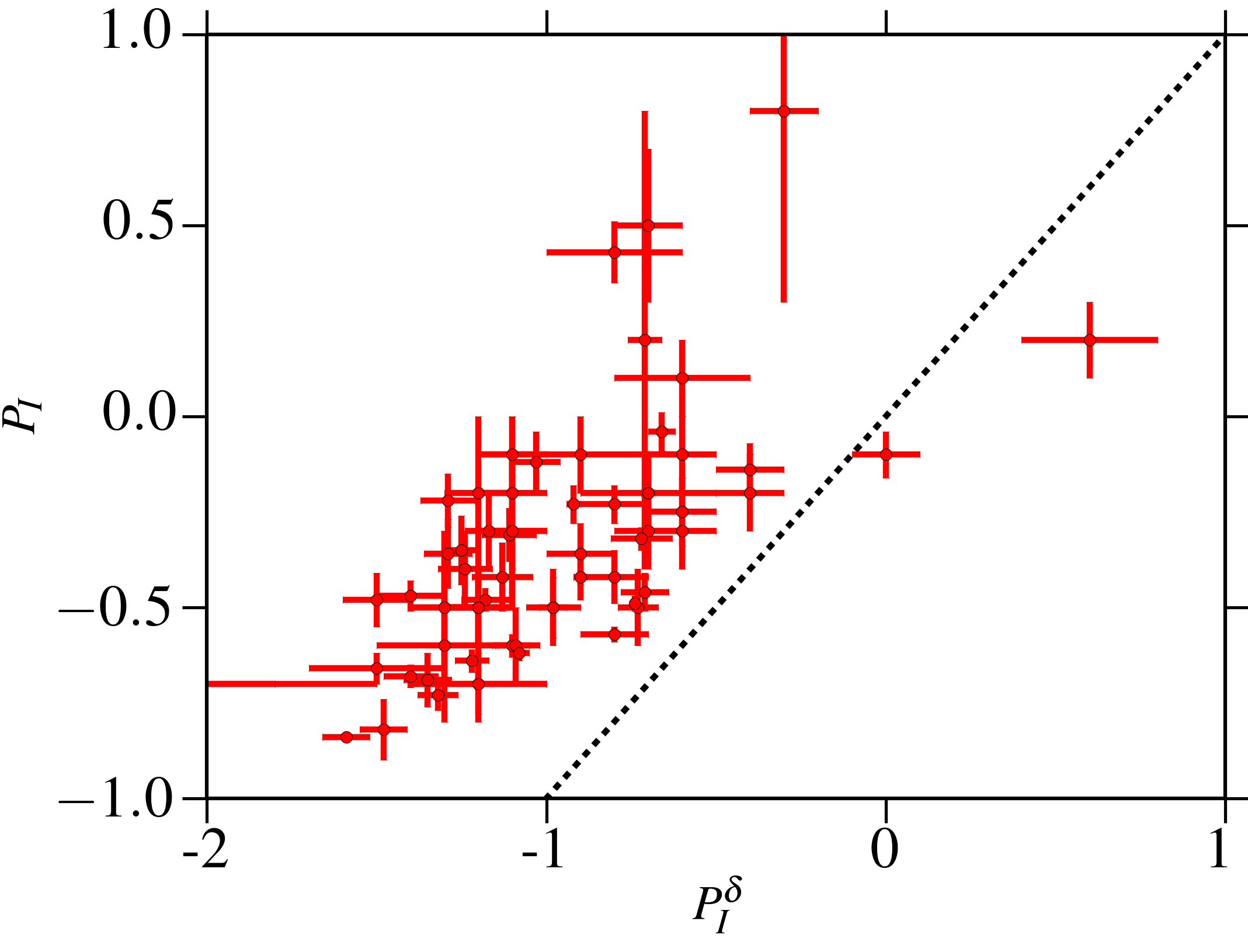}
				\caption{Comparison of power law index for \wgp{} ($P_I$) and \wdp{} ($P_I^{\delta}$) for various samples used in this work. The dotted line shows the $x=y$ relation. \wdp{} has a systematically steeper slope than \wgp{} for separations below $1$\hmpc.}
				\label{fig:wgp_wdp_indx}
			\end{center}
		\end{figure}

		Figs.~\ref{fig:wgp_wdp_amp} and~\ref{fig:wgp_wdp_indx} show the comparisons of power-law parameters (Eq.~\ref{eqn:powerlaw_fit}) fit to \wdp{} and \wgp{} for different samples used in this work. \wdp{} has 
		systematically higher amplitude and steeper power-law index than \wgp{}, which implies that \wgp{} is more flattened compared to \wdp{} below $1$\hmpc. The flattening of \wgp{} at small scales is likely due to the effects of 
		non-linear bias of the subhalo sample used as density tracer.
		However, as observed in Fig.~\ref{F:fig_sgdm}, the ratio of \wdp/\wgp{} changes for different mass threshold samples,
		which means that there could be some differences due to \ia{} signal as well. Subhalos are biased 
		tracers of density field and it is conceivable that \ia{} signal at small scales can change when subhalos are used as the density tracers (for example, there may not be enough subhalos around a galaxy at small scales to 
		fairly measure the \ia{} signal). This can have important implications for observational studies of \ia{}, where we can only use galaxies as biased tracers of  the density field, so the small scale \ia{} could be underestimated. 

\subsection{Comparison with luminous galaxy intrinsic alignments}
\begin{figure}
\begin{center}
\includegraphics[width=3.2in]{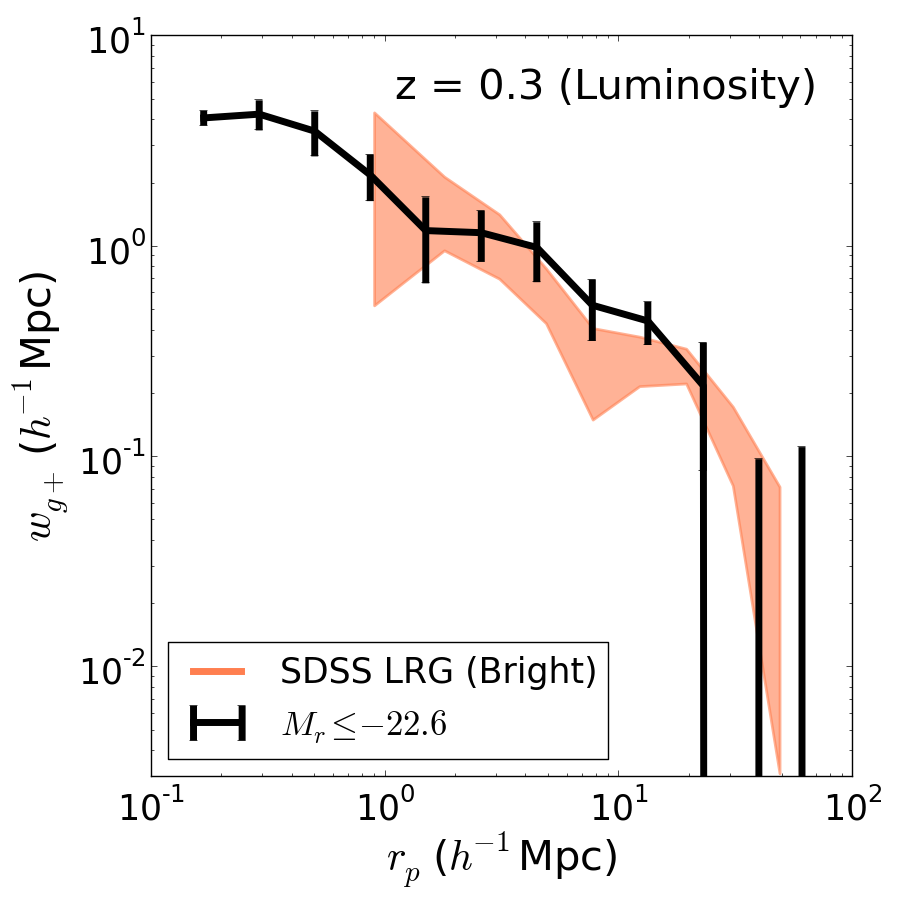}
\caption{\label{F:fig_wgp_ca} $w_{g+}$ correlation function for
  galaxies selected according to $r$-band luminosity (such that $M_{r}\leq -22.6$)and comparison with observational
  results using SDSS LRG sample. Note that the bias of the density tracer sample has
  been taken into account in order to make a fair comparison, by dividing $w_{g+}$ with the large scale linear bias.} 
\end{center}
\end{figure}

\begin{figure*}
\begin{center}
\includegraphics[width=3.2in]{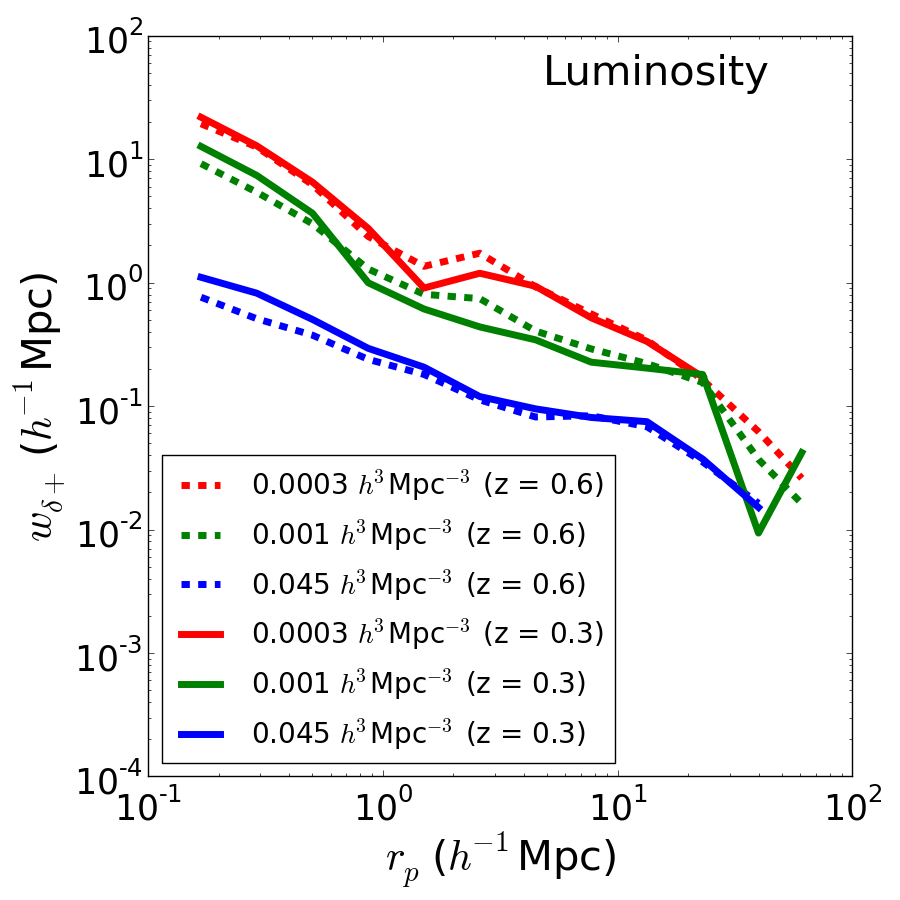}
\includegraphics[width=3.2in]{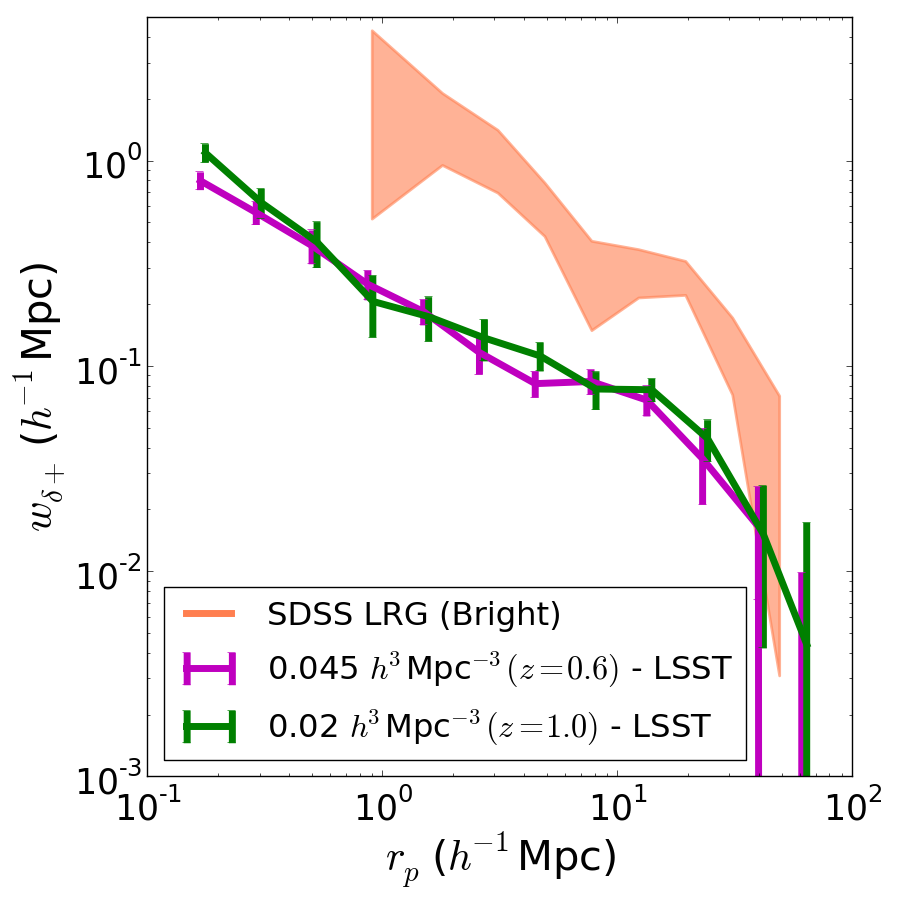}
\caption{\label{F:fig_wgp_nca} {\em Left:} $w_{\delta +}$ correlation
  function at redshifts $z = 0.6$ and $0.3$ for galaxies selected by a
  luminosity threshold to match three values of comoving abundance as
  labeled on the plot. {\em Right:} Prediction of $w_{\delta +}$ for
  galaxies that will be used for lensing in the LSST survey, made by matching the estimated comoving abundances at
  $z=1.0$ and $z=0.6$.  The shaded regions show jackknife errorbars.}
\end{center}
\end{figure*}

In Fig.~\ref{F:fig_wgp_ca}, we show the $w_{g+}$ correlation
function for the subhalos selected by luminosity in $r$-band such that the absolute AB magnitudes satisfy $M_{r} \leq -22.6$. The error bars shown here are obtained using the jackkinfe technique. The
observational measurements are obtained from  an SDSS LRG sample in the redshift range $0.27 < z < 0.35$, with luminosity cuts as defined in \cite{2007MNRAS.381.1197H}. The galaxies from the simulation are selected to match the luminosity threshold of the Bright LRG sample ($M_{r} \leq -22.6$), against which we compare our results. The amplitude of the predicted $w_{g+}$ for this sample is in good agreement with the
observational results for the LRG sample\footnote{The bias of the density sample in both simulations and observations has been taken into account in this comparison.}. The mass
range of the galaxies from the simulation roughly corresponds to a subhalo mass threshold of $M > 10^{13}\hMsun$ which is indeed the appropriate halo mass range
for LRGs. However, there is an important caveat in this comparison. The LRG sample has color cuts and so, unlike the simulated galaxies, the LRG samples are not perfectly luminosity selected. Hence, it is difficult to make an exact comparison of the  the  amplitude of correlation function in spite of selecting the same luminosity thresholds. If we ignore the amplitude, which is likely to be a nuisance 
parameter that gets marginalized over in a typical intrinsic alignment
mitigation scheme, what is more important is that the scaling with
transverse separation is consistent with that in real data, as is the
scaling with mass that was shown earlier in this work. This
confirms that MB-II can provide reasonable templates for intrinsic
alignment models to be used in real data analysis.

\subsection{Predictions for future weak lensing surveys}\label{ia_comvabnpred}
Using the SDSS $r$-band luminosity of galaxies in the simulation, we
can make predictions for the $w_{\delta +}$ correlation function for
upcoming surveys. However, we do not separate the galaxies by their color. So, the IA signals shown here also include the type dependence. Here we focus on \wdp{} rather than \wgp{} since the intrinsic
alignments contamination of cosmic shear signals is caused by the
entire matter density field. In the left panel of Fig.~\ref{F:fig_wgp_nca}, we plotted the
$w_{\delta +}$
correlation function for galaxy samples selected on the basis of a
luminosity threshold with increasing comoving abundance at redshifts 
$z = 0.3$ and $0.6$. Our results suggest that the amplitude of the
$w_{\delta +}$ correlation function decreases with increasing comoving
abundance at both redshifts, with the shape of the correlation function changing as well
(such that the 1-halo to 2-halo transition is no longer
evident for lower luminosity samples, perhaps because they occupy
host halos with a wide range of masses). 

In the right panel of Fig.~\ref{F:fig_wgp_nca}, we show the $w_{\delta +}$ signals at $z=0.6$ and $z=1.0$ that help us to predict the \ia{} for the galaxies that will be used to measure lensing in the upcoming LSST survey. At redshift $z=1.0$, the comoving abundance of $0.02~
(\hmpc)^{-3}$ corresponds to the estimated number density of galaxies
in the LSST. Similarly, at redshift $z=0.6$, the estimated
comoving abundance is $0.045~(\hmpc)^{-3}$. The galaxy number
densities mentioned here are based on the results from
\cite{2013MNRAS.434.2121C}. From the observational measurements of intrinsic
alignments using SDSS LRGs (Fig.~\ref{F:fig_wgp_ca}), we know the value of
$w_{g+}$ which would be a good match to the signal obtained from a luminosity based comoving number density threshold of
$3 \times 10^{-4}(\hmpc)^{-3}$ (left panel of Fig.~\ref{F:fig_wgp_nca}). For galaxies in the LSST sample, our
results predict that the intrinsic alignments decrease by a factor of
$\sim 18$ for scales below $1\hmpc$. At large scales, based on the NLA
model fits tabulated in Appendix~\ref{appn:fit_results}, we predict
that the amplitude of the signal decreases by a factor of $\sim5$ at
$z=0.6$ compared to the measured signal using LRGs.  

\begin{figure*}
			\begin{center}
			{\includegraphics[width=2.0\columnwidth]{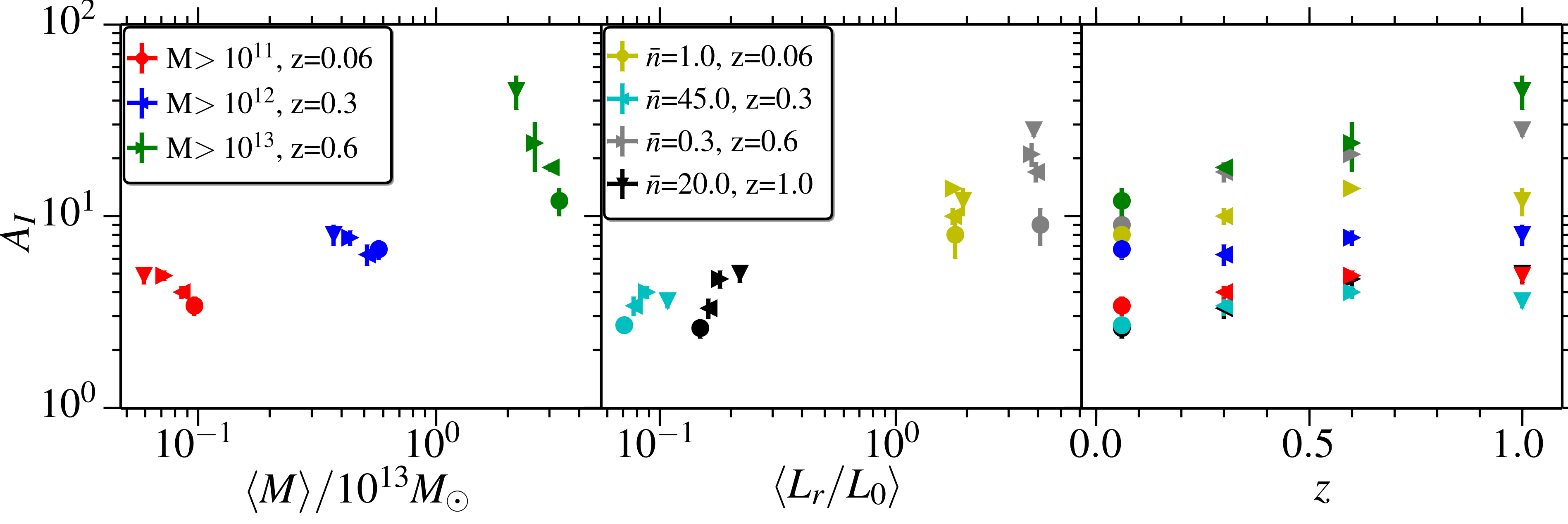} }
			\caption{NLA amplitude, $A_I$, as a function of different sample properties. The horizontal axis indicates the average mass, luminosity or redshift of different samples. Points are colored by sample definition: comoving 
			abundance ($\bar{n}$) in units of $10^{-3} h^{3}$Mpc$^{-3}$ based on a luminosity threshold, or the mass threshold of the sample (not average mass), while markers are set according to the redshift.
		}
		\label{fig:ia_amps_various}
			\end{center}
		\end{figure*}

Fig.~\ref{fig:ia_amps_various} shows the evolution of NLA amplitude $A_I$, for different samples defined by mass threshold and comoving abundance. We observe clear evolution with mass and luminosity with more massive and luminous objects having stronger alignments. We also observe mild evolution in redshift which is inconsistent with NLA assumption that \ia{} are setup at time of galaxy formation, if we assume that all our galaxies formed at $z\gg1$. This assumption is however likely to break down over the broad redshift and mass range of our sample, due to growth of structure as well as dynamical evolution of galaxies which will bring the \ia{} signal down, consistent with our results. As in Sec.~\ref{sec:fit_results}, to quantify the luminosity and redshift evolution of \ia{} amplitude we fit a power law defined in Eq.~\eqref{eqn:L_evolution} to luminosity threshold samples and a similar power law in average mass and redshift as defined in Eq.~\eqref{eqn:M_evolution} to mass threshold samples.
	\begin{equation}
			A_I=A\left(\frac{\langle M \rangle}{10^{13}h^{-1}\Msun}\right)^{\alpha_M} (1+z)^{\alpha_z}\label{eqn:M_evolution}
		\end{equation}
Since $A$ and $\alpha_z$ are same for both luminosity and mass threshold samples we fit both luminosity and mass threshold samples simultaneously to get all the parameters. Parameters are given in Table~\ref{tab:amp_fit_threshold}. We note that our samples defined by threshold cuts are correlated and hence the values given in Table~\ref{tab:amp_fit_threshold} should not be directly compared with observational results, where samples are usually defined in luminosity or mass bins. The purpose of our fits given here is to give scaling relations for overall expected \ia{} for sources that will be used to measure lensing in surveys like LSST and Euclid, for which the source samples will likely be derived from taking most of the galaxies above some flux cut.

\begin{table}
	\caption{Results of fitting different parameters (for mass and luminosity threshold samples only) to find their mass and luminosity evolution (Eq.~\ref{eqn:M_evolution} and~\ref{eqn:L_evolution}). Different columns are the parameters that go into Eq.~\eqref{eqn:M_evolution} and~\eqref{eqn:L_evolution} while different rows are for different \ia{} model parameters, with $A_I$ being the NLA amplitude, $P_A$ and $P_I$ are power law fits (Eq.~\ref{eqn:powerlaw_fit}) to \wgp{} and $P_A^\delta$ and $P_I^\delta$ are power law fits to \wdp{}.
	}
	\label{tab:amp_fit_threshold}
	\begin{tabular}{|@{}c@{}|@{}c|c|c|c@{}|}\hline
		Parameter & $A$ & $\alpha_M$ &$\alpha_L$ & $\alpha_z$ \\ \hline
$A_I$ & 7.7$\pm$0.5 & 0.35$\pm$0.03 & 0.48$\pm$0.03 & 0.69$\pm$0.16\\ 
$P_A$ & 0.65$\pm$0.04 & 0.57$\pm$0.03 & 0.7$\pm$0.04 & -0.2$\pm$0.2\\ 
$P_I$ & -0.42$\pm$0.05 & -0.09$\pm$0.06 & 0.1$\pm$0.09 & 0.1$\pm$0.3\\ 
$P_A^\delta$ & 1.23$\pm$0.12 & 0.53$\pm$0.04 & 0.58$\pm$0.04 & -1.0$\pm$0.2\\ 
$P_I^\delta$ & -1.05$\pm$0.06 & 0.14$\pm$0.02 & 0.11$\pm$0.02 & 0.0$\pm$0.1\\ 	
\end{tabular}
\end{table}

\section{Conclusions}\label{ia_conc}
In this paper, we used the MB-II cosmological hydrodynamic simulation
to study the intrinsic alignments of galaxies using the Ellipticity-Direction (ED) and the projected shape correlation function ($w_{g+}$). We are able to directly
measure the shapes of the stellar matter component of the galaxies and use
these to estimate the two-point correlation functions which can be
compared with intrinsic alignment measurements from
observations. The use of hydrodynamic simulations, which include the
physics of galaxy formation, has an advantage over $N$-body simulations
in that we do not have to make assumptions about the occupation of
halos with galaxies and their alignments with the host halo. We also
have information on the luminosities of galaxies in the simulation,
which is useful for comparisons with observations and making
predictions of intrinsic alignments for upcoming surveys.

It is necessary to adopt a definition for the shapes of
dark matter and stellar matter components in subhalos. We investigated
the variation in the distribution of axis ratios of shapes obtained
using iterative and non-iterative forms of the unweighted and reduced
inertia tensor. The axis ratios and orientations of the shapes
obtained using unweighted iterative and non-iterative inertia tensor
are very similar. For comparison with observations, it might be useful
to use the reduced form of inertia tensor which gives more weight to
particles in the inner regions of a subhalo. The non-iterative reduced
inertia tensor produces shapes that are biased towards being very
spherical and hence is not considered. The axis ratios of shapes
defined by dark matter subhalos obtained using the iterative reduced
inertia tensor have slightly larger axis ratios when compared with
those obtained using the unweighted inertia tensor, which is in
agreement with the findings of \cite{2012MNRAS.420.3303B}. For shapes defined by stellar matter, the reduced inertia tensor produces shapes which are slightly more oblate. 

We can also
define a luminosity weighted unweighted and reduced inertia tensors
for shapes of stellar matter. We concluded that the shapes obtained
using the unweighted inertia tensor are similar when the star particle is
weighted by its luminosity or mass. However, we observe noticeable
changes in the distribution of axis ratios for shapes obtained using
the reduced form of the inertia tensor when we weight each particle by
its luminosity. This is not surprising, as it
indicates that the mass to light ratio is not constant in the inner
regions of galaxy, which is expected.
However, our results suggest
this effect of luminosity-weighting does not affect the intrinsic alignment signals which are consistent with the ED and $w_{g+}$ determined using shapes from mass-weighted inertia tensor.

To investigate the color dependence of intrinsic alignments, the galaxies in the simulation are
roughly divided into red and blue types by choosing a median value of the rest-frame color,
$M_{u}-M_{r}$. By comparing the $w_{g+}$ correlation function for red
and blue galaxies, we concluded that there is no significant difference in the ED and $w_{g+}$ correlation functions for red and blue galaxies.

We measured the dependence of
the two-point correlation functions, ED (position angle statistic)
and $w_{g+}$ (projected shape correlation function), on the mass and
redshift. The $w_{g+}$ correlation function is more relevant for
comparison with many observational results and for contamination of
upcoming weak lensing measurements by intrinsic alignments, given that
it includes the overall galaxy shape. For both ED and $w_{g+}$, the
amplitude of the correlation function is smaller for shapes defined by the
reduced form of the inertia tensor. By plotting the correlation functions
for galaxy samples selected in the mass bins, $10^{11-12}\hMsun$, $10^{12-13}\hMsun$ and $10^{13-15}\hMsun$, 
we concluded
that the amplitude of the correlation function increases strongly with increasing
mass. We also consider the redshift dependence of ED and $w_{g+}$
correlation functions. For the ED correlation function, the
amplitude of the correlation function decreases at low redshifts, which
indicates that the shape defined by the stellar component tends to get
slightly less correlated with the density field traced by
subhalos. Our findings for the mass and redshift dependence of ED
correlation function using the shapes of stellar matter are similar to
the conclusions of \cite{2008MNRAS.389.1266L} based on $N$-body
simulations. However, we do not notice a significant redshift
dependence for the $w_{g+}$ correlation function for fixed mass
threshold samples. 

The simulation also
allows us to directly study the intrinsic alignment in centrals and
satellite galaxies, as it is possible to split our subhalos into
centrals and satellites. Previously, the intrinsic alignments in
centrals and satellites has been modeled analytically using the halo
model \citep{2010MNRAS.402.2127S}. Here, we concluded that in low  mass galaxies, the satellites have larger intrinsic
alignment when compared to centrals at small scales (i.e., in the
language of the halo model, the satellite galaxies have a stronger
one-halo term than the centrals). At large scales,
the intrinsic alignment signal for satellite galaxies goes down and is
smaller than those for central galaxies (centrals have a larger
two-halo term than satellites). We do not observe statistically significant differences in the intrinsic alignments of centrals and satellites in more massive galaxies.
 
We also fit non-linear alignment model (NLA) in the range
$6\mpch<r_p<25\mpch$ and study the evolution of with mass, luminosity
and redshift.  The NLA amplitude $A_I$ increases with mass and luminosity,
qualitatively consistent with LRGs observations though our scalings
are different from LRGs observations, possibly due to our focus on lower luminosity galaxies.We also fit a simple power law model to study \ia{} at
small scales, and observe that \ia{} signal gets lower and more
flattened as we go to lower mass and luminosities. We observe
that \ia{} get more flattened for \wgp{} as compared to \wdp{}, which
implies that sub halos don't allow a fair measurement of \ia{} signal
at small scales. This has important implications for observations,
where we can only use galaxies to trace the density field.
 
 Finally, we are able to make predictions for the intrinsic alignments
 for upcoming surveys at redshifts $z=1.0$ and $z=0.6$ by calculating
 the $w_{\delta +}$ correlation function (cross-correlation of
 projected shapes with density field traced by dark matter particles).
 For these predictions, we select galaxy samples based on a threshold
 in luminosity such that the comoving abundance matches the expected
 number density of galaxies at the given redshifts. We concluded that,
 as expected, the amplitude of $w_{\delta +}$ correlation decreases as
 we go to larger comoving abundances. This result is important as we
 already have the observationally measured result for $w_{g +}$
 using data from  the SDSS LRG sample. Using our results from simulation,
 we predict that for galaxies that will be used to measure lensing in the LSST survey, the
 IA signal decreases by a factor of $\sim5$--$18$ depending on the
 radial separation (from $\sim 30$ down to $\sim 0.5$\hmpc) compared
 to the measured value for LRGs. This differs from the conclusion of
 \cite{2014arXiv1406.4668C}, where they detected no intrinsic
 alignment signal in their sample of reddest galaxies at $z=1.2$. The
 difference can be due to the fact that \cite{2014arXiv1406.4668C}
 define shapes with a spin statistic which is suitable for blue galaxies. As
 mentioned in their paper, spins do not fully capture the shape of a
 galaxy or the effects of intrinsic alignments on the two-point shear statistics. It is also to be noted that their hydrodynamics is implemented based on AMR code. As our approach is based on SPH, it will be interesting to directly compare the intrinsic alignments of galaxies using a similarly defined observable to understand the differences due to numerical implementation. 

In future work, we will compare the results of our two-point
correlation function with predictions from a dark matter only
simulation run with the same initial conditions, in order to
understand the importance of the physics of galaxy formation and
processes such as feedback on the intrinsic alignments. We will also try to
apply additional post-processing techniques to match the color of
galaxies in our simulation to those from observational results.
However, the results in this work suggest that high-resolution and
large-volume SPH simulations such as MB-II will be a powerful tool for
predicting and mitigating intrinsic alignments in future weak lensing
surveys.

\section*{Acknowledgments}
We thank Benjamin Joachimi, Alina Kiessling, Cristob\'al Sif\'on, and Jonathan
Blazek for providing helpful feedback on this work.
RM's work on this project is supported in part by the Alfred P. Sloan Foundation.
The simulations used in this work were run on NSF XSEDE HPC facilities
at NICS. We acknowledge support from Moore foundation which enabled us
to perform the data analysis at the McWilliams Center of Cosmology at
CMU. TDM has been funded by the National Science Foundation (NSF)
PetaApps, OCI-0749212 and by NSF AST-1009781 and ACI-1036211.

\bibliographystyle{mn2e} \bibliography{sub_v2.bib}

\appendix
	\section{Fitting Results}\label{appn:fit_results}
		Here we present results of fitting the NLA and power law models to different samples for which \wgp{} and \wdp{} were measured. These are helpful to produce IA signals that scale with mass, luminosity and transverse signals according to predictions from the MBII simulation. At linear and quasi-linear scales ($6\mpch<r_p<25\mpch$) we fit \wdp{} and \wgp{} 
		simultaneously for amplitude $A_I$ and subhalo linear bias $b_D$ in \wgp{}. $b_D$ values are not shown in tables but we get $b_D$ consistent with values expected for $\xi_{gg}$ and $\xi_{mm}$ measurements. 
		The power-law was fitted separately to \wgp{} and \wdp{} for $r_p<1$\mpch, with two free parameters, amplitude $P_A$ and index $P_I$. A subscript $\delta$ on power-law parameters denotes a fit to \wdp{}. Power-law 
		parameters evolve with mass 
		and luminosity, with the function becoming more shallow for lower mass and luminosity.  As discussed in Sec.~\ref{sec:fit_results}, there are also 
		differences in power law fits to \wdp{} and \wgp{}, with the function being more shallow for \wgp{}. See Sec.~\ref{sec:fit_results} for more a  detailed discussion.
		
\begin{figure}
		\begin{center}
			\includegraphics[width=1.0\columnwidth]{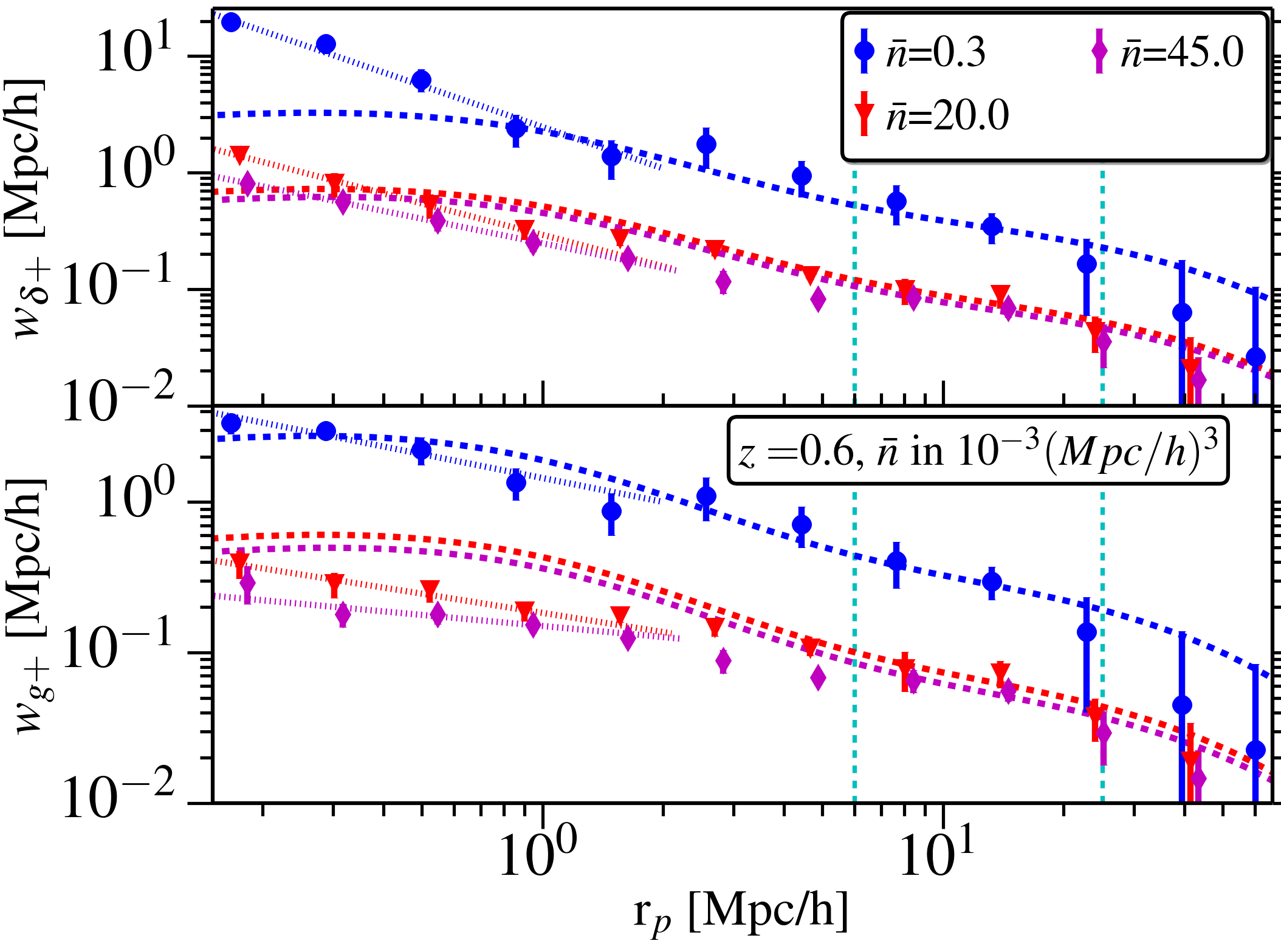}
			\caption{Intrinsic alignment correlation functions, $w_{\delta +}$ and $w_{g+}$, for different samples defined on the basis of comoving number density threshold, at redshift $z=0.6$. There is clear evolution with number density, where samples with lower number density and 
			hence more luminous galaxies have higher \ia{}. As discussed in main text, this has important implications for future weak lensing surveys such as Euclid and LSST.}
			\label{fig:lcomv_z0.6}
		\end{center}
	\end{figure}

		Table~\ref{tab:fit_results_lcomv} presents results for different samples defined by their comoving abundance. Fig.~\ref{fig:lcomv_z0.6} shows the intrinsic alignments signal for some of the samples at $z=0.6$. The \ia{} 
		amplitude generally 
		increases with decreasing comoving abundance, consistent with the fact that more massive and brighter objects have stronger \ia{}. 
		
\begin{figure}
		\begin{center}
			\includegraphics[width=1.0\columnwidth]{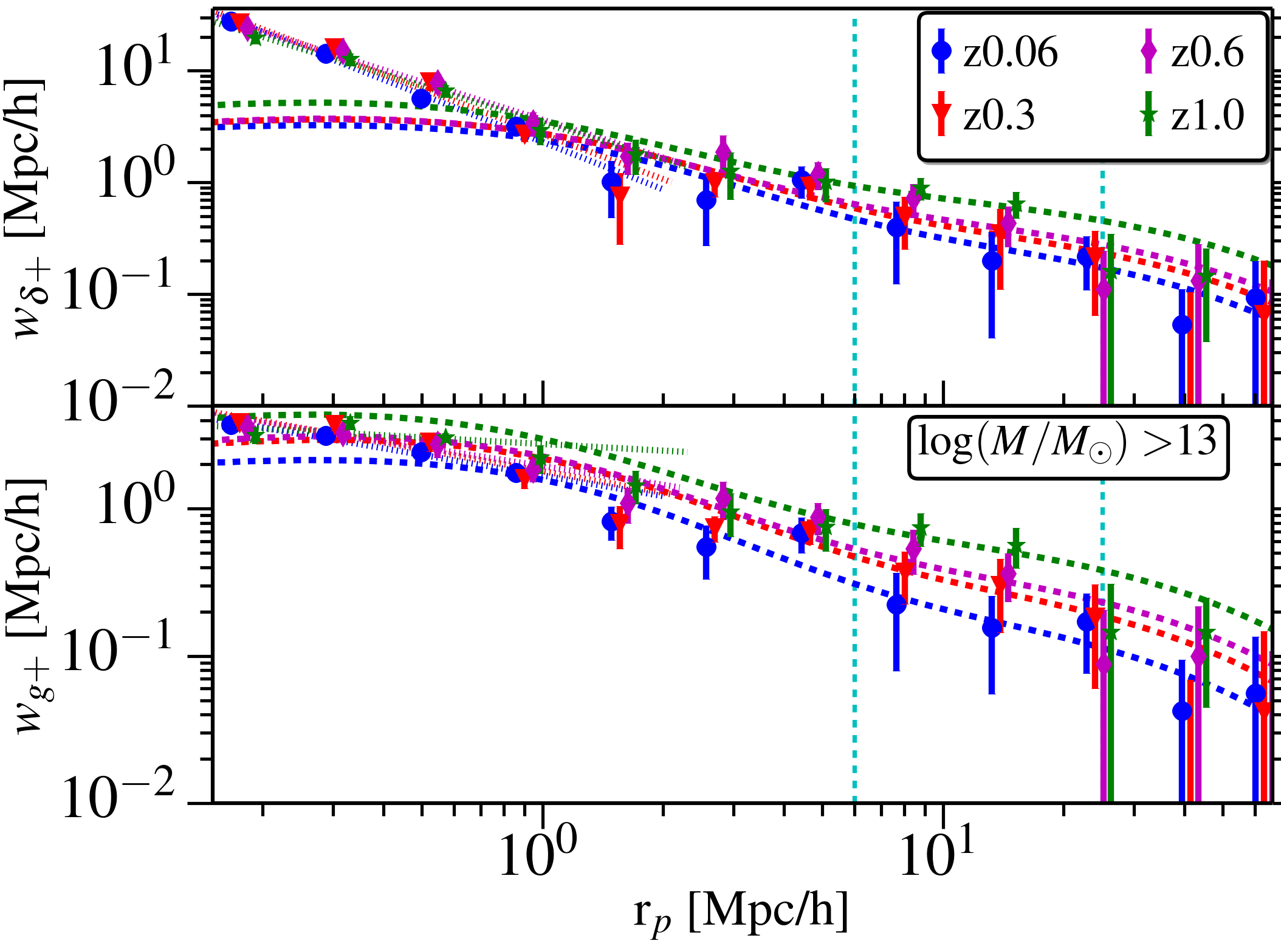}
			\caption{Intrinsic alignment correlation functions, $w_{\delta +}$ and $w_{g+}$, for the mass threshold sample, $M>10^{13}h^-1\Msun$, at redshifts, $z = 1.0$, $0.6$, $0.3$, and $0.06$. We see some redshift evolution as \wdp{} and \wgp{} magnitude increases at higher redshift.}
			\label{fig:M13_z_many}
		\end{center}
	\end{figure}

		Table~\ref{tab:fit_results_mass} presents results for different samples defined by subhalo mass threshold. Average subhalo mass are given for each sample. Fig.~\ref{fig:M13_z_many} shows the signal for $M>10^{13}
		h^{-1}\Msun$ sample at different redshift. Samples with more massive subhalos show stronger \ia{}, along with some redshift evolution as discussed in the main text.
		
\begin{figure}
		\begin{center}
			\includegraphics[width=1.0\columnwidth]{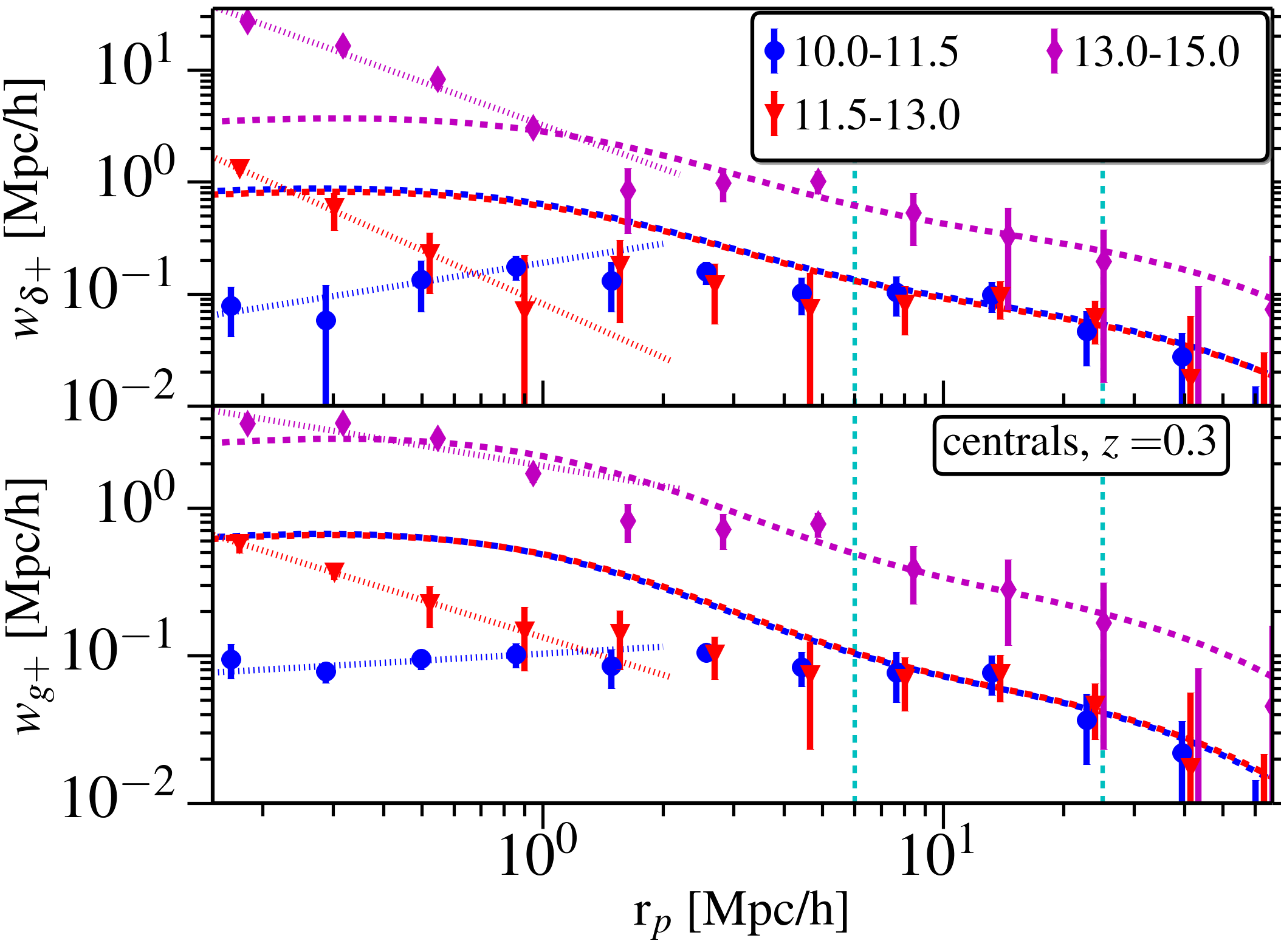}
			\caption{Intrinsic alignment correlation functions, $w_{\delta +}$ and $w_{g+}$, for central subhalos in different mass bins, at redshift $z=0.3$. We detect both large scale and small scale \ia{} for central sub halos, with more massive sub halos also showing stronger 
			alignments. The downturn in the lowest mass bin at small scales indicates a transition to the 1-halo term.}
			\label{fig:central_z0.3}
		\end{center}
	\end{figure}

\begin{figure}
		\begin{center}
			\includegraphics[width=1.0\columnwidth]{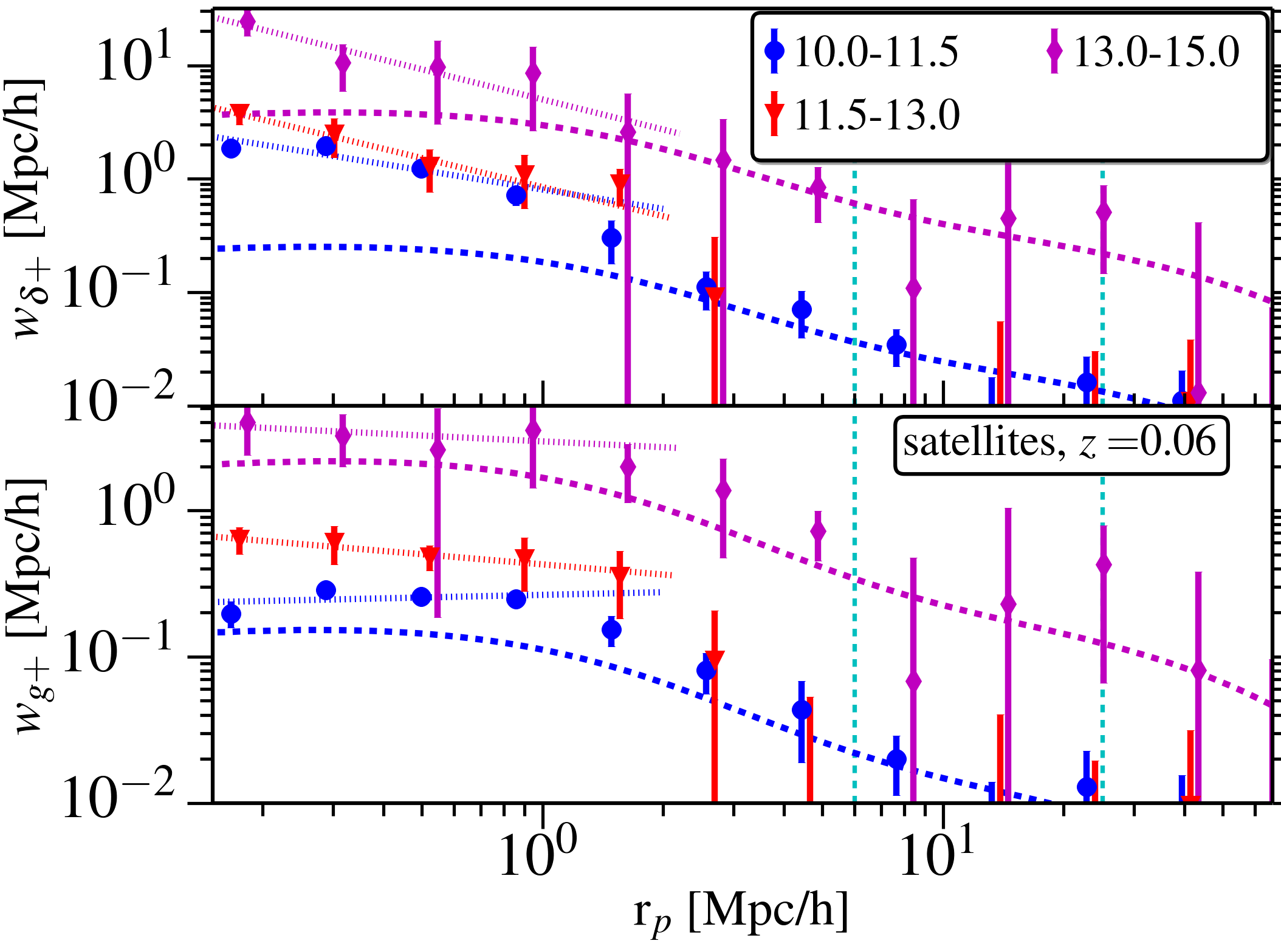}
			\caption{Intrinsic alignment correlation functions, $w_{\delta +}$ and $w_{g+}$, for satellite subhalos in different mass bins,, $M1$, $M2$ and $M3$, at redshift $z=0.3$. Satellites show no significant alignments at large scales, though small scale alignment is very strong, consistent with the radial 
			alignment of satellites.}
			\label{fig:sat_z0.06}
		\end{center}
	\end{figure}

		Table~\ref{tab:fit_results_central} and Table~\ref{tab:fit_results_sat} present results for satellite and central subhalos, with sample selection using 
		different mass thresholds. Fig.~\ref{fig:central_z0.3} and Fig.~\ref{fig:sat_z0.06} also show signal for some of the samples. We observe clear large scale alignments for 
		central subhalos, also with clear mass evolution. Satellite subhalos on the other hand show very little or no alignments at large scales with $A_I$ consistent with zero or at least much smaller than that for central 
		subhalos at the same redshift and in the same mass range. These results are consistent with the halo model, as satellites show radial alignments within the halo and hence their large scale signal is much weaker.
		
		Table~\ref{tab:fit_results_lum} presents results for samples defined by luminosity bins. We observe evolution of \ia{} with luminosity, with more luminous objects having stronger alignments and there is also some redshift evolution observed in two of the three luminosity bins. See section \ref{sec:fit_results} for more detailed discussion.
		\begin{table*}
			\caption{Model fits to samples defined by a  luminosity threshold, including all galaxies above some lower luminosity limit such that a given comoving abundance is achieved. $A_I$ is the NLA model amplitude, 
			$P_A$ and $P_I$ are the power law parameters. The power-law is fit separately to \wgp{} and \wdp{}, with superscript $\delta$ indicating the fits to \wdp{}. $\langle L /L_0\rangle$ gives average luminosity for the sample, normalized by pivot luminosity $L_0$, corresponding to r-band magnitude $M_{r,0}=-22$. 
			}
		\label{tab:fit_results_lcomv}
			\begin{tabular}{|c|c|c|c|c|c|c|c|}\hline
$\bar{n}$& $z$ & $A_I$ & $P_A$ & $P_I$ & $P_A^\delta$ & $P_I^\delta$ & $\langle L /L_0\rangle$ \\ \hline
1.0$\times10^{-4}$ & 0.06 & 18$\pm$2 & 2.8$\pm$0.2 & -0.22$\pm$0.07 & 4.6$\pm$0.5 & -1.29$\pm$0.08 & 7.8  \\ 
1.0$\times10^{-4}$ & 0.3 & 36$\pm$4 & 3$\pm$1 & -0.2$\pm$0.2 & 5$\pm$1 & -1.2$\pm$0.1 & 7.5  \\ 
1.0$\times10^{-4}$ & 0.6 & 30$\pm$10 & 2.8$\pm$0.2 & -0.31$\pm$0.07 & 4.9$\pm$0.6 & -1.11$\pm$0.08 & 7.0  \\ 
1.0$\times10^{-4}$ & 1.0 & 50$\pm$10 & 3.2$\pm$0.3 & -0.12$\pm$0.08 & 3.6$\pm$0.3 & -1.03$\pm$0.07 & 6.7  \\ 
3.0$\times10^{-4}$ & 0.06 & 9$\pm$2 & 1.9$\pm$0.2 & -0.35$\pm$0.09 & 2.8$\pm$0.2 & -1.25$\pm$0.05 & 4.1  \\ 
3.0$\times10^{-4}$ & 0.3 & 17$\pm$2 & 1.6$\pm$0.4 & -0.4$\pm$0.1 & 2.5$\pm$0.3 & -1.24$\pm$0.08 & 3.9  \\ 
3.0$\times10^{-4}$ & 0.6 & 21$\pm$3 & 1.4$\pm$0.2 & -0.5$\pm$0.1 & 2.5$\pm$0.5 & -1.2$\pm$0.1 & 3.8  \\ 
3.0$\times10^{-4}$ & 1.0 & 28$\pm$2 & 1.7$\pm$0.4 & -0.3$\pm$0.2 & 1.8$\pm$0.3 & -1.1$\pm$0.1 & 3.8  \\ 
1.0$\times10^{-3}$ & 0.06 & 8$\pm$2 & 0.89$\pm$0.03 & -0.48$\pm$0.03 & 1.4$\pm$0.2 & -1.18$\pm$0.07 & 1.8  \\ 
1.0$\times10^{-3}$ & 0.3 & 10$\pm$1 & 0.8$\pm$0.2 & -0.6$\pm$0.2 & 1.3$\pm$0.4 & -1.3$\pm$0.2 & 1.7  \\ 
1.0$\times10^{-3}$ & 0.6 & 13.9$\pm$0.3 & 0.8$\pm$0.1 & -0.6$\pm$0.1 & 1.4$\pm$0.2 & -1.09$\pm$0.07 & 1.8  \\ 
1.0$\times10^{-3}$ & 1.0 & 12$\pm$2 & 0.8$\pm$0.1 & -0.5$\pm$0.1 & 0.8$\pm$0.1 & -1.2$\pm$0.1 & 1.9  \\ 
2.0$\times10^{-2}$ & 0.06 & 2.6$\pm$0.3 & 0.21$\pm$0.01 & -0.23$\pm$0.05 & (37.0$\pm$0.9)$\times 10^{-2}$ & -0.92$\pm$0.02 & 1.5$\times10^{-1}$  \\ 
2.0$\times10^{-2}$ & 0.3 & 3.3$\pm$0.4 & 0.14$\pm$0.01 & 0.43$\pm$0.08 & 0.36$\pm$0.08 & -0.8$\pm$0.2 & 1.6$\times10^{-1}$  \\ 
2.0$\times10^{-2}$ & 0.6 & 4.7$\pm$0.5 & 0.18$\pm$0.01 & -0.42$\pm$0.06 & (28.0$\pm$0.9)$\times 10^{-2}$ & -0.9$\pm$0.02 & 1.8$\times10^{-1}$  \\ 
2.0$\times10^{-2}$ & 1.0 & 5.0$\pm$0.5 & (14.9$\pm$0.8)$\times 10^{-2}$ & -0.5$\pm$0.08 & 0.18$\pm$0.01 & -0.98$\pm$0.04 & 2.2$\times10^{-1}$  \\ 
4.5$\times10^{-2}$ & 0.06 & 2.7$\pm$0.2 & 0.19$\pm$0.01 & -0.04$\pm$0.05 & 0.36$\pm$0.02 & -0.66$\pm$0.04 & 7.1$\times10^{-2}$  \\ 
4.5$\times10^{-2}$ & 0.3 & 3.4$\pm$0.4 & 0.1$\pm$0.01 & 0.5$\pm$0.2 & 0.31$\pm$0.05 & -0.7$\pm$0.1 & 7.8$\times10^{-2}$  \\ 
4.5$\times10^{-2}$ & 0.6 & 4.0$\pm$0.3 & 0.15$\pm$0.01 & -0.2$\pm$0.1 & (23.2$\pm$0.6)$\times 10^{-2}$ & -0.7$\pm$0.02 & 8.8$\times10^{-2}$  \\ 
4.5$\times10^{-2}$ & 1.0 & 3.6$\pm$0.3 & 0.11$\pm$0.01 & -0.5$\pm$0.1 & 0.16$\pm$0.01 & -0.73$\pm$0.06 & 1.1$\times10^{-1}$  \\ 

				\hline 
 		\end{tabular}
		\end{table*}
		\begin{table*}
			\caption{Model fits to samples defined by mass threshold subhalo mass.$\langle M/h^{-1}\Msun\rangle$ is the average subhalo mass with in the sample. See Table~\ref{tab:fit_results_lcomv} for description of different parameters.}
			\label{tab:fit_results_mass}
			\begin{tabular}{|c|c|c|c|c|c|c|c|}\hline
$\log(M/h^{-1}\Msun)$ & $z$ & $A_I$ & $P_A$ & $P_I$ & $P_A^\delta$ & $P_I^\delta$ &  $\langle M/h^{-1}\Msun\rangle$ \\ \hline
$>$11 & 0.06 & 3.4$\pm$0.4 & 0.15$\pm$0.02 & -0.36$\pm$0.08 & 0.2$\pm$0.04 & -0.9$\pm$0.1 & 9.6$\times10^{11}$  \\ 
$>$11 & 0.3 & 4.0$\pm$0.3 & (14.8$\pm$0.4)$\times 10^{-2}$ & -0.49$\pm$0.02 & (26.0$\pm$0.3)$\times 10^{-2}$ & (-74.0$\pm$0.7)$\times 10^{-2}$ & 8.5$\times10^{11}$  \\ 
$>$11 & 0.6 & 4.9$\pm$0.3 & 0.14$\pm$0.02 & -0.5$\pm$0.1 & 0.16$\pm$0.02 & -0.98$\pm$0.08 & 7.2$\times10^{11}$  \\ 
$>$11 & 1.0 & 4.9$\pm$0.5 & (12.5$\pm$0.3)$\times 10^{-2}$ & -0.6$\pm$0.03 & (11.6$\pm$0.9)$\times 10^{-2}$ & -1.1$\pm$0.05 & 5.9$\times10^{11}$  \\ 
$>$12 & 0.06 & 6.7$\pm$0.8 & 0.36$\pm$0.02 & -0.68$\pm$0.03 & 0.4$\pm$0.05 & -1.4$\pm$0.08 & 5.7$\times10^{12}$  \\ 
$>$12 & 0.3 & 6.3$\pm$0.8 & 0.39$\pm$0.04 & -0.7$\pm$0.08 & 0.6$\pm$0.2 & -1.2$\pm$0.2 & 5.1$\times10^{12}$  \\ 
$>$12 & 0.6 & 7.7$\pm$0.7 & 0.41$\pm$0.02 & -0.64$\pm$0.03 & 0.53$\pm$0.04 & -1.22$\pm$0.05 & 4.3$\times10^{12}$  \\ 
$>$12 & 1.0 & 8$\pm$1 & 0.41$\pm$0.04 & -0.69$\pm$0.07 & 0.39$\pm$0.04 & -1.35$\pm$0.07 & 3.7$\times10^{12}$  \\ 
$>$13 & 0.06 & 12$\pm$2 & 1.68$\pm$0.08 & -0.47$\pm$0.04 & 2.3$\pm$0.3 & -1.4$\pm$0.1 & 3.3$\times10^{13}$  \\ 
$>$13 & 0.3 & 17.9$\pm$0.9 & 1.8$\pm$0.4 & -0.5$\pm$0.2 & 2.5$\pm$0.5 & -1.3$\pm$0.1 & 3.0$\times10^{13}$  \\ 
$>$13 & 0.6 & 24$\pm$7 & 1.9$\pm$0.2 & -0.42$\pm$0.09 & 3.3$\pm$0.4 & -1.13$\pm$0.09 & 2.6$\times10^{13}$  \\ 
$>$13 & 1.0 & 45$\pm$9 & 2.7$\pm$0.5 & -0.1$\pm$0.1 & 2.9$\pm$0.4 & -1.1$\pm$0.1 & 2.2$\times10^{13}$  \\  
				\hline 
 		\end{tabular}
		\end{table*}
		\begin{table*}
			\caption{Model fits to central galaxy intrinsic alignment correlation functions.$\langle M/h^{-1}\Msun\rangle$ is the average subhalo mass with in the sample. See Table~\ref{tab:fit_results_lcomv} for description of different parameters.}
			\label{tab:fit_results_central}
			\begin{tabular}{|c|c|c|c|c|c|c|c|}\hline
$\log(M/h^{-1}M_\odot)$ & $z$ & $A_I$ & $P_A$ & $P_I$ & $P_A^\delta$ & $P_I^\delta$ & $\langle M/h^{-1}\Msun\rangle$\\ \hline
$\in [10.0,11.5]$ & 0.06 & 3.7$\pm$0.4 & 0.09$\pm$0.02 & 0.1$\pm$0.2 & 0.17$\pm$0.04 & 1.9$\pm$0.9 & 1.3$\times10^{11}$  \\ 
$\in [10.0,11.5]$ & 0.3 & 4.2$\pm$0.5 & 0.1$\pm$0.01 & 0.2$\pm$0.1 & 0.19$\pm$0.02 & 0.6$\pm$0.2 & 1.3$\times10^{11}$  \\ 
$\in [10.0,11.5]$ & 0.6 & 4.7$\pm$0.3 & (10.6$\pm$0.6)$\times 10^{-2}$ & -0.1$\pm$0.06 & 0.14$\pm$0.03 & (-0.0$\pm$0.1)$\times 10^{0}$ & 1.2$\times10^{11}$  \\ 
$\in [10.0,11.5]$ & 1.0 & 4.1$\pm$0.4 & (8.3$\pm$0.7)$\times 10^{-2}$ & -0.2$\pm$0.1 & 0.07$\pm$0.01 & -0.4$\pm$0.1 & 1.1$\times10^{11}$  \\ 
$\in [11.5,13.0]$ & 0.06 & 3.1$\pm$0.8 & 0.14$\pm$0.02 & -0.7$\pm$0.1 & 0.03$\pm$0.03 & -2.1$\pm$0.6 & 1.3$\times10^{12}$  \\ 
$\in [11.5,13.0]$ & 0.3 & 4.0$\pm$0.6 & (12.7$\pm$0.3)$\times 10^{-2}$ & -0.84$\pm$0.01 & (7.5$\pm$0.9)$\times 10^{-2}$ & -1.59$\pm$0.07 & 1.2$\times10^{12}$  \\ 
$\in [11.5,13.0]$ & 0.6 & 5.8$\pm$0.5 & 0.13$\pm$0.03 & -0.7$\pm$0.2 & 0.03$\pm$0.02 & -2.1$\pm$0.3 & 1.2$\times10^{12}$  \\ 
$\in [11.5,13.0]$ & 1.0 & 5.5$\pm$0.8 & 0.15$\pm$0.02 & -0.82$\pm$0.08 & 0.1$\pm$0.01 & -1.48$\pm$0.07 & 1.1$\times10^{12}$  \\ 
$\in [13.0,15.0]$ & 0.06 & 11$\pm$2 & 1.6$\pm$0.1 & -0.48$\pm$0.07 & 2.1$\pm$0.4 & -1.5$\pm$0.1 & 3.4$\times10^{13}$  \\ 
$\in [13.0,15.0]$ & 0.3 & 18$\pm$1 & 1.8$\pm$0.4 & -0.5$\pm$0.2 & 2.8$\pm$0.5 & -1.3$\pm$0.1 & 3.1$\times10^{13}$  \\ 
$\in [13.0,15.0]$ & 0.6 & 24$\pm$5 & 1.7$\pm$0.3 & -0.5$\pm$0.1 & 2.7$\pm$0.4 & -1.2$\pm$0.1 & 2.6$\times10^{13}$  \\ 
$\in [13.0,15.0]$ & 1.0 & 49$\pm$9 & 2.6$\pm$0.4 & -0.2$\pm$0.2 & 2.8$\pm$0.4 & -1.1$\pm$0.1 & 2.2$\times10^{13}$  \\ 
				\hline 
 		\end{tabular}
		\end{table*}
		\begin{table*}
			\caption{Model fits to satellite galaxy intrinsic alignment correlation functions.$\langle M/h^{-1}\Msun\rangle$ is the average subhalo mass with in the sample. See Table~\ref{tab:fit_results_lcomv} for description of different parameters.}
			\label{tab:fit_results_sat}
			\begin{tabular}{|c|c|c|c|c|c|c|c|}\hline
$\log(M/h^{-1}\Msun)$ & $z$ & $A_I$ & $P_A$ & $P_I$ & $P_A^\delta$ & $P_I^\delta$ & $\langle M/h^{-1}\Msun\rangle$ \\ \hline
$\in [10.0,11.5]$ & 0.06 & 0.9$\pm$0.3 & 0.26$\pm$0.03 & 0.1$\pm$0.1 & 0.8$\pm$0.2 & -0.6$\pm$0.2 & 6.5$\times10^{10}$  \\ 
$\in [10.0,11.5]$ & 0.3 & 1.7$\pm$0.1 & 0.21$\pm$0.03 & -0.25$\pm$0.09 & 0.6$\pm$0.1 & -0.6$\pm$0.1 & 6.2$\times10^{10}$  \\ 
$\in [10.0,11.5]$ & 0.6 & 1.6$\pm$0.2 & 0.18$\pm$0.03 & -0.1$\pm$0.1 & 0.34$\pm$0.07 & -0.6$\pm$0.1 & 6.1$\times10^{10}$  \\ 
$\in [10.0,11.5]$ & 1.0 & 2.0$\pm$0.4 & 0.14$\pm$0.01 & -0.3$\pm$0.1 & 0.3$\pm$0.07 & -0.6$\pm$0.1 & 5.7$\times10^{10}$  \\ 
$\in [11.5,13.0]$ & 0.06 & -1$\pm$1 & 0.42$\pm$0.03 & -0.23$\pm$0.05 & 0.8$\pm$0.1 & -0.8$\pm$0.1 & 1.0$\times10^{12}$  \\ 
$\in [11.5,13.0]$ & 0.3 & 4.1$\pm$0.7 & (32.1$\pm$0.8)$\times 10^{-2}$ & -0.57$\pm$0.02 & 0.9$\pm$0.2 & -0.8$\pm$0.1 & 1.0$\times10^{12}$  \\ 
$\in [11.5,13.0]$ & 0.6 & 3$\pm$1 & 0.66$\pm$0.06 & -0.14$\pm$0.07 & 1.5$\pm$0.2 & -0.4$\pm$0.1 & 9.9$\times10^{11}$  \\ 
$\in [11.5,13.0]$ & 1.0 & (0.0$\pm$0.3)$\times 10^{1}$ & 0.37$\pm$0.05 & -0.2$\pm$0.1 & 0.7$\pm$0.2 & -0.7$\pm$0.2 & 9.0$\times10^{11}$  \\ 
$\in [13.0,15.0]$ & 0.06 & 10$\pm$10 & 2.9$\pm$0.6 & -0.1$\pm$0.1 & 5$\pm$2 & -0.9$\pm$0.3 & 2.2$\times10^{13}$  \\ 
$\in [13.0,15.0]$ & 0.3 & 20$\pm$20 & (0.0$\pm$0.2)$\times 10^{0}$ & -3$\pm$2 & (0.0$\pm$0.3)$\times 10^{7}$ & (0.0$\pm$0.2)$\times 10^{6}$ & 1.7$\times10^{13}$  \\ 
$\in [13.0,15.0]$ & 0.6 & 40$\pm$30 & 5$\pm$1 & 0.8$\pm$0.5 & 11$\pm$2 & -0.3$\pm$0.1 & 2.3$\times10^{13}$  \\ 
$\in [13.0,15.0]$ & 1.0 & -40$\pm$20 & 3$\pm$1 & 0.2$\pm$0.6 & 6.3$\pm$0.3 & -0.71$\pm$0.05 & 2.1$\times10^{13}$  \\ 
				\hline 
	 		\end{tabular}
		\end{table*}
		
		\begin{table*}
			\caption{Model fits to \ia{} measurements for samples defined in luminosity bins. See Table~\ref{tab:fit_results_lcomv} for description of different parameters.}
			\label{tab:fit_results_lum}
			\begin{tabular}{|c|c|c|c|c|c|c|c|}\hline
$M_r$ & $z$ & $A_I$ & $P_A$ & $P_I$ & $P_A^\delta$ & $P_I^\delta$ & $\langle L/L_0\rangle$ \\ \hline
$\leq -22.6$ & 0.06 & 12$\pm$2 & 2.0$\pm$0.2 & -0.36$\pm$0.09 & 3.1$\pm$0.3 & -1.29$\pm$0.07 & 4.5  \\ 
$\leq -22.6$ & 0.3 & 20$\pm$2 & 1.9$\pm$0.4 & -0.3$\pm$0.1 & 3.0$\pm$0.3 & -1.17$\pm$0.07 & 4.4  \\ 
$\leq -22.6$ & 0.6 & 24$\pm$4 & 1.6$\pm$0.2 & -0.5$\pm$0.1 & 2.7$\pm$0.5 & -1.2$\pm$0.1 & 4.1  \\ 
$\leq -22.6$ & 1.0 & 27$\pm$2 & 1.7$\pm$0.4 & -0.3$\pm$0.2 & 1.9$\pm$0.3 & -1.1$\pm$0.1 & 3.8  \\ 
$\in [-22.6,-20.3]$ & 0.06 & 6$\pm$1 & 0.37$\pm$0.02 & -0.66$\pm$0.04 & 0.4$\pm$0.1 & -1.5$\pm$0.2 & 5.2$\times10^{-1}$  \\ 
$\in [-22.6,-20.3]$ & 0.3 & 6$\pm$1 & 0.32$\pm$0.04 & -0.7$\pm$0.1 & 0.5$\pm$0.1 & -1.2$\pm$0.2 & 4.9$\times10^{-1}$  \\ 
$\in [-22.6,-20.3]$ & 0.6 & 5.6$\pm$0.2 & 0.31$\pm$0.01 & -0.62$\pm$0.02 & 0.49$\pm$0.02 & -1.08$\pm$0.03 & 4.9$\times10^{-1}$  \\ 
$\in [-22.6,-20.3]$ & 1.0 & 4.6$\pm$0.5 & 0.21$\pm$0.01 & -0.73$\pm$0.04 & 0.2$\pm$0.02 & -1.32$\pm$0.06 & 5.0$\times10^{-1}$  \\ 
$\in [-20.3,-18.0]$ & 0.06 & 1.5$\pm$0.3 & (14.6$\pm$0.5)$\times 10^{-2}$ & -0.32$\pm$0.03 & 0.34$\pm$0.05 & -0.72$\pm$0.09 & 6.6$\times10^{-2}$  \\ 
$\in [-20.3,-18.0]$ & 0.3 & 2.9$\pm$0.3 & 0.16$\pm$0.01 & -0.42$\pm$0.07 & 0.29$\pm$0.05 & -0.8$\pm$0.1 & 6.6$\times10^{-2}$  \\ 
$\in [-20.3,-18.0]$ & 0.6 & 4.2$\pm$0.5 & 0.13$\pm$0.02 & -0.3$\pm$0.1 & 0.18$\pm$0.03 & -0.7$\pm$0.1 & 6.5$\times10^{-2}$  \\ 
$\in [-20.3,-18.0]$ & 1.0 & 3.8$\pm$0.3 & (11.0$\pm$0.6)$\times 10^{-2}$ & -0.46$\pm$0.05 & 0.15$\pm$0.02 & -0.71$\pm$0.07 & 6.3$\times10^{-2}$  \\
			\hline
	 		\end{tabular}
		\end{table*}


\end{document}